\documentclass{sig-alternate-05-2015}
\newtheorem{thm}{Theorem}

\newtheorem{lemma}{Lemma}
\newtheorem{mydef}{Definition}
\DeclareMathOperator*{\argmin}{arg\,min}
\DeclareMathOperator*{\argmax}{arg\,max}

\usepackage{subfig}
\usepackage{algorithm}
\usepackage{algorithmic}
\usepackage{graphicx}
\usepackage{url}
\usepackage{multirow}

\begin{document}
%
% --- Author Metadata here ---
%\conferenceinfo{KDD'}{'16, August 13--17, 2016, San Francisco, CA USA}
%\CopyrightYear{2007} % Allows default copyright year (20XX) to be over-ridden - IF NEED BE.
%\crdata{0-12345-67-8/90/01}  % Allows default copyright data (0-89791-88-6/97/05) to be over-ridden - IF NEED BE.
% --- End of Author Metadata ---

\title{Spectral Algorithms for Temporal Graph Cuts}

\numberofauthors{3} %  in this sample file, there are a *total*
\author{
\alignauthor Arlei Silva\\
\affaddr{University of California}\\
\affaddr{Santa Barbara, CA, USA}\\
\email{arlei@cs.ucsb.edu}
\alignauthor Ambuj Singh\\
\affaddr{University of California}\\
\affaddr{Santa Barbara, CA, USA}\\
\email{ambuj@cs.ucsb.edu}
\alignauthor Ananthram Swami\\
\affaddr{Army Research Laboratory}\\
\affaddr{Adelphi, MD, USA}\\
\email{ananthram.swami.civ@mail.mil}
% 1st. author
% There's nothing stopping you putting the seventh, eighth, etc.
% author on the opening page (as the 'third row') but we ask,
% for aesthetic reasons that you place these 'additional authors'
% in the \additional authors block, viz.
% Just remember to make sure that the TOTAL number of authors
% is the number that will appear on the first page PLUS the
% number that will appear in the \additionalauthors section.
}
\maketitle

\maketitle`
\begin{abstract}
The sparsest cut problem consists of identifying a small set of edges that breaks the graph into balanced sets of vertices. The normalized cut problem balances the total degree, instead of the size, of the resulting sets. Applications of graph cuts include community detection and computer vision. However, cut problems were originally proposed for static graphs, an assumption that does not hold in many modern applications where graphs are highly dynamic. In this paper, we introduce the sparsest and normalized cut problems in temporal graphs, which generalize their standard definitions by enforcing the smoothness of cuts over time. We propose novel formulations and algorithms for computing temporal cuts using spectral graph theory, multiplex graphs, divide-and-conquer and low-rank matrix approximation. Furthermore, we extend our formulation to dynamic graph signals, where cuts also capture node values, as graph wavelets. Experiments show that our solutions are accurate and scalable, enabling the discovery of dynamic communities and the analysis of dynamic graph processes. \\
\textbf{Categories and Subject Descriptors:} H.2.8 [\textbf{Database Management}]: Database applications $-$ \textit{data mining} \\
\textbf{General Terms:} Algorithms, Experimentation 
\end{abstract}

\section{Introduction}
\label{sec::introduction}

Temporal graphs represent how a graph changes over time, being ubiquitous in data mining applications. Users in social networks present a dynamic behavior, leading to the evolution of communities \cite{backstrom2006group}. In hyperlinked environments, such as blogs, the rise of new topics of interest drive modifications in content and link structure \cite{kumar2005bursty}. Communication, epidemics and mobility are other scenarios where temporal graphs can enable the understanding of complex dynamic processes. 
However, several key concepts and algorithms for static graphs have not been generalized to temporal graphs. 

This paper focuses on cut problems in temporal graphs, which consist of finding a small sets of edges (or cuts) that break the graph into balanced sets of vertices. Two traditional graph cut problems are the \textit{sparsest cut} \cite{leighton1988approximate,hagen1992new} and the \textit{normalized cut} \cite{shi2000normalized,chung1997spectral}. In sparsest cuts, the resulting partitions are balanced in terms of size, while in normalized cuts, the balance is in terms of total degree (or volume) of the resulting sets. Graph cuts have applications in community detection, image segmentation, clustering, and VLSI design. Moreover, the computation of graph cuts based on eigenvectors of graph-related matrices is one of the earliest results in spectral graph theory \cite{chung1997spectral}, a subject with great impact in information retrieval \cite{langville2011google}, graph sparsification \cite{spielman2011spectral}, and machine learning \cite{lafferty2005diffusion}. It is rather surprising that there are no existing algorithms for cuts in temporal graphs. 

One of our motivations to study graph cuts in this new setting is the emerging field of \textit{Signal Processing on Graphs (SPG)} \cite{shuman2013emerging,sandryhaila2014big}. SPG is a framework for the analysis of data residing on vertices of a graph, as a generalization of traditional signal processing. %In temporal graphs, signal processing can support the identification of dynamic processes based on vertex attributes (e.g. traffic congestions or events driving Wikipedia access). %The most popular approach in SPG, known as Graph Fourier Transform, represents graph signals as linear combinations of eigenvectors of the graph Laplacian \cite{shuman2013emerging}. 
In a recent work \cite{silva2016graph}, the authors proposed a data-driven partitioning-based scheme for SPG, highlighting connections with sparse cuts. Temporal cuts can generalize these results to dynamic graphs.

\textbf{Contributions of this paper.} We propose formulations of sparsest and normalized cuts in a sequence of graph snapshots. The idea is to extend classical definitions of these problems while enforcing the smoothness (or stability) of cuts over time. Our formulations can be understood using a multiplex view of the temporal graph, where temporal edges connect the same vertex in different snapshots (see Figure \ref{fig::temporal_graph_and_multiplex_view}). Therefore, both sparsity and smoothness are represented in the same space, as edges across partitions (see Figure \ref{fig::example_alternative_temporal_cuts}).

\begin{figure}
\centering
\subfloat[Time I]
{
\includegraphics[keepaspectratio, width=0.15\textwidth,trim={4.5cm 3cm 6cm 2.5cm},clip]{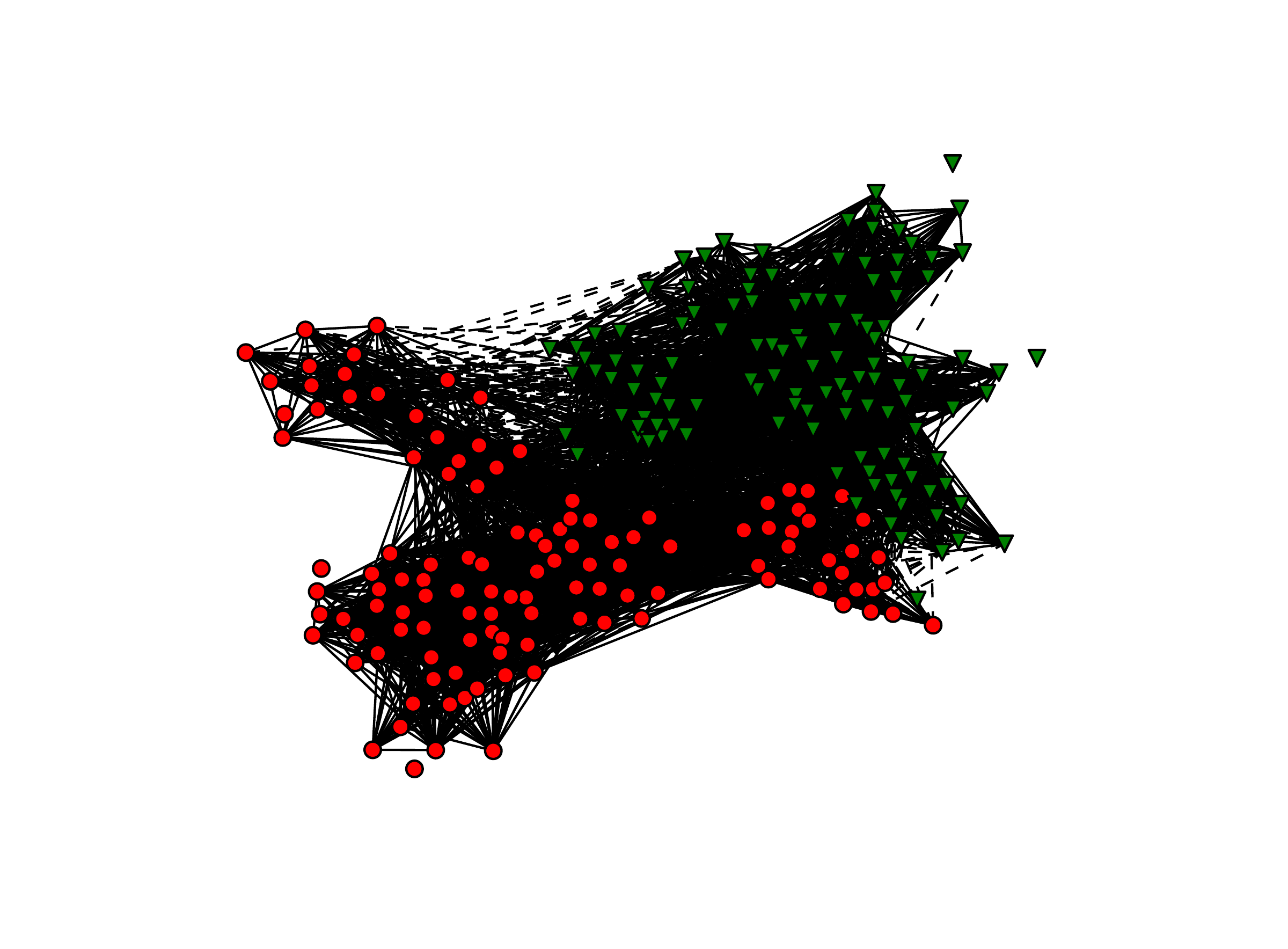}
}
\subfloat[Time II]
{
\includegraphics[keepaspectratio, width=0.15\textwidth,trim={4.5cm 3cm 6cm 2.5cm},clip]{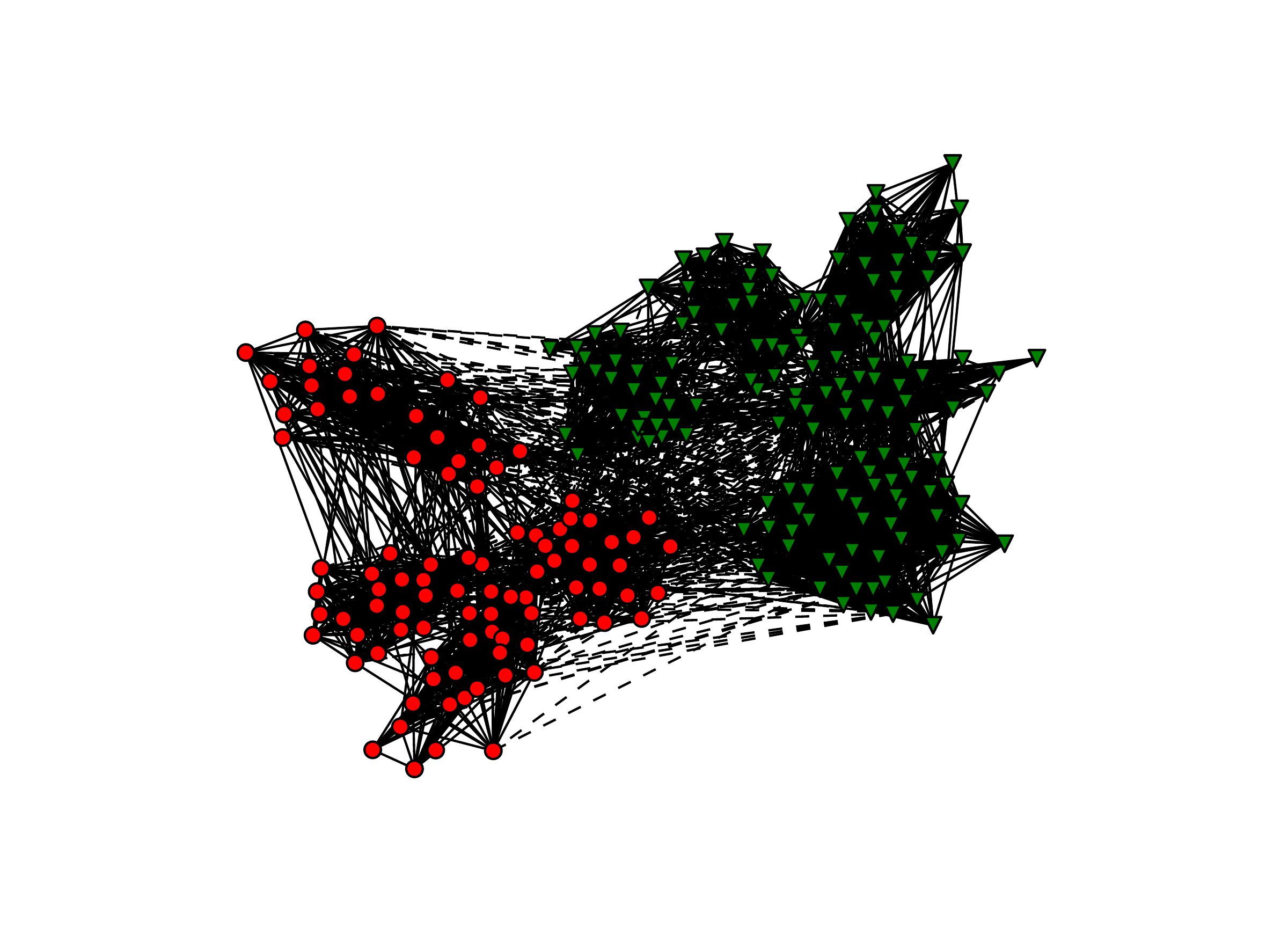}
}
\subfloat[Time III]
{
\includegraphics[keepaspectratio, width=0.15\textwidth,trim={4.5cm 3cm 6cm 2.5cm},clip]{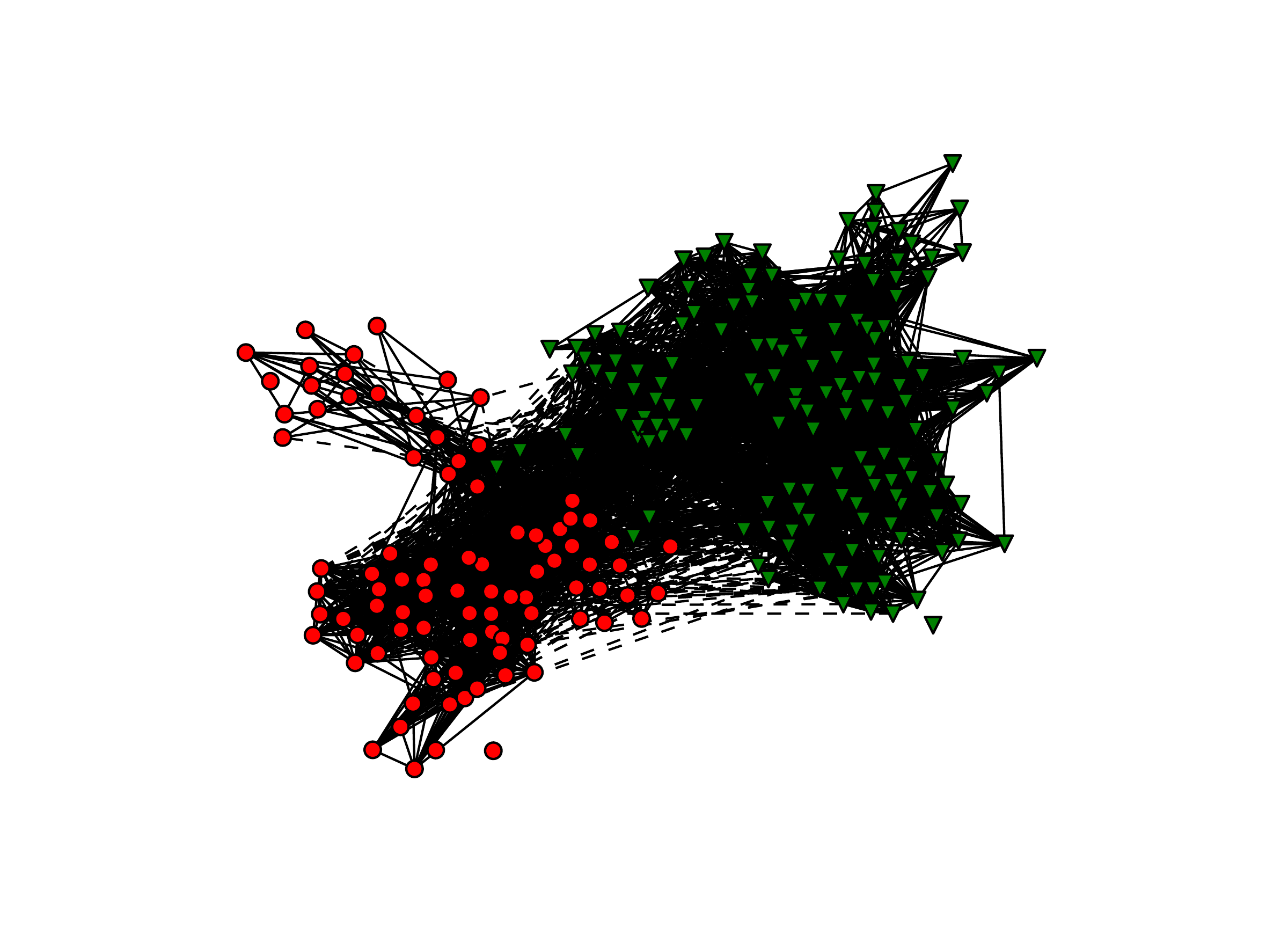}
}
\caption{Temporal graph cut for a primary school network. The cut, represented as node colors, reflects the network dynamics, capturing major changes in the children's interactions. Figure is better seen in color. \label{fig::example_primary_school}}
\end{figure}

Figure \ref{fig::example_primary_school} shows a sparse temporal graph cut for a primary school network~\cite{stehle2011high}, where (6-12 years old) children from a French school are connected based on proximity measured by sensors. The vertices are naturally organized into communities resulting from classes/ages. However, there is a significant amount of interaction across classes (e.g. during lunch). We have divided the total duration of the experiment (176 minutes) into three equal parts. Major changes in the contact network can be noticed during the experiment, causing several vertices to move across partitions. The temporal cut is able to capture such trends while keeping the remaining node assignments mostly unchanged.

Traditional spectral solutions---which compute approximated cuts as rounded eigenvectors of the Laplacian matrix---do not generalize to our setting. Thus, we propose new algorithms, still within the framework of spectral graph theory, for the computation of temporal cuts. We further exploit important properties of this formulation to design efficient approximation algorithms for temporal cuts combining \textit{divide-and-conquer} and low-rank matrix approximation.

In order to model not only structural changes, but also dynamic data embedded on graphs, we apply temporal cuts as data-driven wavelet bases for graph signals. Our approach exploits smoothness in both space and time, illustrating how the techniques presented in this paper provide a powerful and general framework for the analysis of dynamic graphs.  

\textbf{Summary of contributions.} %The main contributions of this paper can be summarized as follows:
\begin{itemize}
\item We generalize sparsest and normalized cuts to temporal graphs; to model graph signals, we further extend temporal cuts as dynamic graph wavelets.
\item We formulate temporal cuts using spectral graph theory and propose efficient approximate solutions via divide-and-conquer and low-rank approximation. 
\item We provide an extensive evaluation of our proposed algorithms for temporal cuts, including applications in community detection and signal processing on graphs.
\end{itemize}

\textbf{Related Work.} Computing graph cuts is a traditional problem in graph theory \cite{leighton1988approximate,arora2009expander}. From a practical standpoint, graph cuts find applications in a diverse set of problems, ranging from image segmentation \cite{shi2000normalized} to community detection \cite{leskovec2009community}. This paper is focused on the sparsest and normalized cut problems, which are of particular interest due to their connections with the spectrum of the Laplacian matrix, mixing time of random walks, geometric embeddings, effective resistance, and graph expanders \cite{hagen1992new,ghosh2014interplay,spielman2004nearly,chung1997spectral}.

Community detection in temporal graphs has attracted great interest in recent years \cite{greene2010tracking,fortunato2010community}. An evolutionary spectral clustering technique was proposed in \cite{chi2007evolutionary}. The idea is to minimize a cost function in the form $\alpha.CS+\beta.CT$, where $CS$ is a snapshot cost and $CT$ is a temporal cost. FacetNet \cite{lin2008facetnet} and estrangement \cite{kawadia2012sequential} apply a similar approach under different clustering models. An important limitation of these solutions is that they perform community assignments in a step-wise manner, being highly subject to local optima. Another related problem is incremental clustering \cite{charikar1997incremental}, including spectral methods \cite{ning2010incremental}. However, in  these methods the main goal is to avoid recomputation in the streaming setting, and not to capture long-term structural changes.

A formulation for temporal modularity that simultaneously clusters multiple snapshots using a multiplex graph is proposed in \cite{bazzi2016community}. A similar idea was applied in \cite{taylor2015eigenvector} to generalize eigenvector centrality. In this paper, we propose generalizations for cut problems across time by studying spectral properties of multiplex graphs \cite{sole2013spectral}. As one of our contributions, we exploit the link between multiplex graphs and block tridiagonal matrices in the design of algorithms to efficiently approximate temporal cuts. Our approach follows the \textit{divide-and-conquer} paradigm and builds upon more general results in numerical computing \cite{cuppen1980divide,gansterer2003computing}.

Signal processing on graphs \cite{shuman2013emerging,sandryhaila2014big} is an interesting application of temporal cuts. Traditional signal processing operations are also relevant when the signal is embedded into sparse irregular spaces. For instance, in machine learning, object similarity can be represented as a graph and labels as signals to solve semi-supervised learning tasks \cite{gavish10,gadde2014active,chen2015discrete,bronstein2016geometric}. We generalize a previous work on computing data-driven graph wavelets to dynamic graphs \cite{silva2016graph}.

\section{Temporal Graph Cuts}

\subsection{Definitions}
\label{sec::definitions}

A temporal graph $\mathcal{G}$ is a sequence of $m$ graph snapshots $\langle G_1, G_2, \ldots G_m\rangle$ where $G_t$ is the snapshot at timestamp $t$. Each $G_t$ is a tuple $(V,E_t,W_t)$ where $V$ is a fixed set with $n$ vertices, $E_t$ is a dynamic set of undirected edges and $W_t:E_t\to \mathbb{R}$ is an edge weighting function. Figure \ref{fig::example_temporal_graph} shows an example of a temporal graph with two snapshots where edge weights are all equal to 1 and hence omitted.

We model temporal graphs as \textit{multiplex graphs}, which connect different graph layers. We denote as $\chi(\mathcal{G})=(\mathcal{V,E,W})$ the (undirected) multiplex view of $\mathcal{G}$, where $\mathcal{V}=\{v_t|v\in V \wedge t \in [1,m]\}$ ($|\mathcal{V}|=nm$) and $\mathcal{E}=E_1 \cup \ldots E_t \cup \{(v_t,v_{t+1})|v\in V \wedge t \in [1,m-1]\}$. Thus the edge set $\mathcal{E}$ also includes 'vertical' edges between nodes $v_t$ and $v_{t+1}$
The edge weighting function $\mathcal{W}:\mathcal{E}\to \mathbb{R}$ is defined as follows:

\begin{equation}
\mathcal{W}(u_r,v_s)=
\begin{cases}
 W_t(u,v), & \text{if}\ u,v \in V_t\\
\beta, & \text{if}\ u=v \wedge r=s-1\\
0,& \text{otherwise}
\end{cases}
\end{equation}

As a result, each vertex $v \in V$ has $m$ corresponding representatives $\{v_1, \ldots v_m\}$ in $\chi(\mathcal{G})$. Besides the intra-layer edges corresponding to the connectivity of each snapshot ($E_t$), temporal edges $(v_t,v_{t+1})$ connect consecutive versions of a vertex $v$ at different layers---which is known as \textit{diagonal coupling} \cite{bazzi2016community}. Intra-layer edge weights are the same as in $\mathcal{G}$ while inter-layer weights are set to $\beta$. Figure \ref{fig::example_multiplex_graph} shows the multiplex view of the graph from Figure \ref{fig::example_temporal_graph} for $\beta=1$. 

\subsubsection{Sparsest Cut}

A graph cut $(X,\overline{X})$ divides $V$ into two disjoint sets: $X \subseteq V$ and $\overline{X}=V-X$. We denote the weight of a cut $ |(X,\overline{X})|=\sum_{v\in X, u\in \overline{X}}W((u,v))$. The \textit{cut sparsity} $\sigma$ is the ratio of the cut weight and the product of the sizes of the sets \cite{leighton1988approximate}:

\begin{equation}
\sigma(X) = \frac{|(X,\overline{X})|}{|X||\overline{X}|}
\label{eqn::cut_ratio_single_graph}
\end{equation}

Here, we extend the notion of cut sparsity to temporal graphs. A temporal cut $\langle(X_1,\overline{X}_1), \ldots (X_m,\overline{X}_m)\rangle$ is a sequence of graph cuts where $(X_t,\overline{X}_t)$ is a cut of the graph snapshot $G_t$. The idea is that in temporal graphs, besides the cut weights and partition sizes, we also care about the smoothness (i.e. stability) of the cuts over time. We formalize the temporal cut sparsity $\sigma$ as follows:

\begin{equation}
\sigma(X_1, \ldots X_m;\beta) = \frac{\sum_{t=1}^m |(X_t,\overline{X}_t)|+\sum_{t=1}^{m-1}\Delta(X_t,\overline{X}_{t+1})}{\sum_{t=1}^m |X_t||\overline{X}_t|}
\label{eqn::cut_ratio_temporal_graph}
\end{equation}
where $\Delta(X,\overline{X}) = \beta|(X_t,\overline{X}_{t+1})|$ is the number of vertices that move from $X_t$ to $\overline{X}_{t+1}$ times the constant $\beta$, which allows different weights to be given to the cut smoothness. %We assume that the temporal graph is connected, which means that its temporal cut ratio is always non-zero.

\begin{figure}
\centering
\subfloat[Temporal graph \label{fig::example_temporal_graph}]{
\includegraphics[keepaspectratio, width=0.15\textwidth]{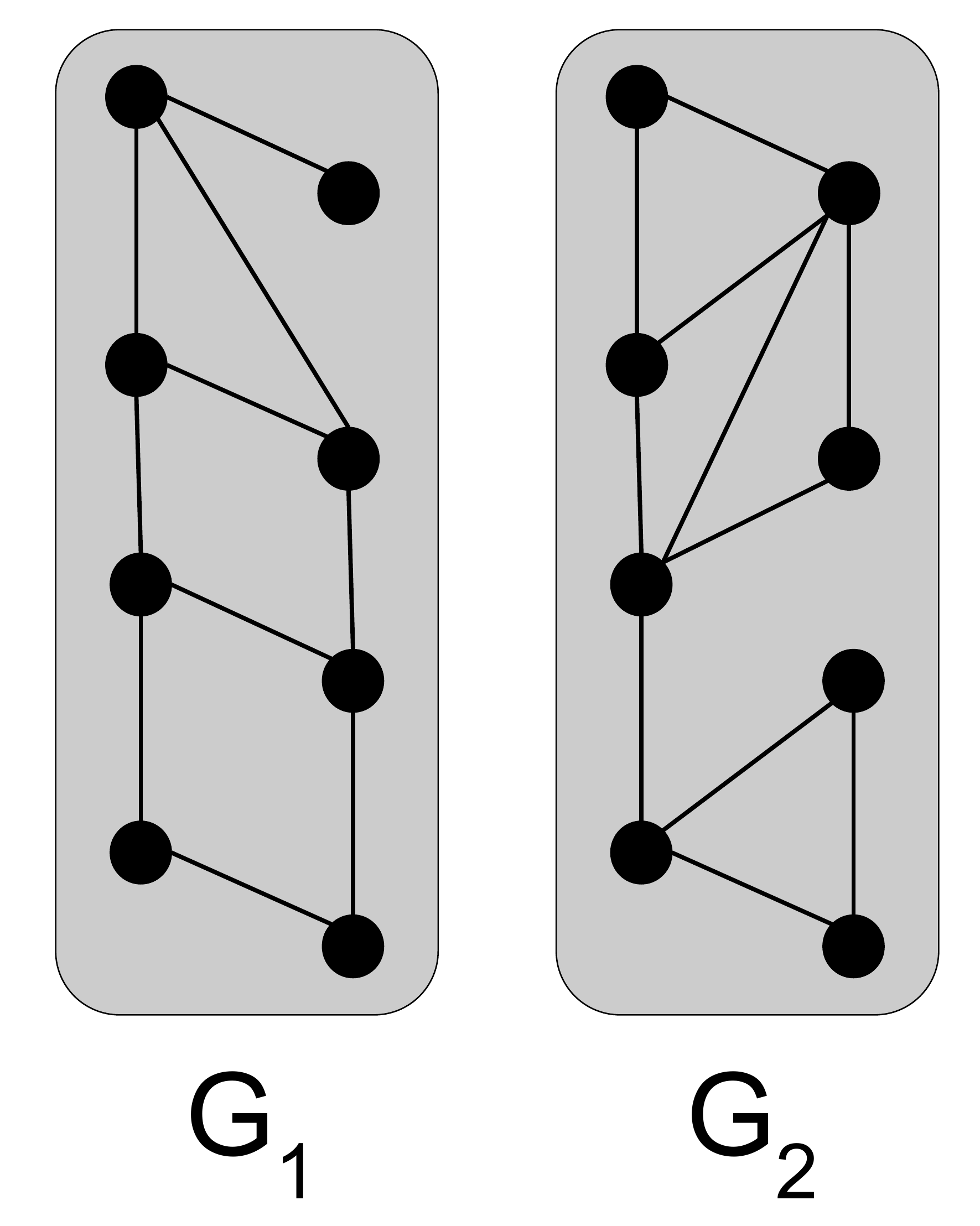}
}
\subfloat[Multiplex view \label{fig::example_multiplex_graph}]{
\includegraphics[keepaspectratio, width=0.15\textwidth]{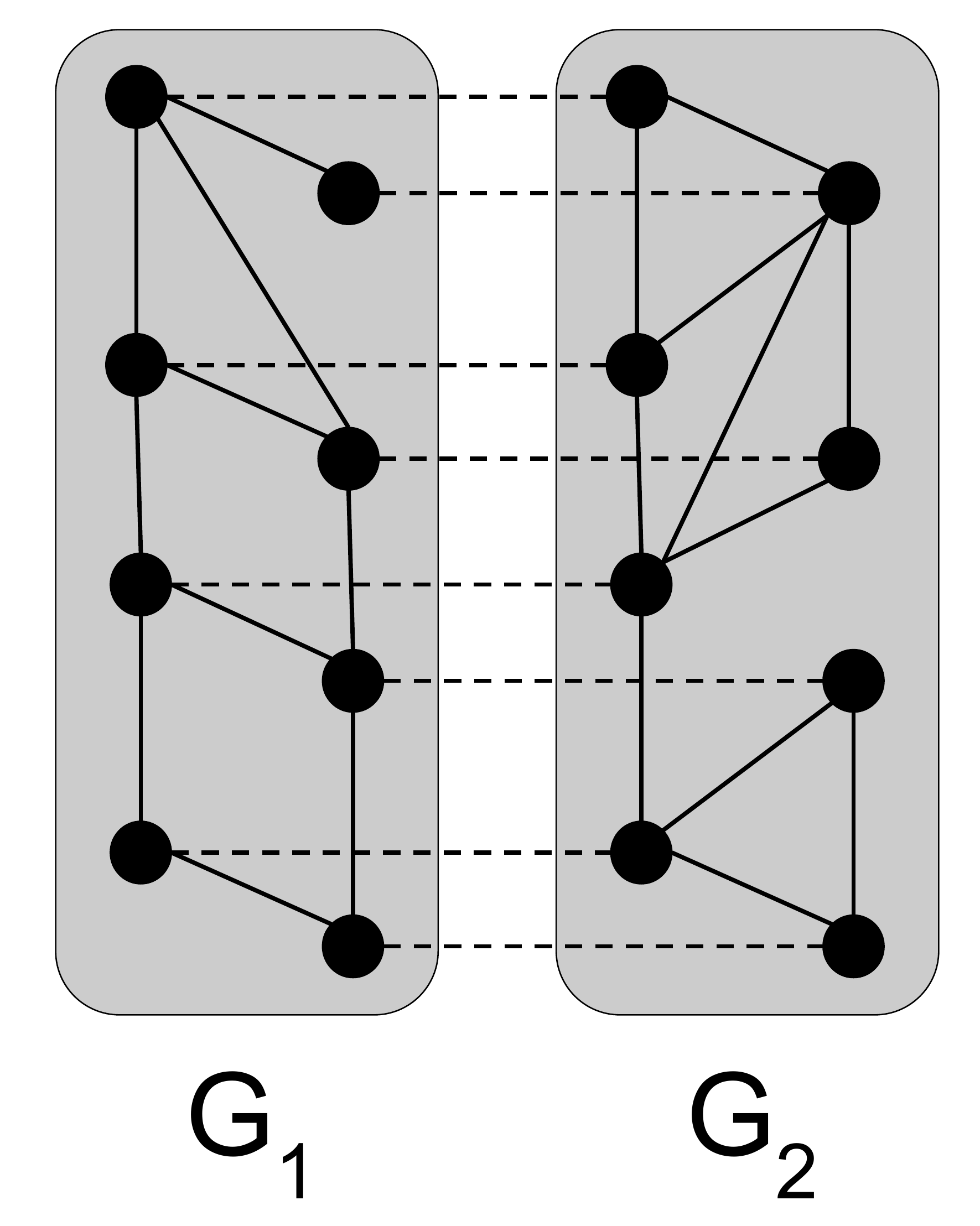}
}
\caption{Temporal graph and its multiplex view, $m=2$. \label{fig::temporal_graph_and_multiplex_view}}
\end{figure}

\begin{figure}
\centering
\subfloat[Temporal cut I \label{fig::example_temporal_cut_one}]{
\includegraphics[keepaspectratio, width=0.15\textwidth]{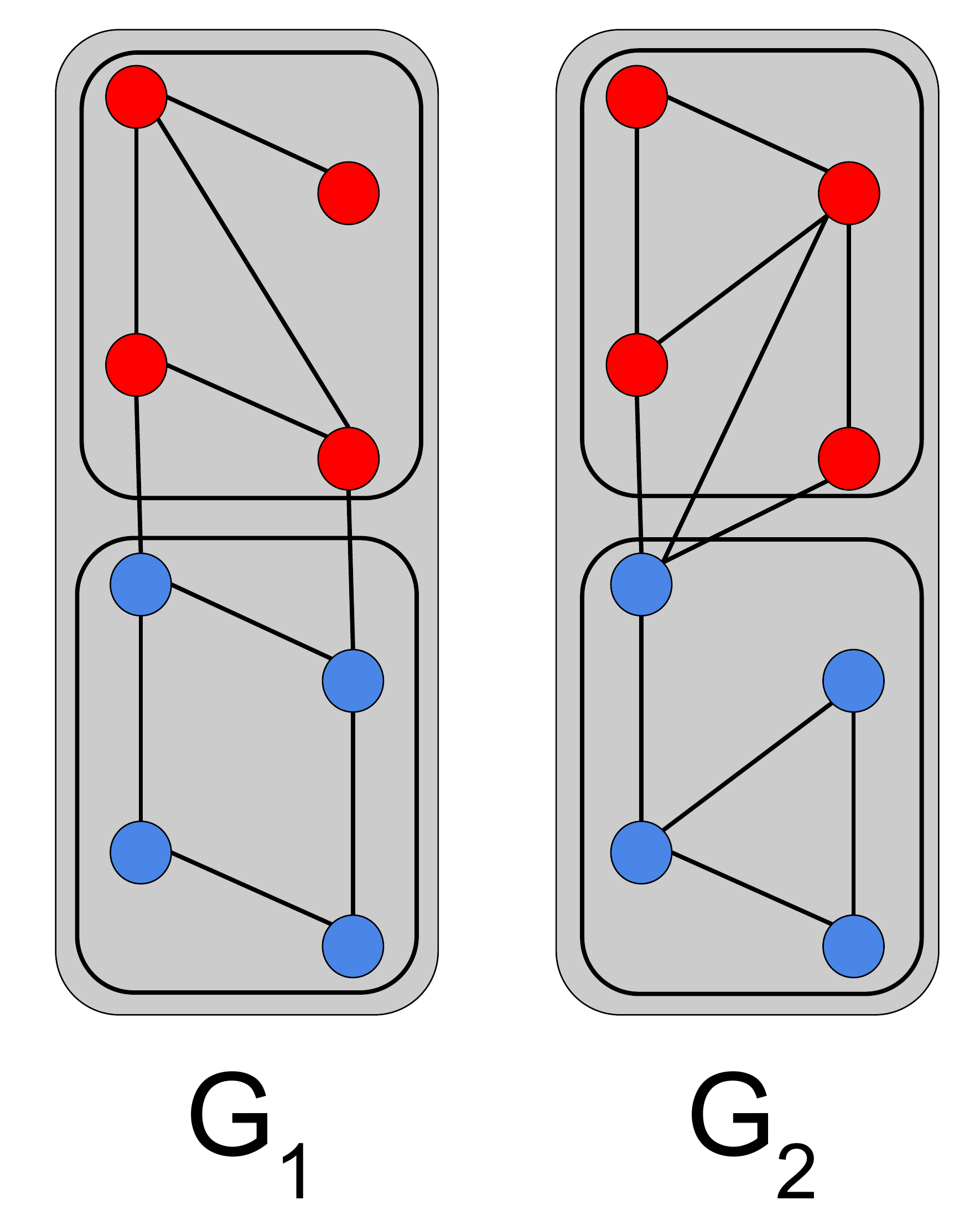}
}
\subfloat[Temporal cut II \label{fig::example_temporal_cut_two}]{
\includegraphics[keepaspectratio, width=0.15\textwidth]{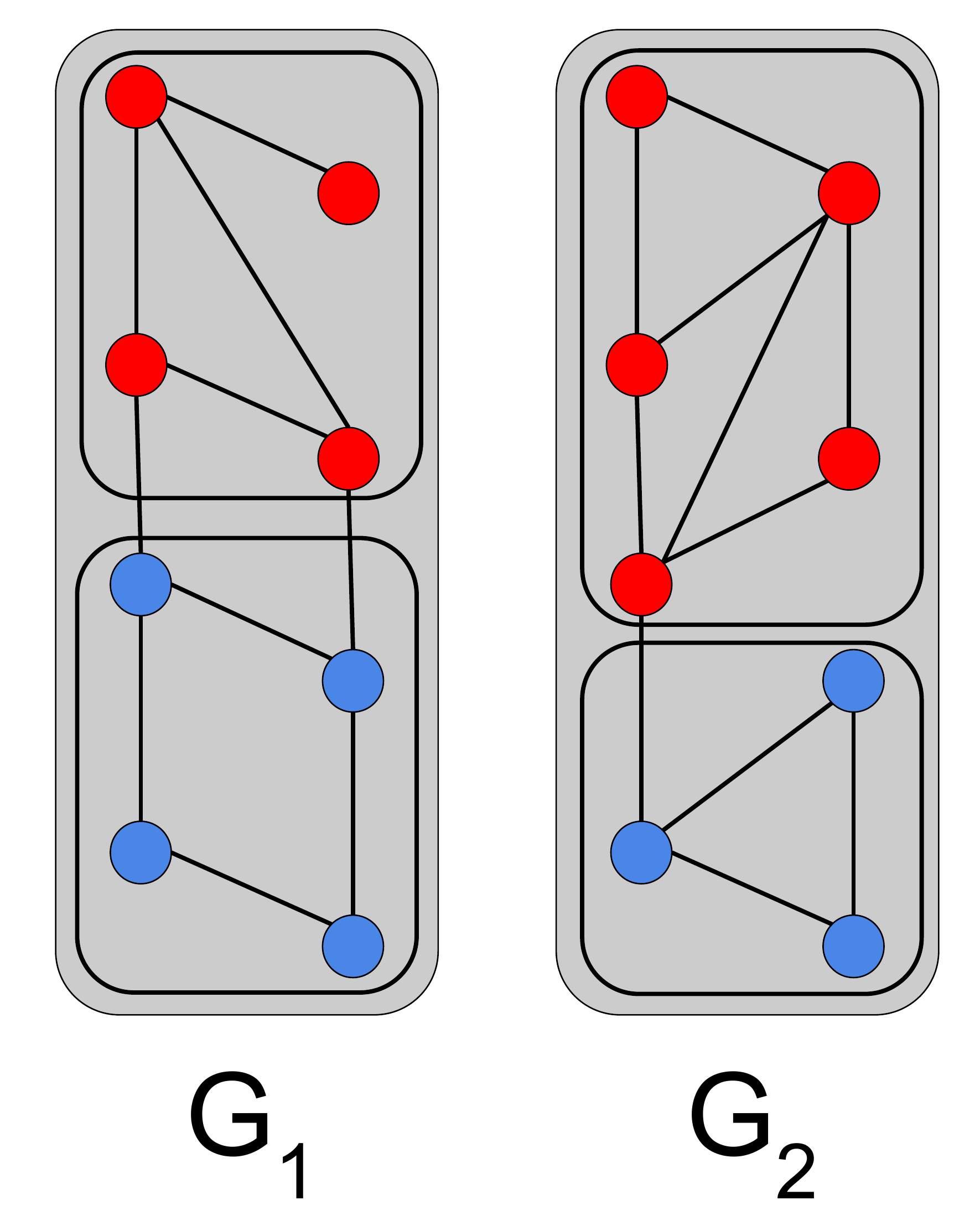}
}
\subfloat[Multiplex cut II \label{fig::example_multiplex_temporal_cut_two}]{
\includegraphics[keepaspectratio, width=0.15\textwidth]{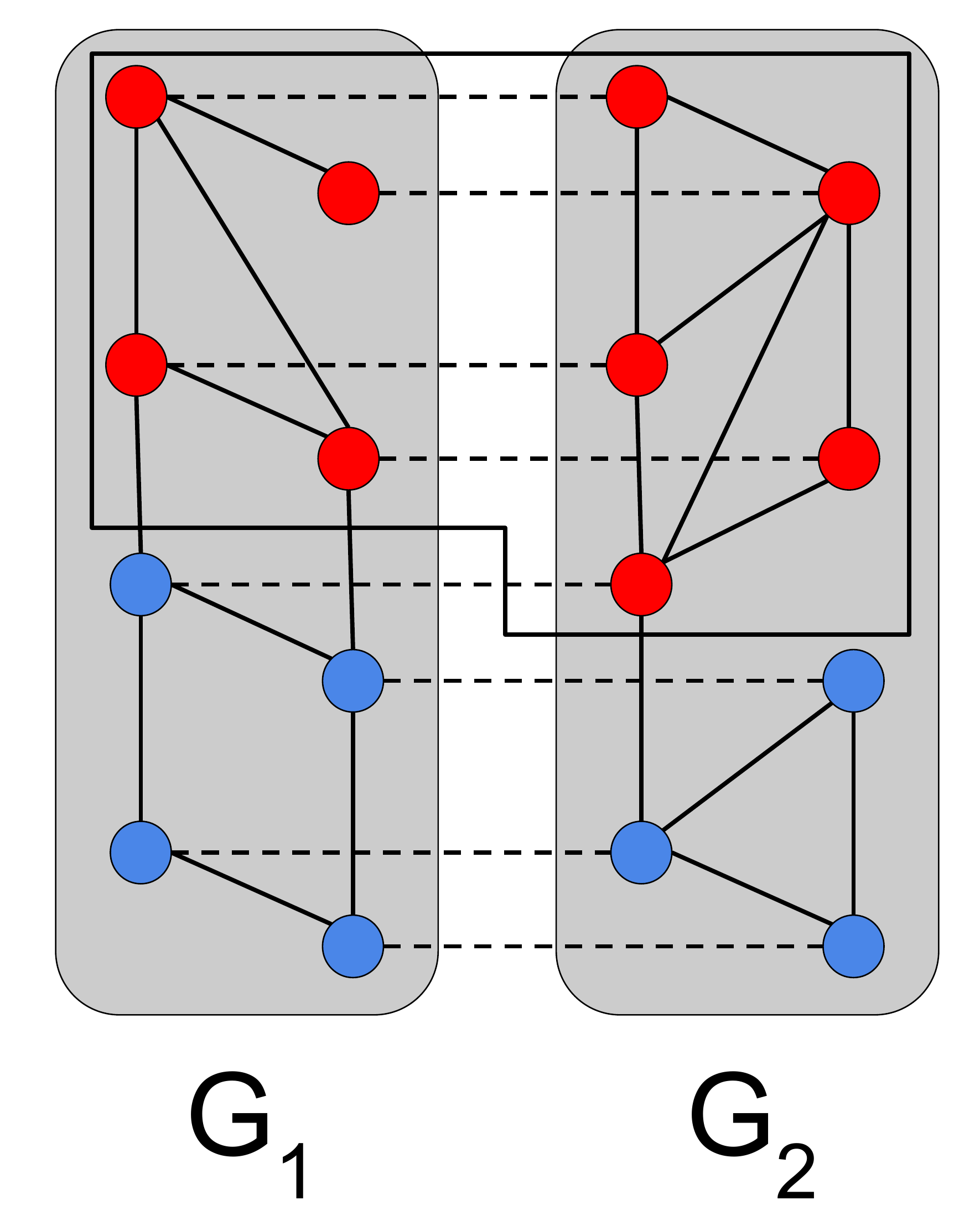}
}
\caption{Two temporal graph cuts (I and II) for the graph shown in Figure \ref{fig::example_temporal_graph} and multiplex view of cut II. For $\beta=1$, cut I has a sparsity ratio $\sigma$ of $(2+3+0)/(4\times4+4\times4)=0.16$ while cut II has a ratio of $(2+1+1)/(4\times4+5\times3)=0.13$. Thus, II is a sparser temporal cut. \label{fig::example_alternative_temporal_cuts}}
\end{figure}

Figure \ref{fig::example_alternative_temporal_cuts} shows two alternative cuts for the temporal graph shown in Figure \ref{fig::example_temporal_graph} (for $\beta=1$). Cut I (Figure \ref{fig::example_temporal_cut_one}) is smooth (no vertex changes partitions) but it has total weight of $5$. Cut II (Figure \ref{fig::example_temporal_cut_two}) is a sparser temporal cut, with weight of $3$ and only one vertex changing partitions. Notice that cut I becomes sparser than cut II for $\beta=2$. We formalize the sparsest cut problem in temporal graphs as follows.

\begin{mydef}
\textbf{Sparsest temporal cut}. The sparsest cut of a temporal graph $\mathcal{G}$, for a constant $\beta$, is defined as:
\begin{equation}
\smash{\arg\min}_{X_1 \ldots X_m}\sigma(X_1,X_m; \beta)
\nonumber
\end{equation}
\label{def::temporal_cut_ratio}
\end{mydef}

The NP-hardness of computing the sparsest temporal cut follows directly from the fact that it is a generalization of sparsest cuts for a single graph, which is also NP-hard \cite{hagen1992new}. %We can compute the the sparsest cut of a single graph by viewing it as a temporal graph with one snapshot. 

An interesting property of the multiplex model is that temporal cuts in  $\mathcal{G}$ become standard---single graph---cuts in $\chi(\mathcal{G})$. We can evaluate the sparsity of a cut in $\mathcal{G}$ by applying the original formulation (Expression \ref{eqn::cut_ratio_single_graph}) to $\chi(\mathcal{G})$, since both edges cut and partition changes in $\mathcal{G}$ become edges cut in $\chi(\mathcal{G})$. As an example, we show the multiplex view of cut II (Figure \ref{fig::example_temporal_cut_two}) in Figure \ref{fig::example_multiplex_temporal_cut_two}. However, notice that not every standard cut in $\chi(\mathcal{G})$ is a valid temporal cut. For instance, cutting all the temporal edges (i.e. separating the two snapshots in our example) would be allowed in the standard formulation, but would lead to an undefined value of sparsity as the denominator in Expression \ref{eqn::cut_ratio_temporal_graph} will be 0. Therefore, we cannot directly apply existing sparsest cut algorithms to $\chi(\mathcal{G})$ and expect to achieve a sparse temporal cut for $\mathcal{G}$.

\subsubsection{Normalized Cut}
A practical limitation of the cut sparsity (Equation \ref{eqn::cut_ratio_single_graph}), is that it favors sparsity over partition size balance. In community detection, this often leads to the discovery of \textit{``whisker communities''} \cite{leskovec2009community}. Normalized cuts \cite{shi2000normalized} take into account the \textit{volume} (i.e. sum of the degrees of vertices) of the resulting partitions, which is less prone to this effect. The normalized version of the cut sparsity is defined as:

\begin{equation}
\phi(X) = \frac{|(X,\overline{X})|}{vol(X).vol(\overline{X})}
\label{eqn::norm_cut_ratio_single_graph}
\end{equation}
where $vol(Y)=\sum_{v\in Y}deg(v)$ and $deg(v)$ is the degree of $v$.

Normalized cuts are related, by a constant factor, to the \textit{graph conductance} \cite{chung1997spectral}. We also generalize the normalized cut sparsity $\phi$ to temporal graphs as follows:

\begin{equation}
\phi(X_1, \ldots X_m;\beta) = \frac{\sum_{t=1}^m |(X_t,\overline{X}_t)|+\sum_{t=1}^{m-1}\Delta(X_t,\overline{X}_{t+1})}{\sum_{t=1}^m vol(\overline{X}_t)vol(X_t)}
\label{eqn::norm_cut_ratio_temporal_graph}
\end{equation}

%Notice that our denominator is not exactly the same as in Equation \ref{eqn::norm_cut_ratio_single_graph}. However, we can re-write each element of the summation as a product to show their relationship:

%\begin{equation*}
%vol(S_t).vol(\overline{S}_t).\left[\frac{|S_t|}{vol(S_t)}+\frac{|\overline{S}_t|}{vol(\overline{S}_t)}\right]
%\end{equation*}

%Intuitively, our expression tries to enforce balance in both volume and partition size, instead of volume alone. This property will be illustrated in our experiments. 

Next, we define the normalized cut problem for temporal graphs.

\begin{mydef}
\textbf{Normalized temporal cut}. The normalized temporal cut of $\mathcal{G}$, for a constant $\beta$, is defined as:
\begin{equation}
\smash{\arg\min}_{X_1 \ldots X_m}\phi(X_1, \ldots X_m;\beta)
\nonumber
\end{equation}
\label{def::norm_temporal_cut_ratio}
\end{mydef}

Computing optimal normalized temporal cuts is NP-hard, since the problem becomes equivalent to the standard definition, also NP-hard, for the case of a single snapshot.
In the next section, we discuss spectral approaches for sparsest and normalized cuts in temporal graphs.

\subsection{Spectral Approaches}
Similar to the single graph case, we also exploit spectral graph theory in order to compute good temporal graph cuts. Let $A$ be an $n\times n$ weighted adjacency matrix of a graph $G(V,E)$, where $A_{u,v}=W(u,v)$ if $(u,v)\in E$ or $0$, otherwise. The degree matrix $D$ is a an $n\times n$ diagonal matrix, with $D_{v,v}=deg(v)$ and $D_{u,v}=0$ for $u \neq v$. The Laplacian matrix of $G$ is defined as $L=D-A$. Let $L_t$ be the Laplacian matrix of the graph snapshot $G_t$ and $I$ be an $n\times n$ identity matrix. We define the Laplacian of the temporal graph $\mathcal{G}$ as the Laplacian of its multiplex view $\chi(\mathcal{G})$:

\begin{equation}
\mathcal{L} = \begin{pmatrix} L_1+\beta I & -\beta I & 0 & \ldots & 0 \\ -\beta I & L_2+2\beta I & -\beta I & \ldots & 0\\ \vdots & & \ddots & \ldots & -\beta I \\ 0 & 0 & \ldots & -\beta I & L_m+\beta I \end{pmatrix} \nonumber
\end{equation}

The matrix $\mathcal{L}$ can be also written using the Kronecker (or tensor) product as $\textbf{diag}(L_1, \ldots L_m)+L^I\otimes I$, where $L^I$ is the Laplacian of a line graph and $I$ is an $n \times n$ identity matrix.
Similarly, we define the degree matrix $\mathcal{D}$ of $\mathcal{G}$ as $\textbf{diag}(D_1,$ $D_2 \ldots D_m)$, where $D_t$ is the degree matrix of $G_t$. Let $C = nI-\textbf{1}_{n\times n}$ be the Laplacian of a complete graph with $n$ vertices, we use $C$ to define another $nm \times nm$ Laplacian matrix $\mathcal{C}=I\otimes C$ associated with $\mathcal{G}$.
%\begin{equation}
%\mathcal{C} = \begin{pmatrix} nI-\textbf{1}_{n\times n}& 0 & \ldots & 0 \\ 0 & nI-\textbf{1}_{n\times n}&  \ldots & 0\\ \vdots &  \ddots & \ldots & 0 \\ 0 &  \ldots & 0 & nI-\textbf{1}_{n\times n}\end{pmatrix} \nonumber
%\end{equation}
%where $\textbf{1}_{n\times n}$ is an $n\times n$ matrix of ones. 
The matrix $\mathcal{C}$ is the Laplacian of a graph composed of $m$ isolated components of size $n$, each of them being a clique. This special matrix will be applied to enforce valid temporal cuts over the snapshots of $\mathcal{G}$. Further, we define a size-$nm$ indicator vector $\textbf{x}$. Each vertex $v \in V$ is represented $m$ times in $\textbf{x}$, one for each snapshot. The value $\textbf{x}[v_t]=1$ if $v_t \in X_t$ and $\textbf{x}[v_t]=-1$ if $v_t \in \overline{X}_t$. In the next sections, we show how to apply this spectral formulation to solve temporal cut problems.

\subsubsection{Sparsest Cut}

The next lemma shows how the matrices $\mathcal{L}$ and $\mathcal{C}$ can be applied to compute the sparsity of a temporal cut in $\mathcal{G}$.

\begin{lemma}
The sparsity of a temporal cut $\langle(X_1,\overline{X}_1),$ $ \ldots$ $(X_m,\overline{X}_m)\rangle$ can be computed as $\frac{\textbf{x}^{\intercal}\mathcal{L}\textbf{x}}{\textbf{x}^{\intercal}\mathcal{C}\textbf{x}}$.
\label{lemm::temporal_cut}
\end{lemma}

%We apply Lemma \ref{lemm::temporal_cut} to compute a relaxed solution for the sparsest cut problem in temporal graphs (Definition \ref{def::temporal_cut_ratio}) in polynomial-time. The idea is to allow for (real) solutions $\textbf{x} \in [-1,1]^{nm}$ that can be found using eigenvalue computation algorithms and then efficiently round these solutions in order to adhere to the original (discrete) constraint $\textbf{x} \in \{-1,1\}^{nm}$. We make use of Lemma \ref{lemm::pseudoinverse_square_root_C} in order to avoid expensive computations involving the matrix $\mathcal{C}$.

%\textit{Please refer to the Appendix of this paper for proofs of all Lemmas and Theorems}. 
\textit{Please refer to the extended version of this paper \cite{long-version} for proofs of all Lemmas and Theorems}. 
Based on Lemma \ref{lemm::temporal_cut}, we can obtain a relaxed solution, $\textbf{x} \in [-1,1]^{nm}$, for the sparsest temporal cut problem (Definition \ref{def::temporal_cut_ratio}) in polynomial-time using an eigenvalue computation algorithm. This solution is later rounded to adhere to the original (discrete) constraint $\textbf{x} \in \{-1,1\}^{nm}$. We make use of Lemma \ref{lemm::pseudoinverse_square_root_C} in order to avoid expensive computations involving the matrix $\mathcal{C}$.

\begin{lemma}
The following property holds for the square-root pseudo-inverse $(\mathcal{C}^+)^{\frac{1}{2}}$ of the matrix $\mathcal{C}$:
\begin{equation}
(\mathcal{C}^+)^{\frac{1}{2}} = \frac{1}{\sqrt{n}} \mathcal{C}
\end{equation}
\label{lemm::pseudoinverse_square_root_C}
\end{lemma}

The following Lemma can be applied in the computation of a relaxed solution for the sparsest temporal cut problem.

\begin{lemma}
A relaxed solution for sparsest temporal cut problem can be computed as:
\begin{equation}
\textbf{y*} = \argmin_{\textbf{y} \in [-1,1]^{nm}, \textbf{y} \perp span\{\textbf{e}_1, \ldots \textbf{e}_m\}} \frac{\textbf{y}^{\intercal}\mathcal{CLC}\textbf{y}}{\textbf{y}^{\intercal}\textbf{y}}
\end{equation}
\begin{equation*}
\text{where } \textbf{e}_i.\lambda_i = \textbf{e}_i.\mathcal{CLC}, \lambda_1 \leq \lambda_2 \leq \ldots \lambda_{nm}
\end{equation*}
\label{lemm::cut_relaxation_one}
\end{lemma}

The ($m$$+$$1$)-th eigenvector of the matrix $(\mathcal{CLC})$ associated with the temporal graph $\mathcal{G}$ provides a relaxed solution for the sparsest temporal cut problem. The matrices $\mathcal{L}$ and $\mathcal{C}$ have $O(mn+\sum_{t}|E_t|)$ and $O(n^2m)$ non-zeros, respectively, and thus computing the matrix product takes $O(n\sum_t |E_t|)$ time. The resulting product has $O(n^2m)$ non-zeros and, as a consequence, computing its ($m$$+$$1$)-th eigenvector takes $O(n^2m^2)$ time in practice, which can be prohibitive in real settings. 
Moreover, the aforementioned lemma does not offer much insight regarding efficient solutions for our problem, as we do not know many properties of the matrix $\mathcal{CLC}$. We will define a more intuitive solution for computing sparse temporal cuts, which makes use of the following lemma.

\begin{lemma}
The matrix $\mathcal{C}$ commutes with any other real symmetric $nm \times nm$ matrix.
\label{lemm::C_commutes}
\end{lemma}

\begin{thm}
A relaxed solution for the sparsest temporal cut problem can, alternatively, be computed as:
\begin{equation}
\textbf{x}* = \argmax_{\textbf{x} \in [-1,1]^{nm}} \frac{\textbf{x}^{\intercal}[(3(n+2\beta)\mathcal{C}-\mathcal{L}]\textbf{x}}{\textbf{x}^{\intercal}\textbf{x}}
\label{eqn::diff_matrix}
\end{equation}
which is the largest eigenvector of $3(n+2\beta)\mathcal{C}-\mathcal{L}$.
\label{thm::diff_matrix}
\end{thm}

The matrix $3(n+2\beta)\mathcal{C}-\mathcal{L}$ is a Laplacian associated with a multiplex graph in which temporal edges have weight $w'(v_t,v_{t+1})$$=-\beta$ and intra-layer edges have weight $w'(u,v)$ $=3(n+2\beta)-w(u,v)$. This leads to a reordering of the spectrum of the original Laplacian $\mathcal{L}$ where cuts containing only temporal edges have negative associated eigenvalues and sparse cuts for each Laplacian $L_t$ become dense cuts for a new Laplacian $[3(n+2\beta))C-L_t]$. In terms of complexity, computing the difference $3(n+2\beta)\mathcal{C}-\mathcal{L}$ takes $O(n^2m)$ time and the largest eigenvector of $[3(n+2\beta)\mathcal{C}-\mathcal{L}]$ can be calculated in $O(n^2m)$. The resulting complexity is a significant improvement over the $O(n^2m^2)$ time taken by the previous solution if the number of snapshots is large.

\subsubsection{Normalized Cut}

We follow the same steps as in the previous section to compute normalized temporal cuts. 

\begin{lemma}
The normalized sparsity of a temporal cut $\langle(X_1,$ $\overline{X}_1),$ $ \ldots$ $(X_m,\overline{X}_m)\rangle$ can be computed as $\frac{\textbf{x}^{\intercal}\mathcal{L}\textbf{x}}{\textbf{x}^{\intercal}\mathcal{D}^{\frac{1}{2}}\mathcal{C}\mathcal{D}^{\frac{1}{2}}\textbf{x}}$.
\label{lemm::norm_temporal_cut}
\end{lemma}

Accordingly, we also define an equivalent of Lemma \ref{lemm::cut_relaxation_one} for the case of normalized cuts.

\begin{lemma}
A relaxed solution for the normalized temporal cut problem can be computed as:
\begin{equation}
\textbf{y*} = \argmin_{\textbf{y} \in [-1,1]^{nm}, \textbf{y} \perp span\{\textbf{e}_1, \ldots \textbf{e}_m\}} \frac{\textbf{y}^{\intercal}\mathcal{C}(\mathcal{D}^+)^{\frac{1}{2}}\mathcal{L}(\mathcal{D}^+)^{\frac{1}{2}}\mathcal{C}\textbf{y}}{\textbf{y}^{\intercal}\textbf{y}}
\label{eqn::norm_cut_relaxation_one}
\end{equation}
\begin{equation*}
\text{where } \textbf{e}_i.\lambda_i = \textbf{e}_i.\mathcal{C}(\mathcal{D}^+)^{\frac{1}{2}}\mathcal{L}(\mathcal{D}^+)^{\frac{1}{2}}\mathcal{C}, \lambda_1 \leq \lambda_2 \leq \ldots \lambda_{nm}
\end{equation*}
\label{lemm::norm_cut_relaxation_one}
\end{lemma}

As a consequence, we can apply Lemma \ref{lemm::norm_cut_relaxation_one} to compute a relaxed normalized temporal cut as the ($m+1$)-th eigenvector of the matrix $\mathcal{C}(\mathcal{D}^+)^{\frac{1}{2}}\mathcal{L}(\mathcal{D}^+)^{\frac{1}{2}}\mathcal{C}$. We finish by providing an equivalent of Theorem \ref{thm::diff_matrix}, for the normalized case.

\begin{thm}
A relaxed solution for the sparsest temporal cut problem can, alternatively, be computed as:
\begin{equation}
\textbf{x}* = \argmax_{\textbf{x} \in [-1,1]^{nm}} \frac{\textbf{x}^{\intercal}[(3(n+2\beta)\mathcal{C}-(\mathcal{D}^+)^{\frac{1}{2}}\mathcal{L}(\mathcal{D}^+)^{\frac{1}{2}}]\textbf{x}}{\textbf{x}^{\intercal}\textbf{x}}
\label{eqn::norm_diff_matrix}
\end{equation}
the largest eigenvector of $3(n+2\beta)\mathcal{C}-(\mathcal{D}^+)^{\frac{1}{2}}\mathcal{L}(\mathcal{D}^+)^{\frac{1}{2}}$.
\label{thm::norm_diff_matrix}
\end{thm}

The interpretation of matrix $3(n+2\beta)\mathcal{C}-(\mathcal{D}^+)^{\frac{1}{2}}\mathcal{L}(\mathcal{D}^+)^{\frac{1}{2}}$ is similar to the one for the sparsest cut case, with temporal edges having negative weights. Moreover, the complexity of computing the largest eigenvector of such matrix is also $O(n^2m)$. This quadratic cost on the size of the graph, for both sparsest and normalized cut problems, becomes prohibitive even for reasonably small graphs. The next section is focused on faster algorithms for temporal graph cuts.

\begin{algorithm}[ht!]
\scriptsize
\caption{Spectral Algorithm \label{alg::spectral_algorithm}}
\begin{algorithmic}[1]
\REQUIRE Temporal graph $\mathcal{G}$, rank $r$, constant $\beta$
\ENSURE Temporal cut $\langle(X_1,\overline{X}_1), \ldots (X_m,\overline{X}_m)\rangle$
\FOR{$G_t \in \mathcal{G}$}
\STATE Compute bottom-$r$ eigendecomposition $U_t'\Lambda_t'^LU_t'^{\intercal} \approx L_t$
\STATE Fix $\Lambda_t \leftarrow \textbf{diag}(0,3(n+2\beta)n-\lambda_n, \ldots 3(n+2\beta)n-\lambda_2)$
\ENDFOR
\STATE $\mathcal{Q} \leftarrow \textbf{diag}(\Lambda_1, \ldots \Lambda_m) - \mathcal{B}$
\FOR{$t \in [1,m-1]$}
\STATE $\mathcal{Q}_{t,t+1} \leftarrow U_t'^{\intercal}U'_{t+1}$
\ENDFOR
\STATE Compute largest eigenvector $\textbf{x*} \leftarrow \arg\max_{\textbf{x}} \textbf{x}^{\intercal}\mathcal{Q}\textbf{x}/\textbf{x}^{\intercal}\textbf{x}$
\STATE \textbf{return} rounded cut $\textbf{sweep}(\mathcal{U}.\textbf{x}*, \mathcal{G},\beta)$
\end{algorithmic}
\end{algorithm}

\subsection{Fast Approximations}
\label{sec::fast_approximation}

By definition, sparse temporal cuts are sparse in each snapshot and smooth across snapshots. Similarly, normalized temporal cuts are composed of a sequence of good normalized snapshot cuts that are stable over time. This motivates \textit{divide-and-conquer} approaches for computing temporal cuts that first find a good cut on each snapshot (\textit{divide}) and then combine them (\textit{conquer}). These solutions have the potential to be much more efficient than the ones based on Theorems \ref{thm::diff_matrix} and \ref{thm::norm_diff_matrix} if the conquer step is fast. However, they could lead to sub-optimal results, as optimal temporal cuts might not be composed of each snapshot's best cuts. Instead, better \textit{divide-and-conquer} schemes can explore multiple snapshot cuts in the conquer step to avoid local optima. Since we are working in the spectral domain, it is natural to take eigenvectors of blocks of $\mathcal{M}_S=3(n+2\beta)\mathcal{C}-\mathcal{L}$ and $\mathcal{M}_N=3(n+2\beta)\mathcal{C}-(\mathcal{D}^+)^{\frac{1}{2}}\mathcal{L}(\mathcal{D}^+)^{\frac{1}{2}}$, as continuous notions of snapshot cuts. This section will describe this general \textit{divide-and-conquer} approach. We will focus our discussion on the sparsest cut problem and then briefly show how it can be generalized to normalized cuts.

The following theorem is the basis of our \textit{divide-and-conquer} algorithm. It relates the spectrum of $\mathcal{M}_S$ to the spectrum of each of its dense diagonal blocks.

\begin{thm}
The eigenvalues of the matrix $\mathcal{M}_S=3(n+2\beta)\mathcal{C}-\mathcal{L}$ are the same as the ones for the matrix $\mathcal{Q}$:
\begin{equation}
\begin{small}
\mathcal{Q} = \mathbf{\Lambda} - 
\beta \begin{pmatrix} I & -U_1^{\intercal}U_2 & 0 & \ldots & 0 \\ -U_2^{\intercal}U_1& 2I & -U_2^{\intercal}U_3& \ldots & 0\\ \vdots &  & \ddots & \ldots & U_{m-1}^{\intercal}U_m\\ 0 & 0 & \ldots & U_m^{\intercal}U_{m-1}& I \end{pmatrix}\nonumber
\end{small}
\end{equation}
where $U_t\Lambda_t U_t^{\intercal}$ is the eigendecomposition of $M_t = (3(n+2\beta)C - L_t)$ and $\mathbf{\Lambda} = \textbf{diag}(\Lambda_1 \ldots \Lambda_m)$. An eigenvector $\textbf{e}_j$ of $\mathcal{M}_S$ is computed as $\mathcal{U}.\textbf{e}_j^{\mathcal{Q}}$, where $\mathcal{U} = \textbf{diag}(U_1 \ldots$ $U_m)$ and $\textbf{e}_j^{\mathcal{Q}}$ is the corresponding eigenvector of $\mathcal{Q}$.
\label{thm::divide_and_conquer}
\end{thm}

The matrix $\mathcal{Q}$ has $O(n^2m)$ non-zeros, being asymptotically as sparse as $\mathcal{M}_S$. However, $\mathcal{Q}$ can be block-wise sparsified using low-rank approximations of the matrices $M_t$. Given a constant $r \leq n$, we approximate each $M_t$ as $U_t\Lambda_t'U_t^{\intercal}$, where $\Lambda_i'$ contains only the top-$r$ eigenvalues of $M_t$. The benefits of such a strategy are the following: (1) The cost of computing the eigendecomposition of $M_t$ changes from $O(n^3)$ to $O(rn^2)$; (2) the cost of multiplying eigenvector matrices decreases from $O(n^3)$ to $O(r^3)$; and (3) the number of non-zeros in $\mathcal{Q}$ is reduced from $O(n^2m)$ to $O(r^2m)$. Similar to the case of general block tridiagonal matrices \cite{gansterer2003computing}, we can show that the error associated with such approximation is bounded by $2\lambda_{r+1}^{max}$, where $\lambda_{r+1}^{max}$ is the largest $(r+1)$-nth eigenvalue of the approximated matrices $M_t$.

We improve our approach even further by speeding-up the eigendecomposition of the matrices $M_t$ without any additional error. 
The idea is to operate directly over the original Laplacians $L_t$, which are expected to be sparse. The eigendecomposition of a sparse matrix with $|E|$ edges can be performed in time $O(n|E|)$ and for real-world graphs $|E| << n^2$. The following Lemma shows how the spectrum of $M_t$ can be efficiently computed based on $L_t$. 

\begin{lemma}
Let $\lambda_1^L, \lambda_2^L \ldots \lambda_n^L$ be the eigenvalues of a (connected) Laplacian matrix $L$ in increasing order with associated eigenvectors $e_1^L, e_2^L \ldots e_n^L$. The eigenvectors $e_i^L$ are also eigenvectors of $3(n+2\beta)C-L$ with associated eigenvalues $\lambda_1=0$ and $\lambda_i=3(n+2\beta)n-\lambda_{n-i+1}$ for $i>0$. \label{lemm::complement_matrix}
\end{lemma}

Algorithm \ref{alg::spectral_algorithm} describes our \textit{divide-and-conquer} approach for approximating the sparsest temporal graph cut. Its inputs are the temporal graph $\mathcal{G}$, the rank $r$ that controls the accuracy of the algorithm, and a constant $\beta$. As a result, it returns a cut $\langle(X_1,\overline{X}_1) \ldots$ $(X_m,\overline{X}_m)\rangle$ that (approximately) minimizes the sparsity ratio defined in Equation \ref{eqn::cut_ratio_temporal_graph}. In the \textit{divide} phase, the top-$r$ eigenvalues/eigenvectors of each matrix $M_t$---related to the bottom-$r$ eigenvalues/eigenvectors of $L_t$---are computed using Lemma \ref{lemm::complement_matrix} (steps 1-5). The \textit{conquer} phase (steps 6-11) consists of building the matrix $\mathcal{Q}$, as described in Theorem \ref{thm::divide_and_conquer}, and then computing its largest eigenvector as a relaxed version of a sparse temporal cut. 
The resulting eigenvector is discretized using a standard sweep algorithm (\textbf{sweep}) over the vertices sorted by their corresponding value of $\textbf{x*}$. The selection criteria for the sweep algorithm is the sparsity ratio given by Equation \ref{eqn::cut_ratio_temporal_graph}.

The time complexity of our algorithm is $O(mr\sum_{t=1}^m |E_t|$ $+mr^3)$. The \textit{divide} step has cost $O(mr\sum_{t=1}^m|E_t|)$, which corresponds to the computation of $r$ eigenvectors/eigenvalues of Laplacian matrices $L_t$ with $O(|E_t|)$ non-zeros each. As each snapshot is processed independently, this part of the algorithm can be easily parallelized. In the \textit{conquer} step, the most time consuming operation is in computing $m-1$ $r\times r$ matrix products in the construction of $\mathcal{Q}$, which takes $O(r^3m)$ time in total. Moreover, our algorithm has space complexity of $O(r^2m)$. This is due to the number of non-zeros in the sparse representation of $\mathcal{Q}$, which is, asymptotically, the largest data structure applied in the computation.

We follow the same general approach discussed in this section to efficiently compute normalized temporal cuts. As in Theorem \ref{thm::divide_and_conquer}, we can compute the eigenvectors of $\mathcal{M}_N$ using divide-and-conquer. However, each block $M_t$ will be in the form $3(n-2\beta)C-(D_t^+)^{\frac{1}{2}}L_t(D_t^+)^{\frac{1}{2}}$. Moreover, similar to Lemma \ref{lemm::complement_matrix}, we can also compute the eigendecomposition of $M_t$ based on $(D_t^+)^{\frac{1}{2}}L_t(D_t^+)^{\frac{1}{2}}$. 

\subsection{Generalizations}
\label{sec::generalizations}

Here, we briefly address several generalizations of temporal cuts that aim to increase the applicability of this work.

\textbf{Arbitrary swap costs:} While we have assumed uniform swap costs $\beta$, generalizing our formulation to arbitrary (non-negative) swap costs for pairs $(v_t,v_{t+1})$ is straightforward. 

\textbf{Multiple cuts:} 
Multi-cuts can be computed based on the top eigenvectors of our temporal cut matrices, as proposed in \cite{shi2000normalized}. Eigenvector values are given to a clustering algorithm (e.g. k-means) to obtain a $k$-way partition.

\begin{figure*}[ht!]
\centering
\subfloat[Synthetic]{
\includegraphics[keepaspectratio, width=0.24\textwidth]{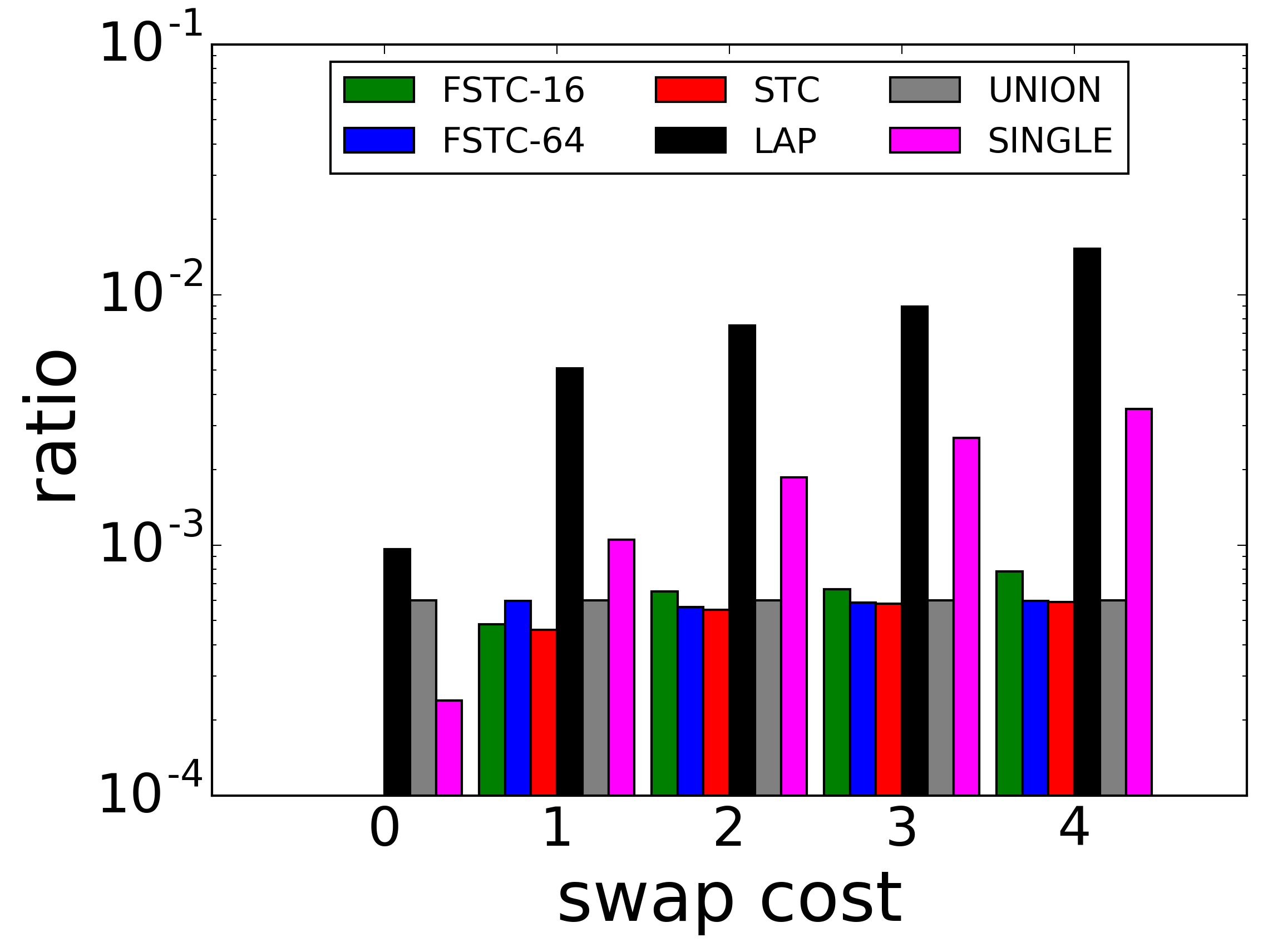}
}
\subfloat[School]{
\includegraphics[keepaspectratio, width=0.24\textwidth]{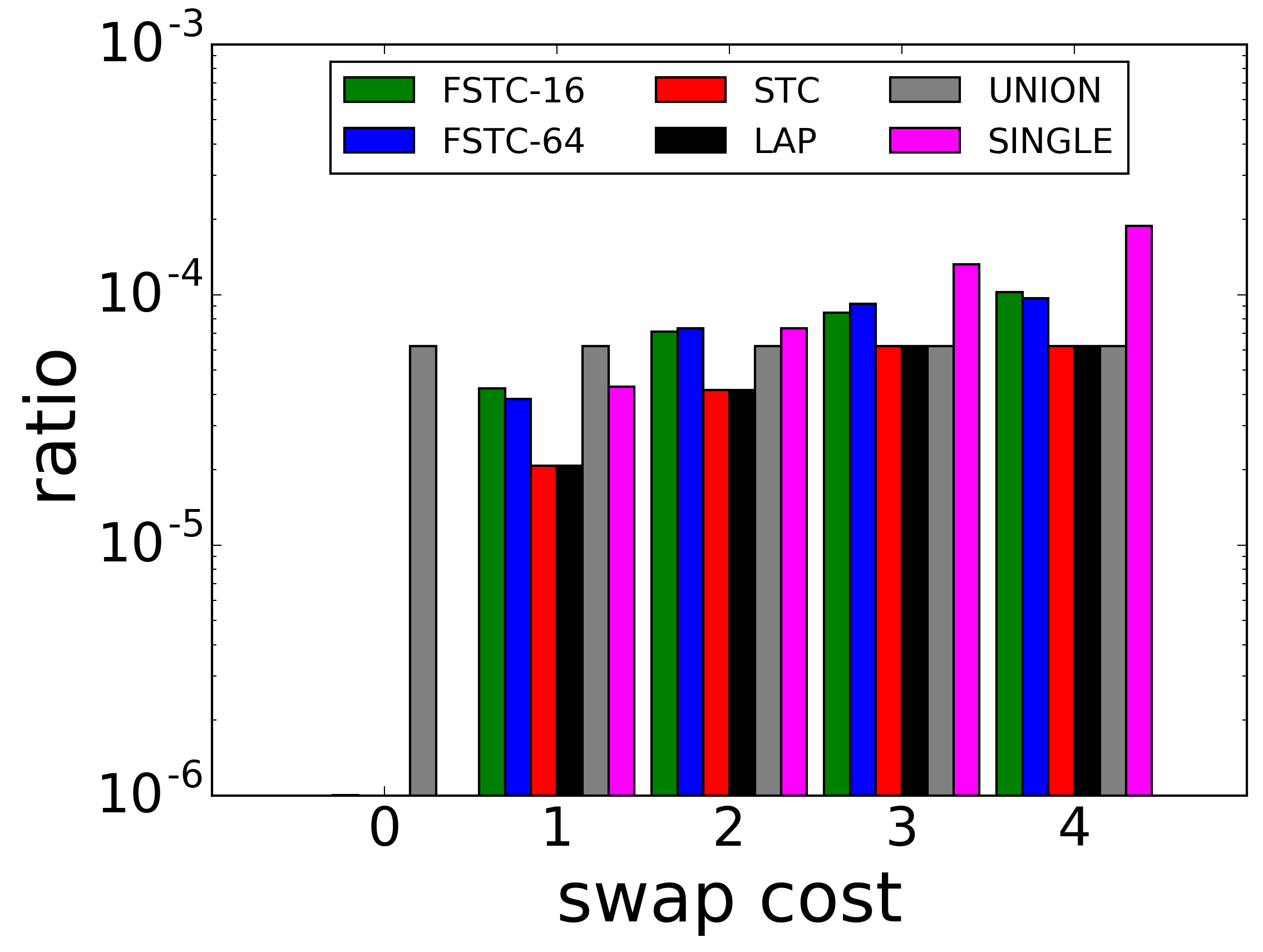}
}
\subfloat[Stock]{
\includegraphics[keepaspectratio, width=0.24\textwidth]{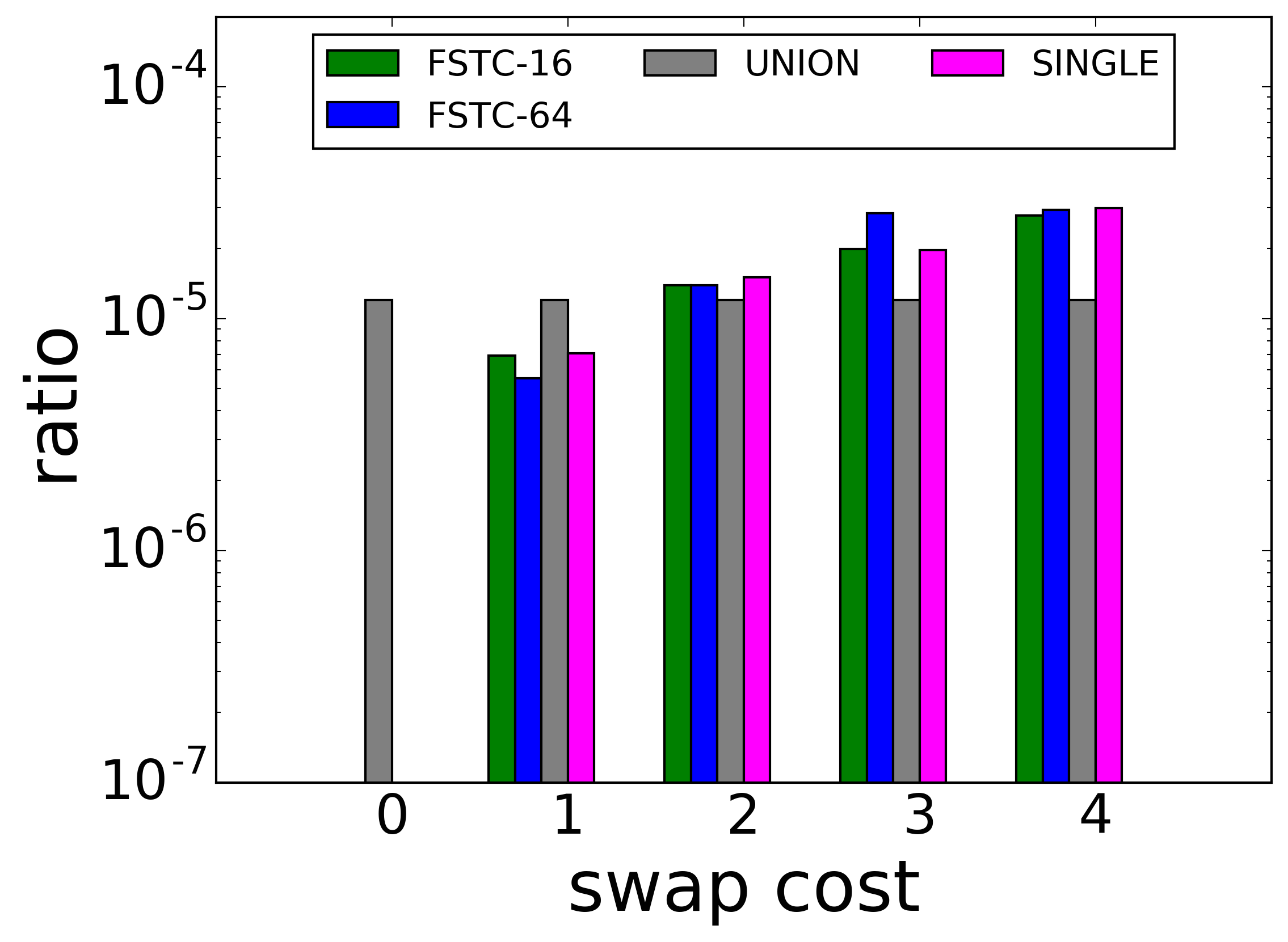}
}
\subfloat[DBLP]{
\includegraphics[keepaspectratio, width=0.24\textwidth]{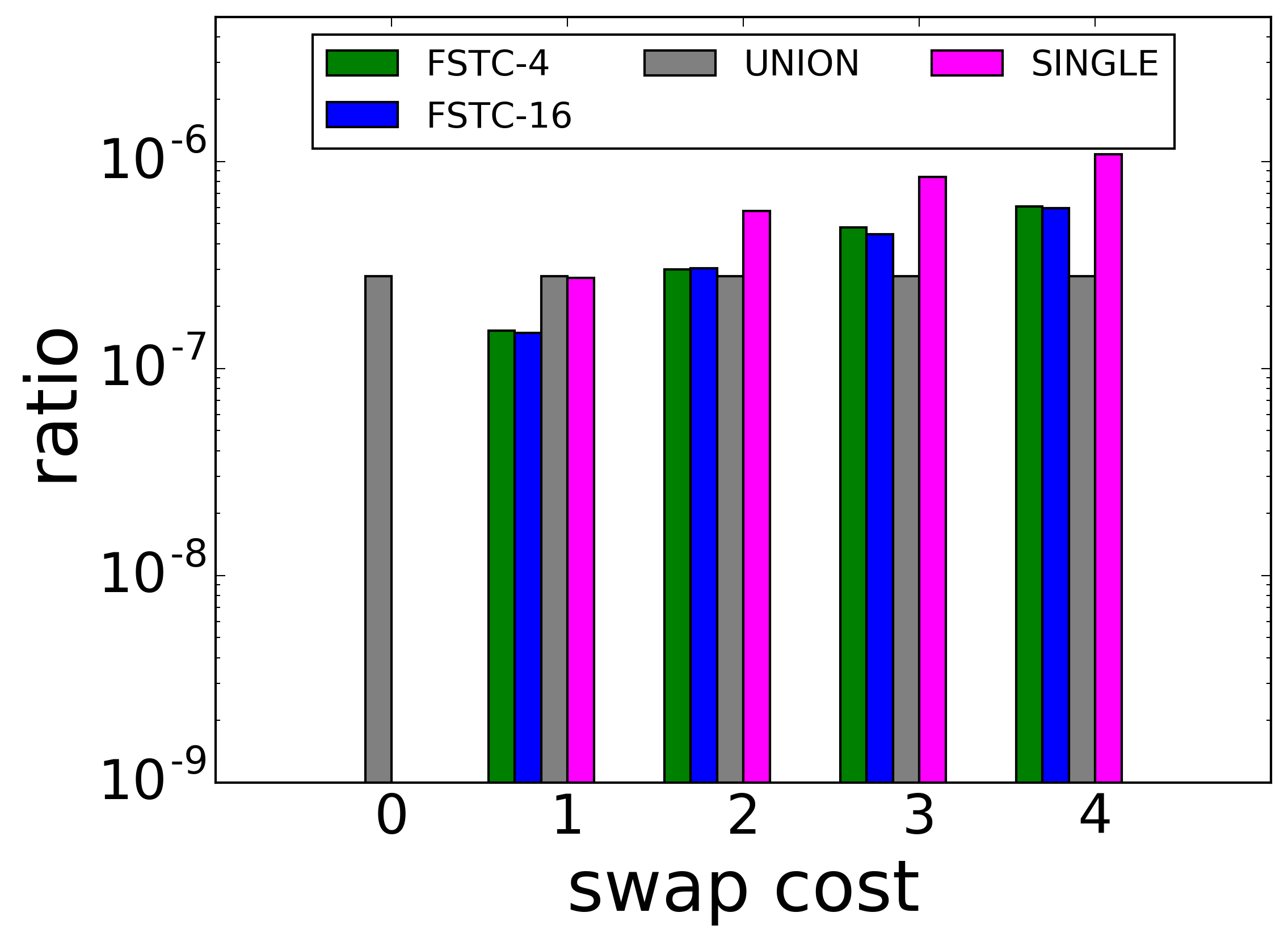}
}

\subfloat[Synthetic]{
\includegraphics[keepaspectratio, width=0.24\textwidth]{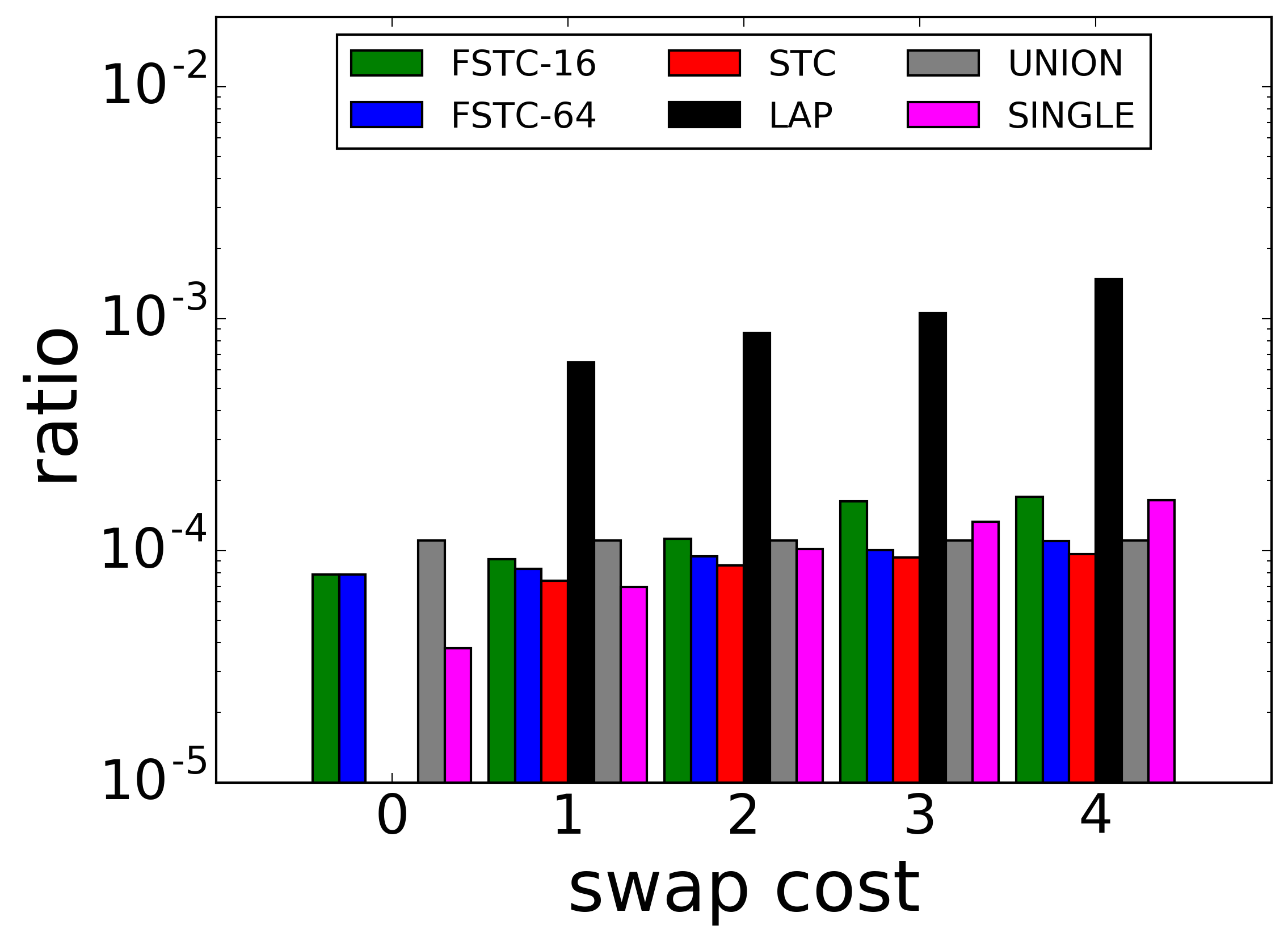}
}
\subfloat[School]{
\includegraphics[keepaspectratio, width=0.24\textwidth]{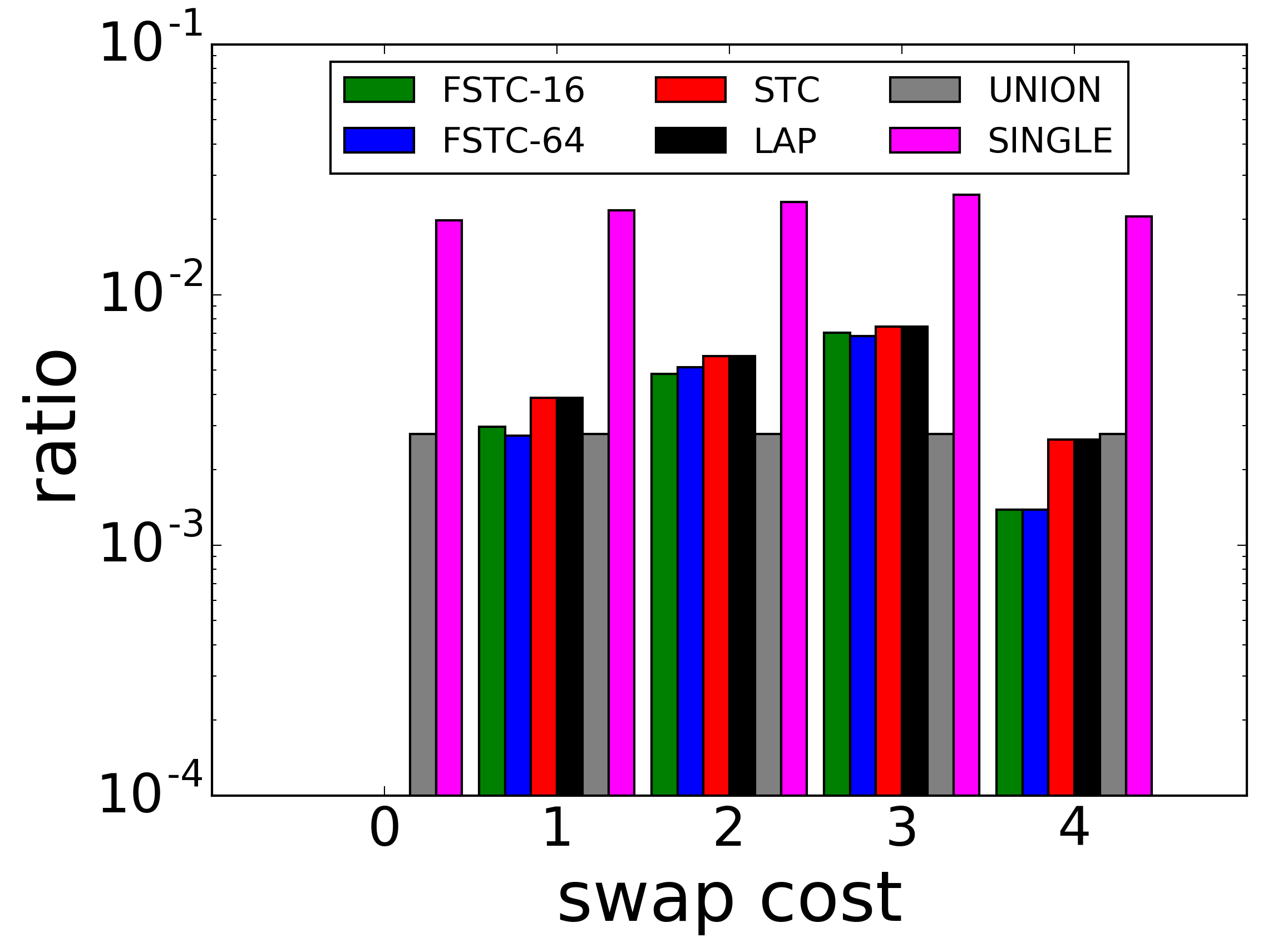}
}
\subfloat[Stock]{
\includegraphics[keepaspectratio, width=0.24\textwidth]{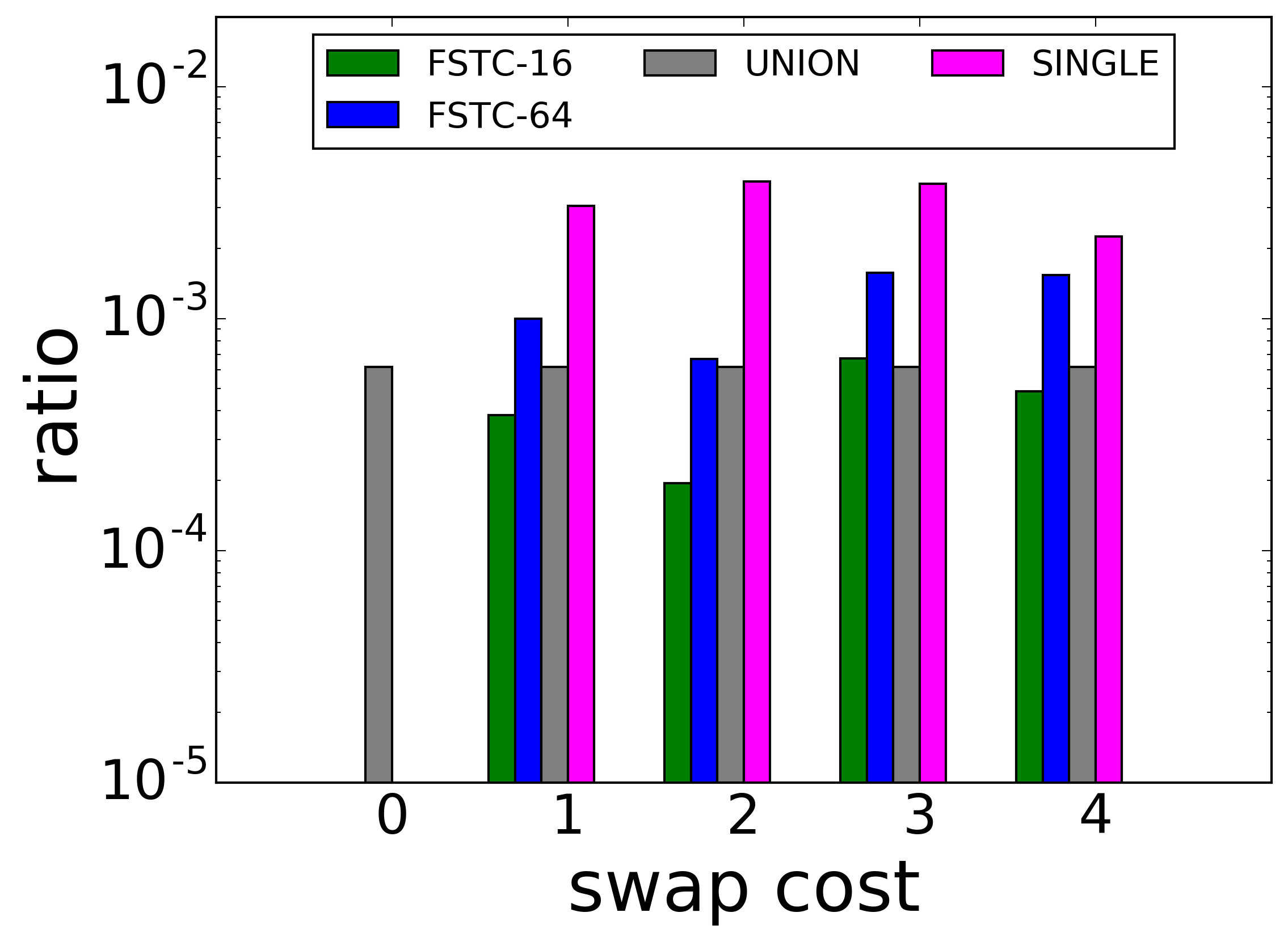}
}
\subfloat[DBLP]{
\includegraphics[keepaspectratio, width=0.24\textwidth]{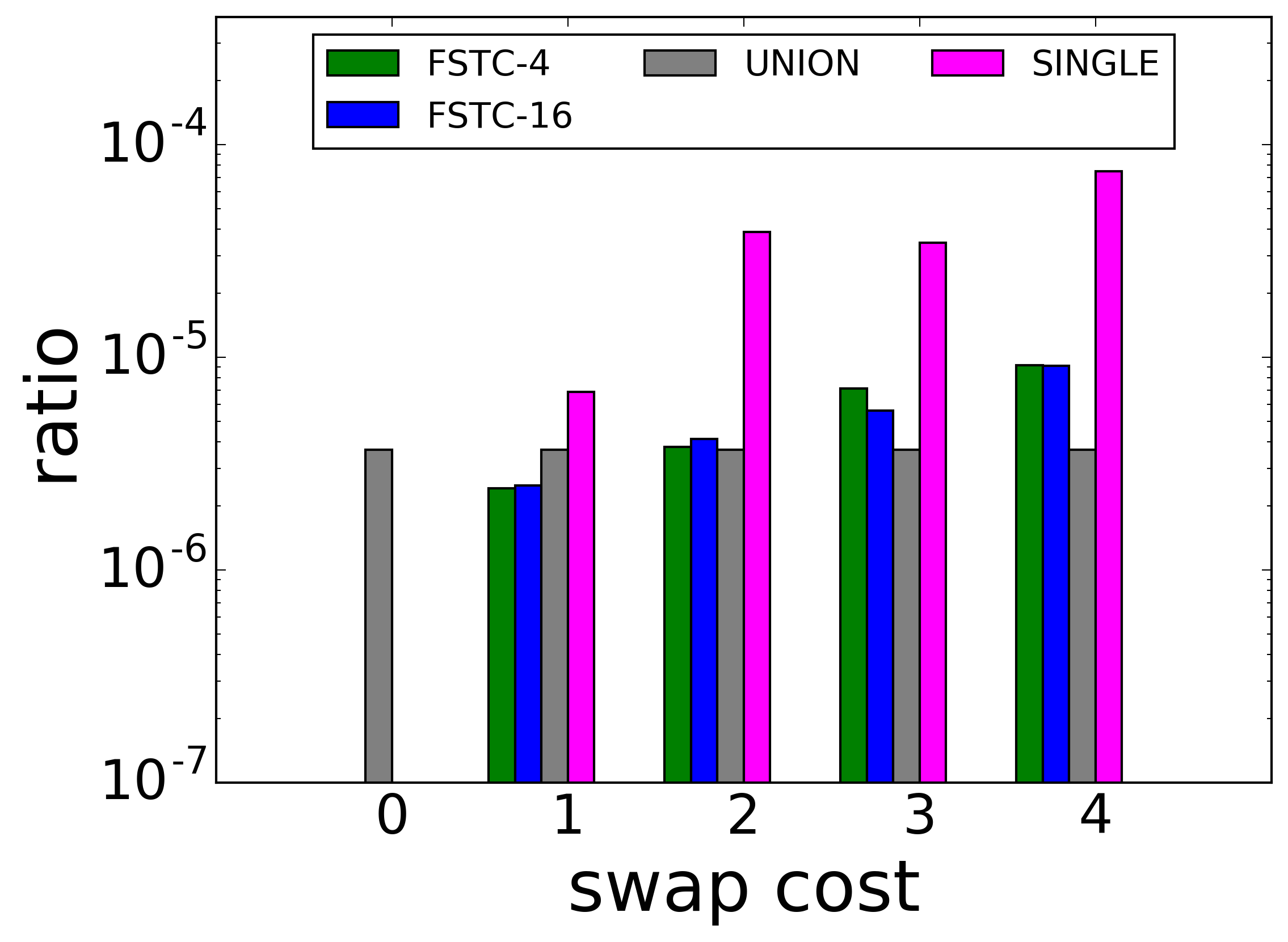}
}

\caption{Quality results (sparsity ratios) for sparsest (a-d) and normalized (e-h) cuts on synthetic and real datasets. \label{fig::ratio_results}}
\end{figure*}

\section{Signal Processing on Graphs}
\label{sec::signal_processing}

We apply graph cuts as data-driven wavelet bases for dynamic signals. The idea is to identify temporal cuts based on both graph structure and signal to compute wavelet coefficients using the resulting partitions. Given a sequence of signals $\langle\textbf{f}_{(1)}, \ldots \textbf{f}_{(m)}\rangle$, $\textbf{f}_{(i)} \in \mathbb{R}^n$, on a temporal graph $\mathcal{G}$, our goal is to discover a temporal cut that is sparse, smooth, and separates vertices with dissimilar signal values. A previous work \cite{silva2016graph} has shown that a relaxation of the $L_2$ energy (or importance) $||a||_2$ of a wavelet coefficient $a$ for a single graph snapshot with signal $\textbf{f}$ can be computed as:

\begin{equation}
\frac{(|X|\sum_{v\in \overline{X}} \textbf{f}[v]-|\overline{X}|\sum_{u\in X} \textbf{f}[u])^2}{|X||\overline{X}|} \propto -\frac{\textbf{x}^{\intercal}CSC\textbf{x}}{\textbf{x}^{\intercal}C\textbf{x}}
\label{eqn::wavelet_energy}
\end{equation}
where $\textbf{x}$ is an indicator vector and $S_{u,v} = (\textbf{f}[v]-\textbf{f}[u])^2$. Sparsity is enforced by adding a Laplacian regularization factor $\alpha \textbf{x}^{\intercal}L\textbf{x}$, where $\alpha$ is a user-defined constant, to the denominator of Equation \ref{eqn::wavelet_energy}. This formulation supports an algorithm for computing graph wavelets, which we extend to dynamic signals. Following the same approach as in Section \ref{sec::definitions}, we apply the multiplex graph representation to compute the energy of a dynamic wavelet coefficient:

\begin{equation}
\frac{\sum_{t=1}^m \Theta_t^2 + \sum_{t=1}^{m-1} \Theta_t\Theta_{t+1}}{\sum_{t=1}^m|X_t||\overline{X}_t|}
\label{eqn::dyn_wavelet_energy}
\end{equation}
where $\Theta_t = |X_t|\sum_{v\in \overline{X}_t} \textbf{f}[v]-|\overline{X}_t|\sum_{u\in X_t} \textbf{f}[u]$. The first term in the numerator of Equation \ref{eqn::dyn_wavelet_energy} is the sum of the numerator of Equation \ref{eqn::wavelet_energy} over all snapshots. The second term acts on sequential snapshots and enforces the partitions to be consistent over time ---i.e. $X_t$'s to be jointly associated with either large or small values. Intuitively, the energy is maximized for partitions that separate different values and are also balanced in size. The next theorem provides a spectral formulation for the energy of dynamic wavelets.

\begin{thm}
The energy of a dynamic wavelet is proportional to $-\frac{\textbf{x}^{\intercal}\mathcal{C}\mathcal{S}\mathcal{C}\textbf{x}}{\textbf{x}^{\intercal}\mathcal{C}\textbf{x}}$, where $\mathcal{S}_{u,v} = (\textbf{f}[u]-\textbf{f}[v])^2$ for values within one snapshot from each other.
\label{thm::dyn_wavelet}
\end{thm}

We apply Theorem \ref{thm::dyn_wavelet} to compute a relaxation of the optimal dynamic wavelet as a regularized eigenvalue problem:

\begin{equation}
\textbf{x}* = \smash{\argmin_{\textbf{x} \in [-1,1]^{nm}}} \frac{\textbf{x}^{\intercal}\mathcal{CSC}\textbf{x}}{\textbf{x}^{\intercal}\mathcal{C}\textbf{x} + \alpha \textbf{x}^{\intercal}\mathcal{L}\textbf{x}}
\label{eqn::dyn_wavelet_relax}
\end{equation}
where $\mathcal{C}$ and $\mathcal{L}$ are the matrices defined in Section \ref{sec::definitions}. The same optimizations discussed in \cite{silva2016graph} can also be applied to efficiently approximate Equation \ref{eqn::dyn_wavelet_relax}. The resulting algorithm has complexity $O(pn\sum_{t}|E_t|+qn^2m^2)$, where $p$ and $q$ are small constants. Similar to Algorithm \ref{alg::spectral_algorithm}, we apply a sweep procedure to obtain a cut from vector $\textbf{x}*$.

\section{Experiments}

\subsection{Datasets} 

\textit{School} is a contact network where vertices represent children from a primary school and edges are created based on proximity detected by sensors \cite{stehle2011high}, with $242$ vertices, $17K$ edges and $3$ snapshots. Edge weights are based on the duration of the contact within an interval. \textit{Stock} is a correlation network of US stocks' end of the day prices\footnote{Source: \url{https://www.quandl.com/data/}}, where stocks are connected if their absolute correlation is at least $.05$ and edge weights are the absolute correlation values. The resulting network has $500$ vertices, $27K$ edges, and $26$ snapshots (one for each year in the interval 1999-2015). \textit{DBLP} is a sample from the DBLP collaboration network. Vertices corresponding to two authors are connected in a given snapshot if they co-authored a paper in the corresponding year. We selected authors who published at least 5 papers one of the following conferences: KDD (data mining), CVPR (computer vision), and FOCS (theory). The resulting temporal network has $3.4K$ vertices, $16.4K$ edges, and $4$ snapshots. 

We also use a synthetic data generator. Its parameters are a graph size $n$, partition size $k < n$, number of hops $h$, and noise level $0 \leq \epsilon \leq 1$. Edges are created based on a $\lceil\sqrt{n}\rceil \times \lceil\sqrt{n}\rceil$ grid, where each vertex is connected to its $h$-hop neighbors. A partition is a sub-grid initialized with $\lceil \sqrt{k}\rceil \times \lceil \sqrt{k}\rceil$ dimensions ($k=n/2$). A value $\pi(v)=1.+N(0,\epsilon)$ is assigned to vertices inside the partition and the remaining vertices receive iid realizations of a Gaussian $N(0,\epsilon)$. Given the node values, the weight $w$ of an edge $(u,v)$ is set as $\exp{(|\pi(v)-\pi(u)|)}$. To produce the dynamics, we move the partition along the main diagonal of the grid. 

To evaluate our wavelets for dynamic signals, we apply our approach to \textit{Traffic} \cite{silva2016graph}, a road network from California with $100$ vertices, $200$ edges, and $12$ snapshots. Average vehicle speeds measured at the vertices of the network were taken as a dynamic signal for the timespan of a Friday in April, 2011. Moreover, we apply the \textit{heat equation} to generate synthetic signals over the \textit{School} network. Different from \textit{Traffic}, which has a static structure, the resulting dataset (\textit{School-heat}) is dynamic in structure and signal.

\subsection{Approximation and Performance}

Two general approaches for computing temporal cuts, for both sparsest and normalized cuts, are evaluated in this section. The first approach, \textit{STC}, is based on Theorems \ref{thm::diff_matrix} and \ref{thm::norm_diff_matrix}, for sparsest and normalized cuts, respectively. The second approach, \textit{FSTC-r}, for a rank $r$, applies the fast approximation described in Section \ref{sec::fast_approximation}. %Our goal is to show that our solutions can efficiently compute temporal cuts in synthetic and real graphs. We omit results for algorithms based on Lemmas \ref{lemm::cut_relaxation_one} and \ref{lemm::norm_cut_relaxation_one}, as they do not finish in reasonable time even for small graphs.

We consider three baselines in this evaluation. \textit{SINGLE} is a heuristic that first discovers the best cut on each snapshot and then combines them into a temporal cut. \textit{UNION} computes the best average cut over all the snapshots---a cut over the union of the snapshots. \textit{LAP} is similar to our approach, but operates directly on the Laplacian matrix $\mathcal{L}$. Notice that each of these baselines can be applied to either sparsest and normalized cuts as long as the appropriate (standard or normalized) Laplacian matrix is used.

Figure \ref{fig::ratio_results} shows the quality results (sparsity ratios) of the methods using real and synthetic data. We vary the swap cost ($\beta$) within a range that enforces local and global (or stable) patterns. The values of $\beta$ shown are normalized to integers for ease of comparison. \textit{STC} and \textit{LAP} took too long to finish for the \textit{Stock} and \textit{DBLP} datasets, and thus such results are omitted. \textit{STC} achieves the best results (smallest ratios) in most of the settings. For the \textit{School} dataset, \textit{LAP} also achieves good results, which is due to the small number of snapshots in the network, which makes sparse cuts in $\mathcal{L}$ coincide with good temporal cuts. \textit{UNION} performs well for \textit{Stock} and \textit{DBLP} with large swap costs, as these settings enforce a fixed cut over time. \textit{SINGLE} achieves good results only when swap costs are close to $0$, when the smoothness of the cuts do not affect the sparsity ratio. Our fast approximation (\textit{FSTC})  achieves good results in most of the settings, specially for $r=64$, being able to adapt to different swap costs. Running time results (see Appendix) show that \textit{FSTC} significantly outperforms \textit{STC} and is competitive with \textit{SINGLE} and \textit{UNION}.

\begin{table}[ht!]
\begin{center}
\subfloat[School \label{table::comm_detec_school}]
{
\begin{small}
\begin{tabular}{|l|l | c | c | c | c |}
\hline
k & Method & \textit{Cut} & \textit{Sparsity} & \textit{N-sparsity} & \textit{Modularity} \\
\hline
\hline
\multirow{4}{*}{2}& \textit{GenLovain} & \textbf{2.6} & \textbf{1.0e-4}& \textbf{5.0e-3} &\textbf{102.0}\\ \cline{2-6}
&\textit{Facetnet} & 6.0 & 3.8e-4& .012&95.7\\ \cline{2-6}
&\textbf{\textit{Sparsest}} & \textbf{2.6} & \textbf{1.0e-4}& \textbf{5.0e-3} & \textbf{102.0}\\ \cline{2-6}
&\textbf{\textit{Norm.}} & \textbf{2.6} & \textbf{1.0e-4}& \textbf{5.0e-3} & \textbf{102.0}\\\hline
\multirow{4}{*}{5}& \textit{GenLovain} & 8.0& 6.8e-4 & 2.7e-2 & \textbf{110.0}\\ \cline{2-6}
&\textit{Facetnet} & 10.0 & 8.4e-4& 3.0e-2 &106.0\\ \cline{2-6}
&\textbf{\textit{Sparsest}} & 8.3 & \textbf{6.4e-4}& 2.6e-2& 109.0\\ \cline{2-6}
&\textbf{\textit{Norm.}} & \textbf{6.1} & 9.9e-4& \textbf{1.8e-2}& \textbf{110.0}\\\hline
\end{tabular}
\end{small}
}

\subfloat[DBLP \label{table::comm_detec_dblp}]
{
\begin{small}
\begin{tabular}{|l|l | c | c | c | c |}
\hline
k & Method & \textit{Cut} & \textit{Sparsity} & \textit{N-sparsity} & \textit{Modularity} \\
\hline
\hline
\multirow{4}{*}{2}& \textit{GenLovain} & 80. & 3.9e-4& 1.3e-5 &\textbf{38,612}\\ \cline{2-6}
&\textit{Facetnet} & 267.0 & 2.6e-3& 8.9e-5 &33,091\\ \cline{2-6}
&\textbf{\textit{Sparsest}} & \textbf{9.0} & \textbf{7.6e-5} & \textbf{3.6e-6} & 38,450\\ \cline{2-6}
&\textbf{\textit{Norm.}} & 19.0 & 1.2e-4& 3.8e-6& 38,516\\\hline
\multirow{4}{*}{5}& \textit{GenLovain} & 174. & 1.3e-3& 4.1e-5 &\textbf{39,342}\\ \cline{2-6}
&\textit{Facetnet} & 501.0 & 7.2e-3& 2.8e-4& 30,116\\ \cline{2-6}
&\textbf{\textit{Sparsest}} & 40.0 & 5.2e-4 & 6.2e-5&38,498\\ \cline{2-6}
&\textbf{\textit{Norm.}} & \textbf{31.0} & \textbf{4.0e-4}& \textbf{1.0e-5} & 39,015\\\hline
\end{tabular}
\end{small}
}
\end{center}
\caption{Community detection results for \textit{Sparsest and Normalized Cuts} (and two baselines) using \textit{School} and \textit{DBLP} datasets. Our methods achieve the best results for most of the metrics and are competitive in terms of modularity.\label{table::comm_detec_results}}
\end{table}
\begin{figure}[ht!]
\centering
\subfloat[I]
{\includegraphics[keepaspectratio, width=0.11\textwidth,trim={3cm 2cm 3.5cm 1.5cm},clip]{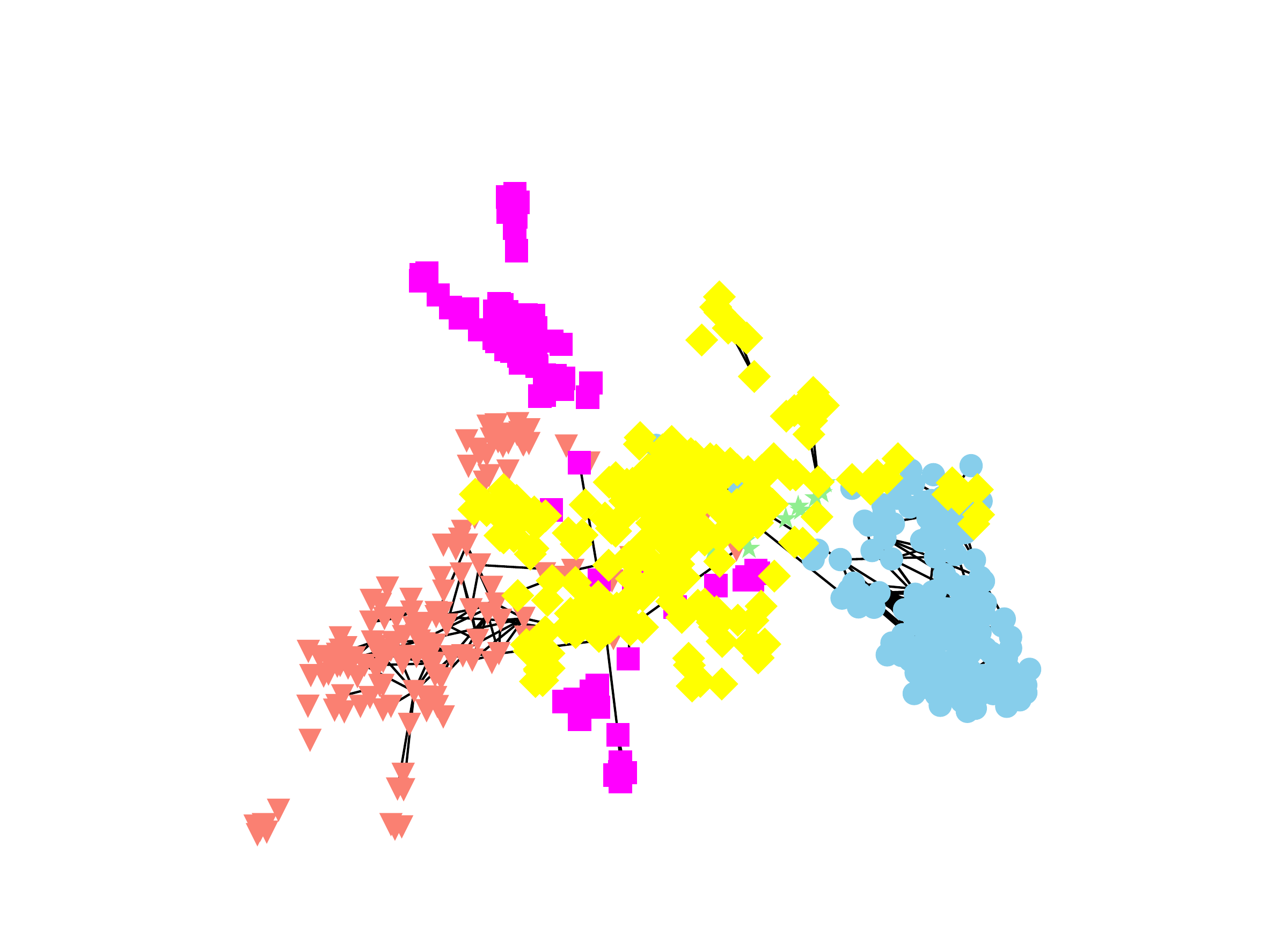}}
\subfloat[II]
{
\includegraphics[keepaspectratio, width=0.11\textwidth,trim={3cm 2cm 3.5cm 1.5cm},clip]{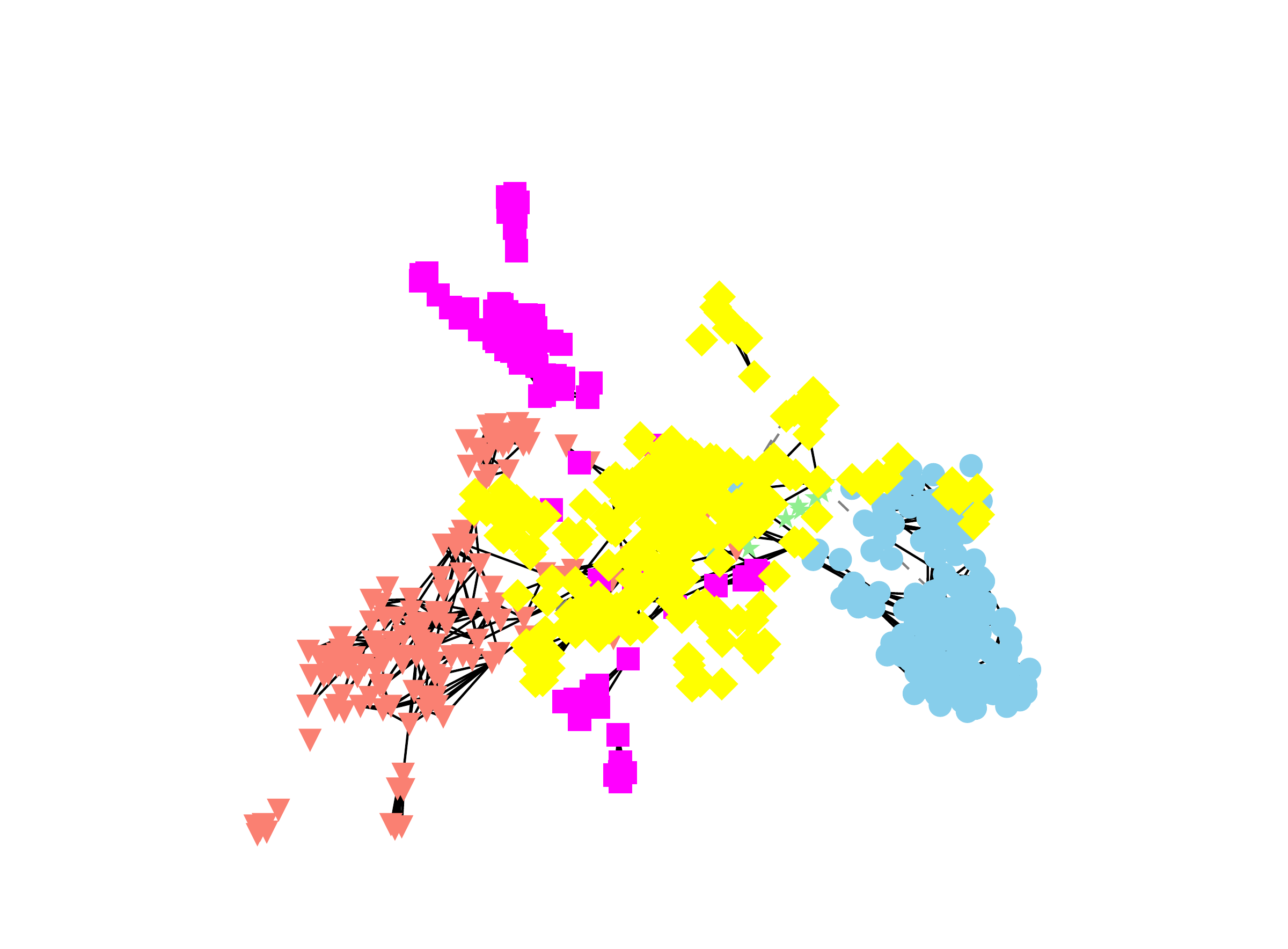}
}
\subfloat[III]
{
\includegraphics[keepaspectratio, width=0.11\textwidth,trim={3cm 2cm 3.5cm 1.5cm},clip]{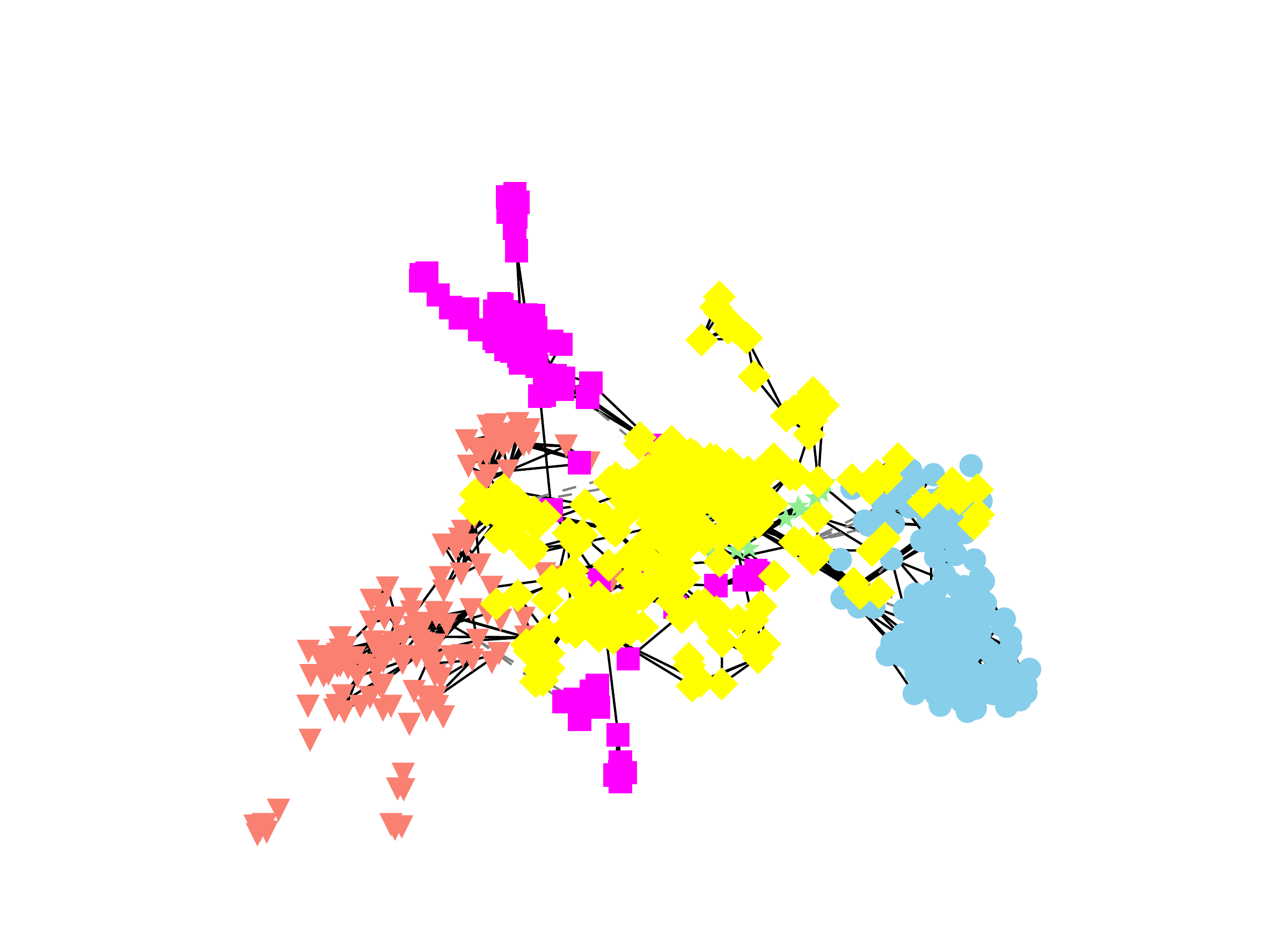}
}
\subfloat[IV]
{
\includegraphics[keepaspectratio, width=0.11\textwidth,trim={3cm 2cm 3.5cm 1.5cm},clip]{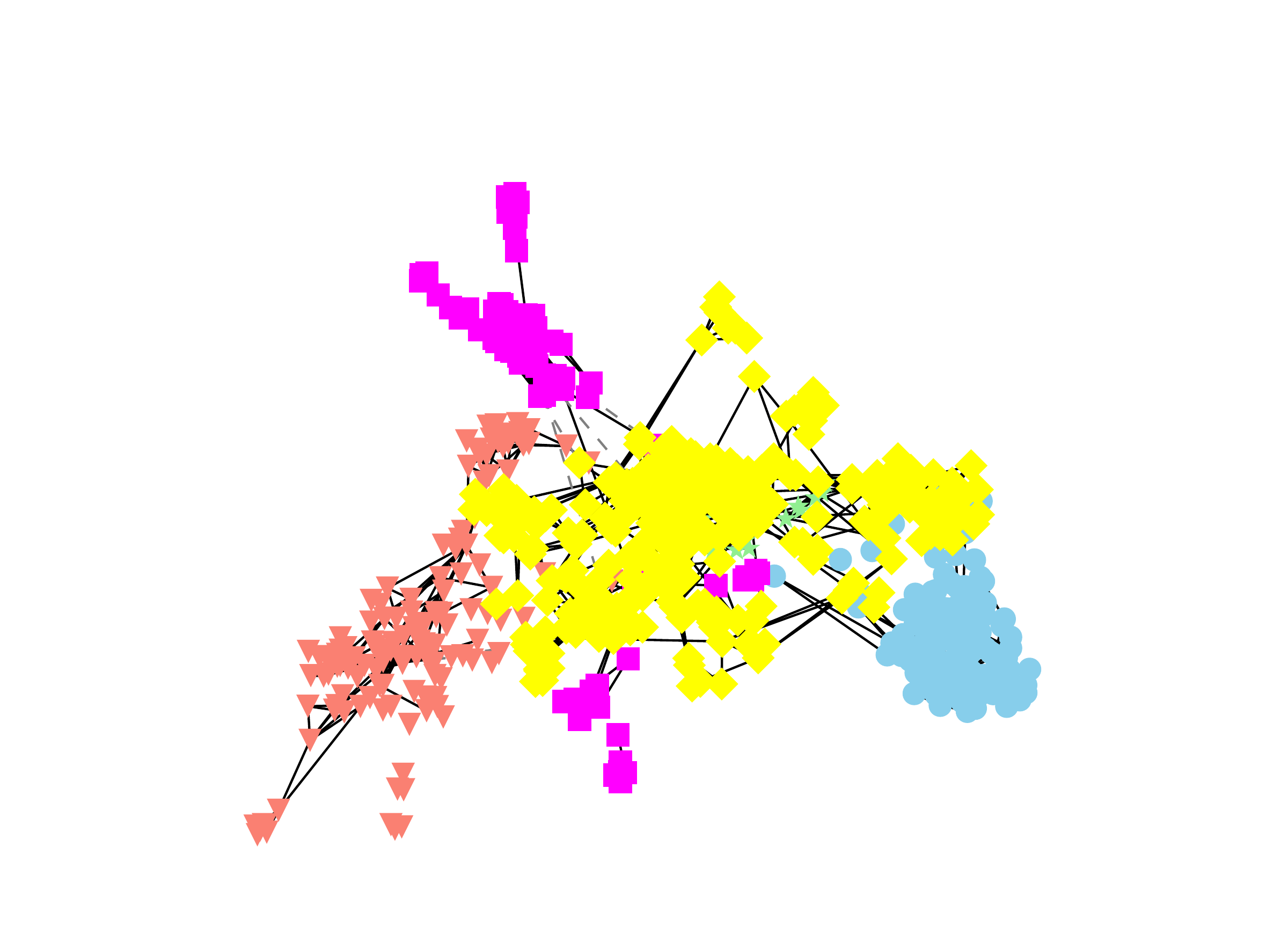}
}

\subfloat[I]
{\includegraphics[keepaspectratio, width=0.11\textwidth,trim={3cm 2cm 3.5cm 1.5cm},clip]{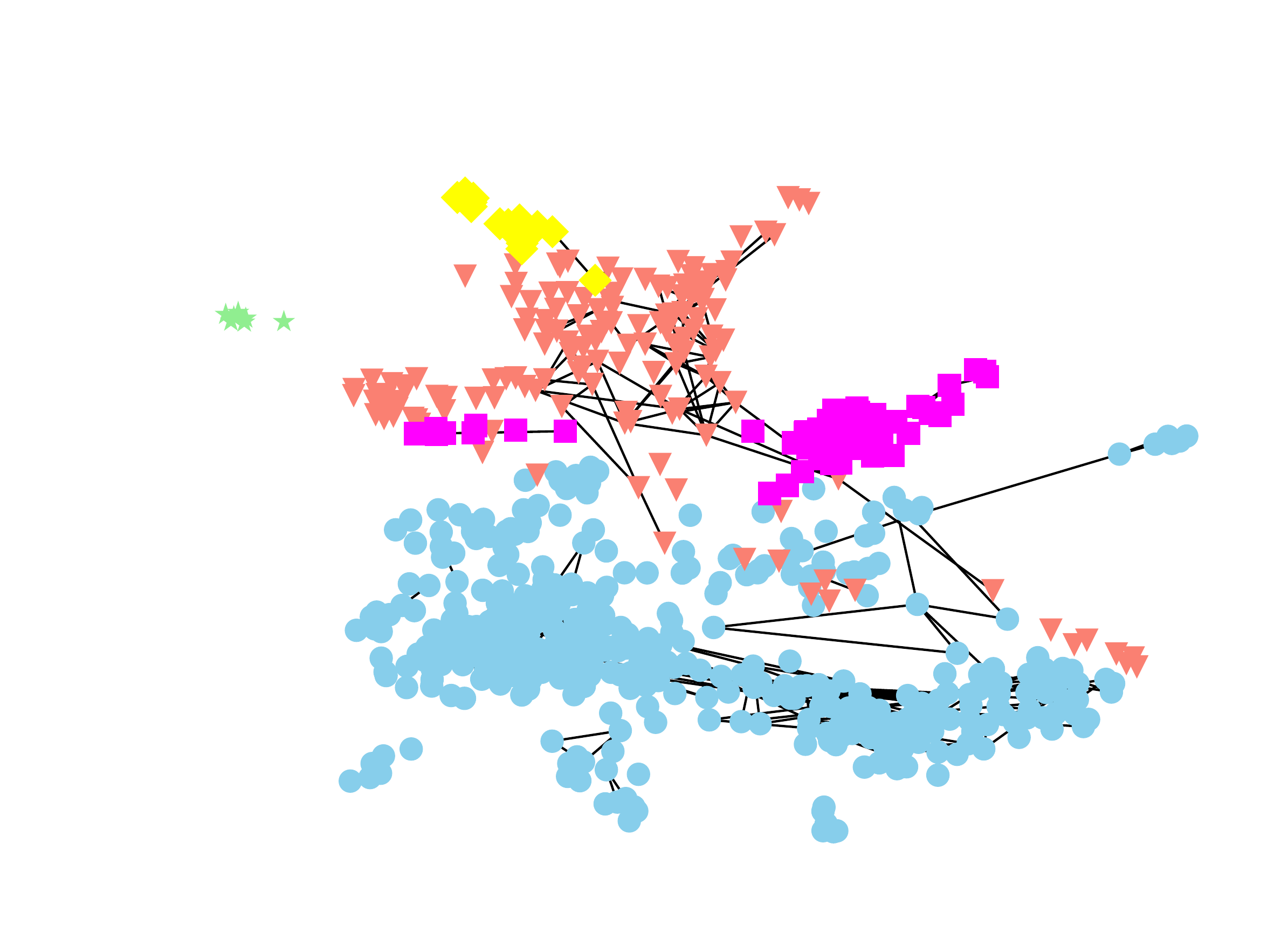}}
\subfloat[II]
{
\includegraphics[keepaspectratio, width=0.11\textwidth,trim={3cm 2cm 3.5cm 1.5cm},clip]{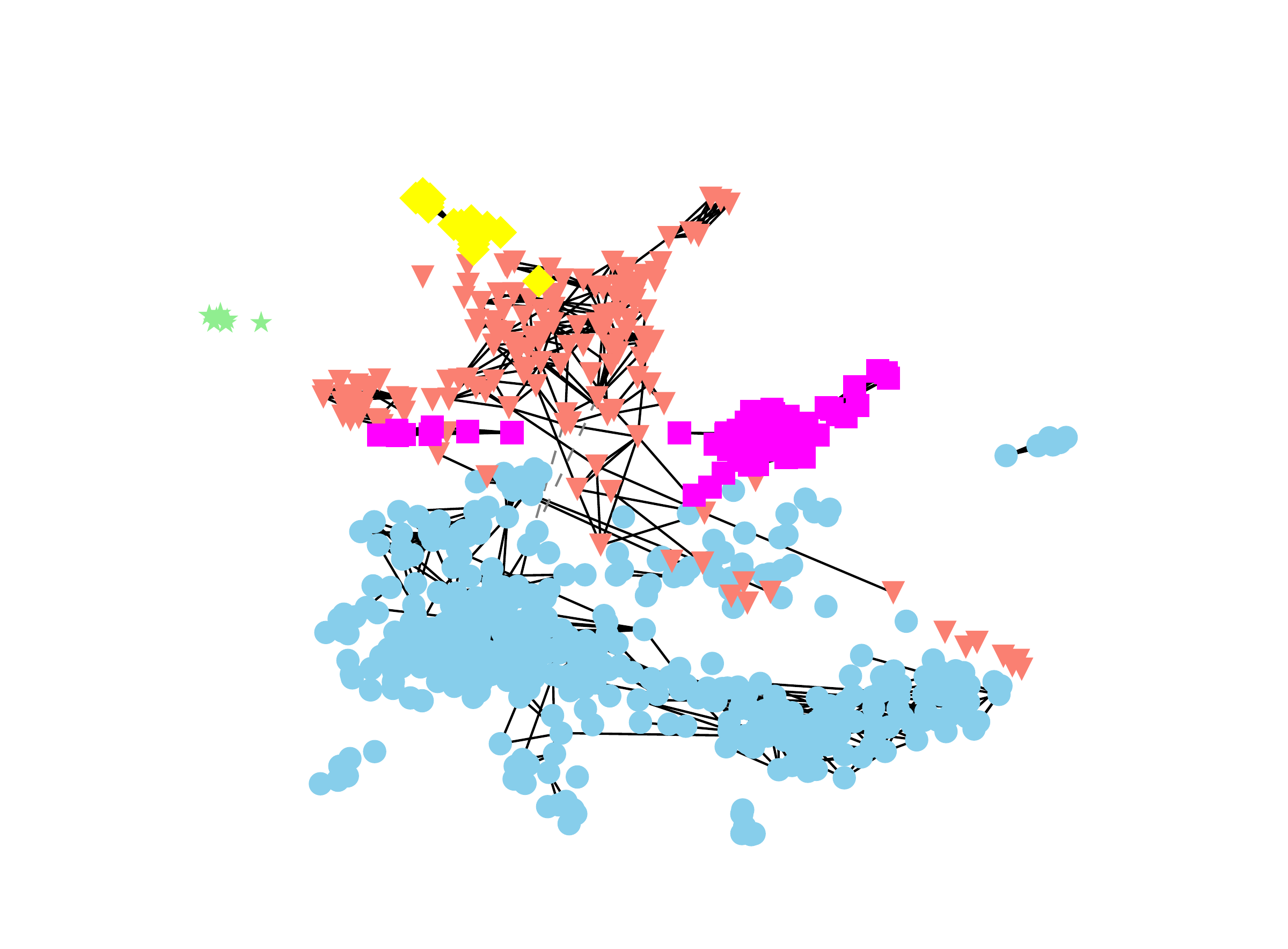}
}
\subfloat[III]
{
\includegraphics[keepaspectratio, width=0.11\textwidth,trim={3cm 2cm 3.5cm 1.5cm},clip]{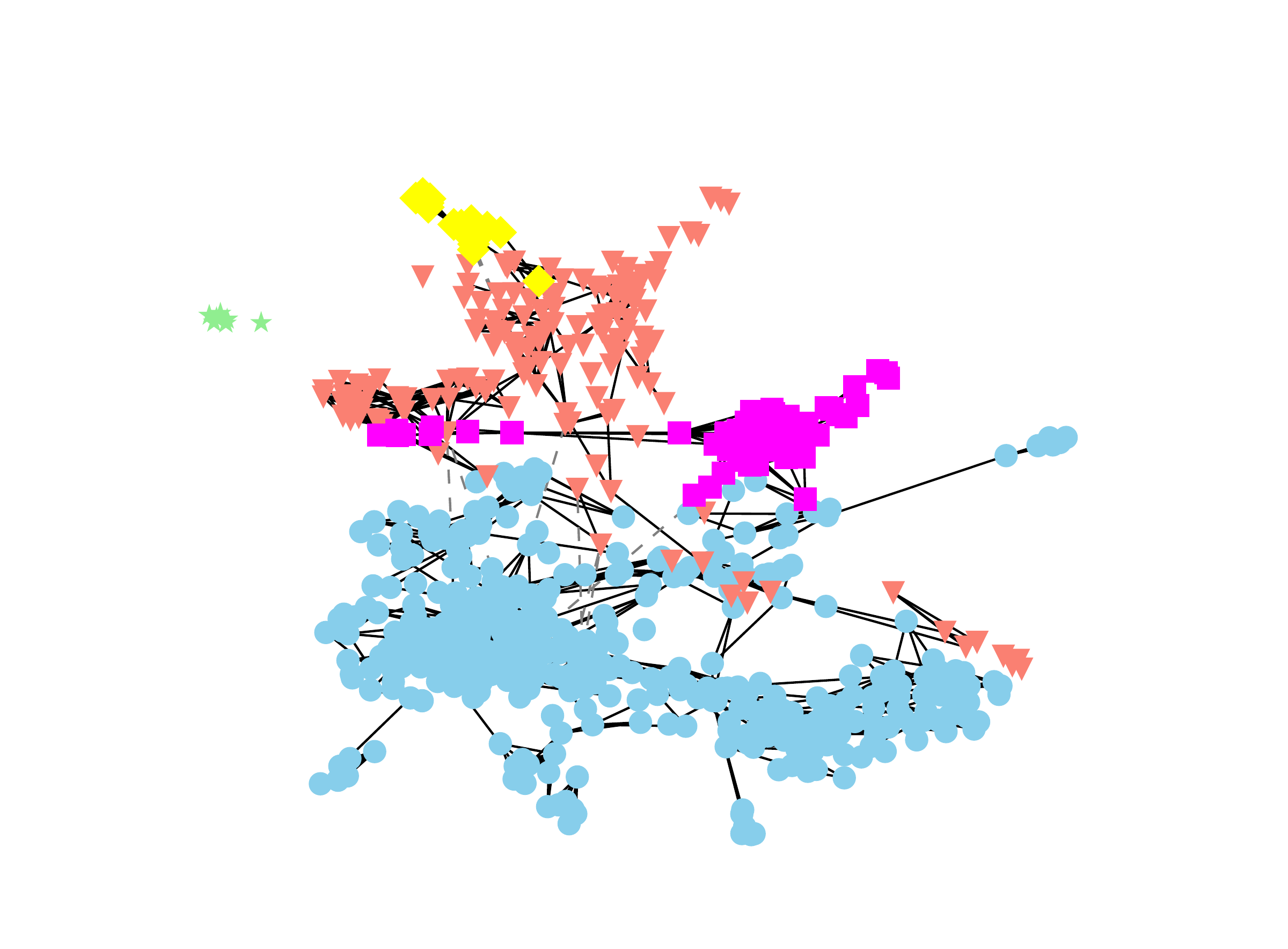}
}
\subfloat[IV]
{
\includegraphics[keepaspectratio, width=0.11\textwidth,trim={3cm 2cm 3.5cm 1.5cm},clip]{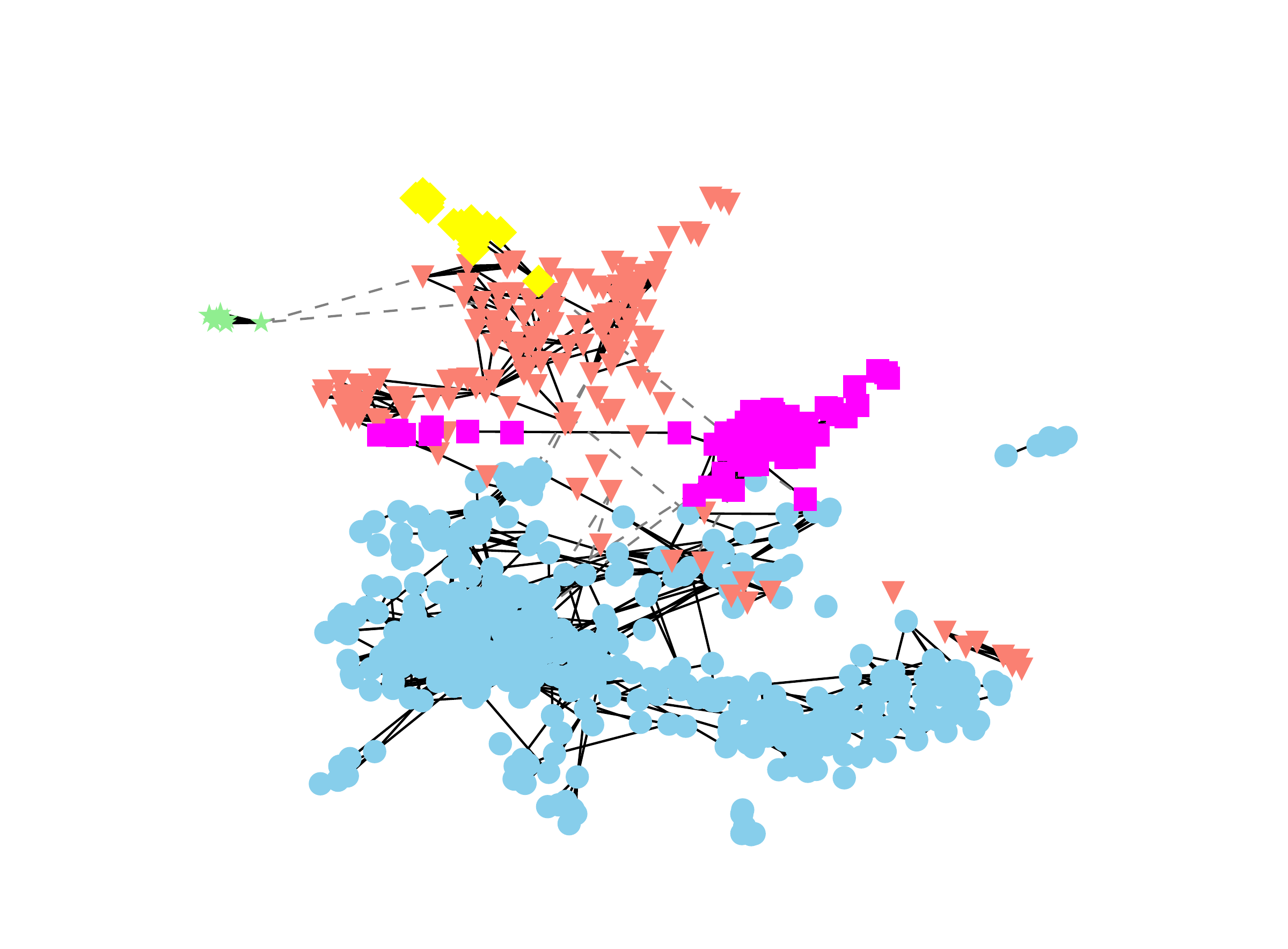}
}
\caption{Dynamic communities discovered using \textit{sparsest} (a-d) and \textit{normalized} (e-h) cuts for the \textit{DBLP} dataset (4 snapshots). %In this example, sparsity produces more dynamic (or changing) communities, while normalized sparsity leads to stable communities over time. 
\textit{Better seen in color}. \label{fig::dyn_comm_dblp}}
\end{figure} 

\subsection{Community Detection}
As discussed in Section \ref{sec::introduction}, dynamic community detection is an interesting application for temporal cuts. Two approaches from the literature, \textit{FacetNet} \cite{lin2008facetnet} and \textit{GenLovain} \cite{bazzi2016community}, are the baselines.  We focus our evaluation on \textit{School} and \textit{DBLP}, which have most meaningful communities. The following metrics are considered for comparison:

\textbf{Cut:} Total weight of the edges across partitions computed as $\sum_{t=1}^{m}\sum_{v,u \in G_t} w(u,v)(1-\delta(c_v,c_u))$, where $c_v$ and $c_u$ are the partitions to which $v$ and $u$ are assigned, respectively.

\textbf{Sparsity:} Sparsity ratio (Equation \ref{eqn::cut_ratio_temporal_graph}) for $k$-cuts \cite{shi2000normalized}: 
\begin{equation*}
\sum_{i=1}^k\frac{\sum_{t=1}^m |(X_{i,t},\overline{X}_{i,t}|+\beta \sum_{t=1}^{m-1}|(X_{i,t},\overline{X}_{i,t+1})|}{\sum_{t=1}^m |X_{i,t}||\overline{X}_{i,t}|}
\end{equation*}

\textbf{N-sparsity:} Normalized $k$-cut ratio (similar to sparsity).

\textbf{Modularity:} Temporal modularity, as defined in \cite{bazzi2016community}.

The baselines parameters were varied within a reasonable range of values and the best results were chosen. For \textit{GenLovain}, we fixed the number of partitions by agglomerating pairs while maximizing modularity \cite{bazzi2016community} and for \textit{FacetNet}, we assign each vertex to its highest weight partition.

\begin{figure}[ht!]
\centering
\subfloat[I \label{fig::dynamic_signal_traffic_alpha_0_0}]
{
\includegraphics[keepaspectratio, width=0.11\textwidth,trim={3cm 2cm 3cm 1cm},clip]{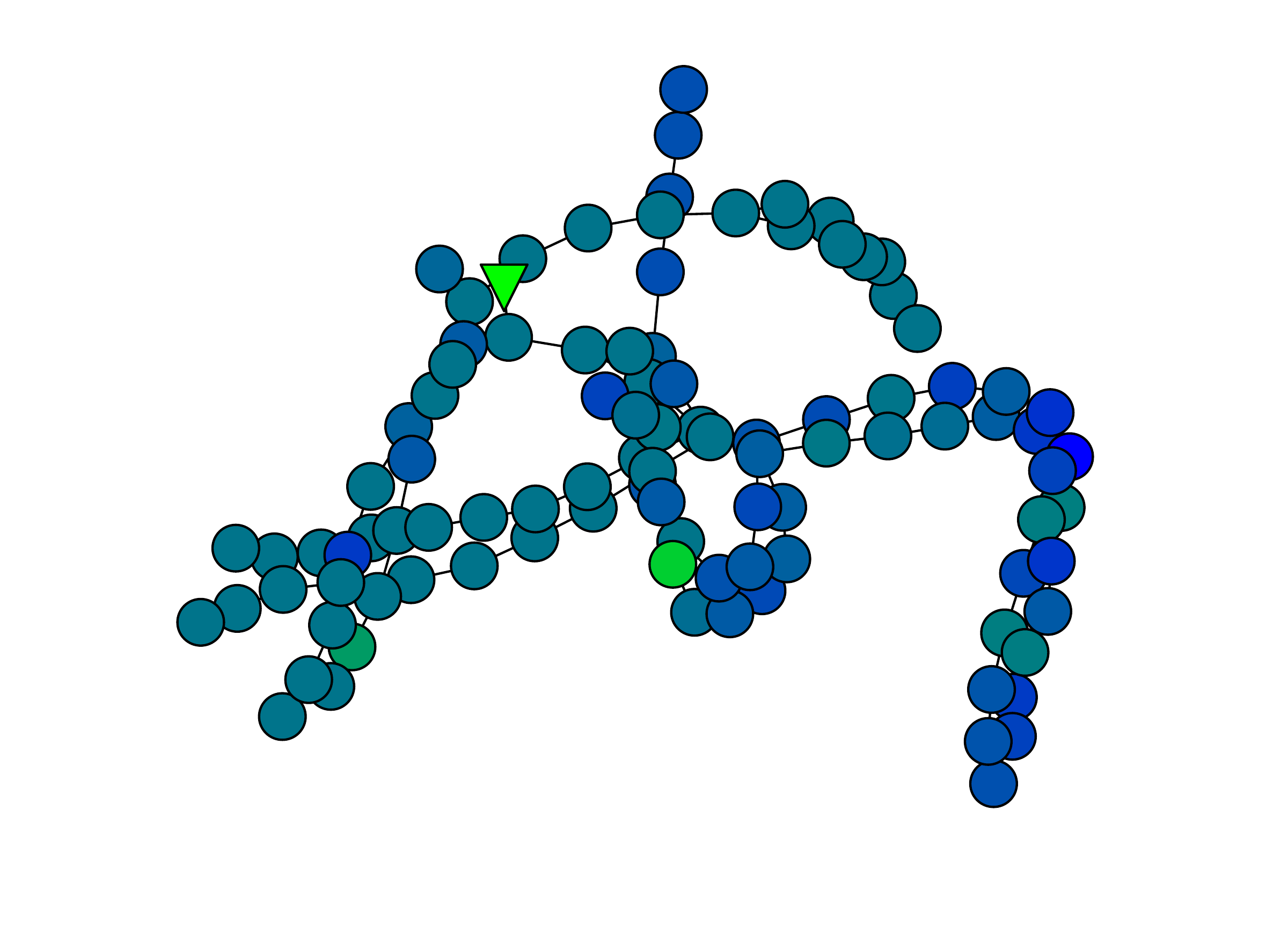}
}
\subfloat[II]
{
\includegraphics[keepaspectratio, width=0.11\textwidth,trim={3cm 2cm 3cm 1cm},clip]{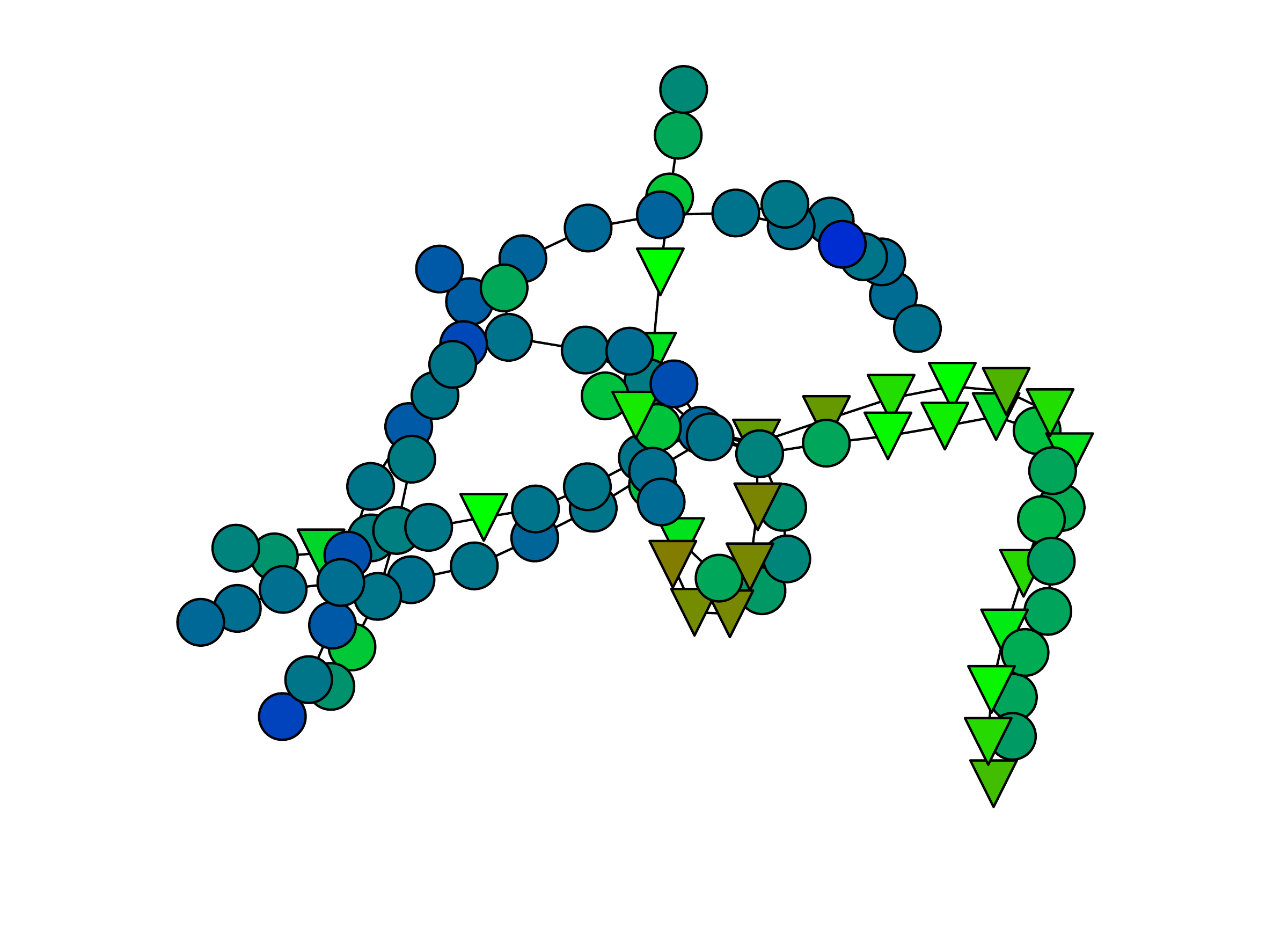}
}
\subfloat[III]
{
\includegraphics[keepaspectratio, width=0.11\textwidth,trim={3cm 2cm 3cm 1cm},clip]{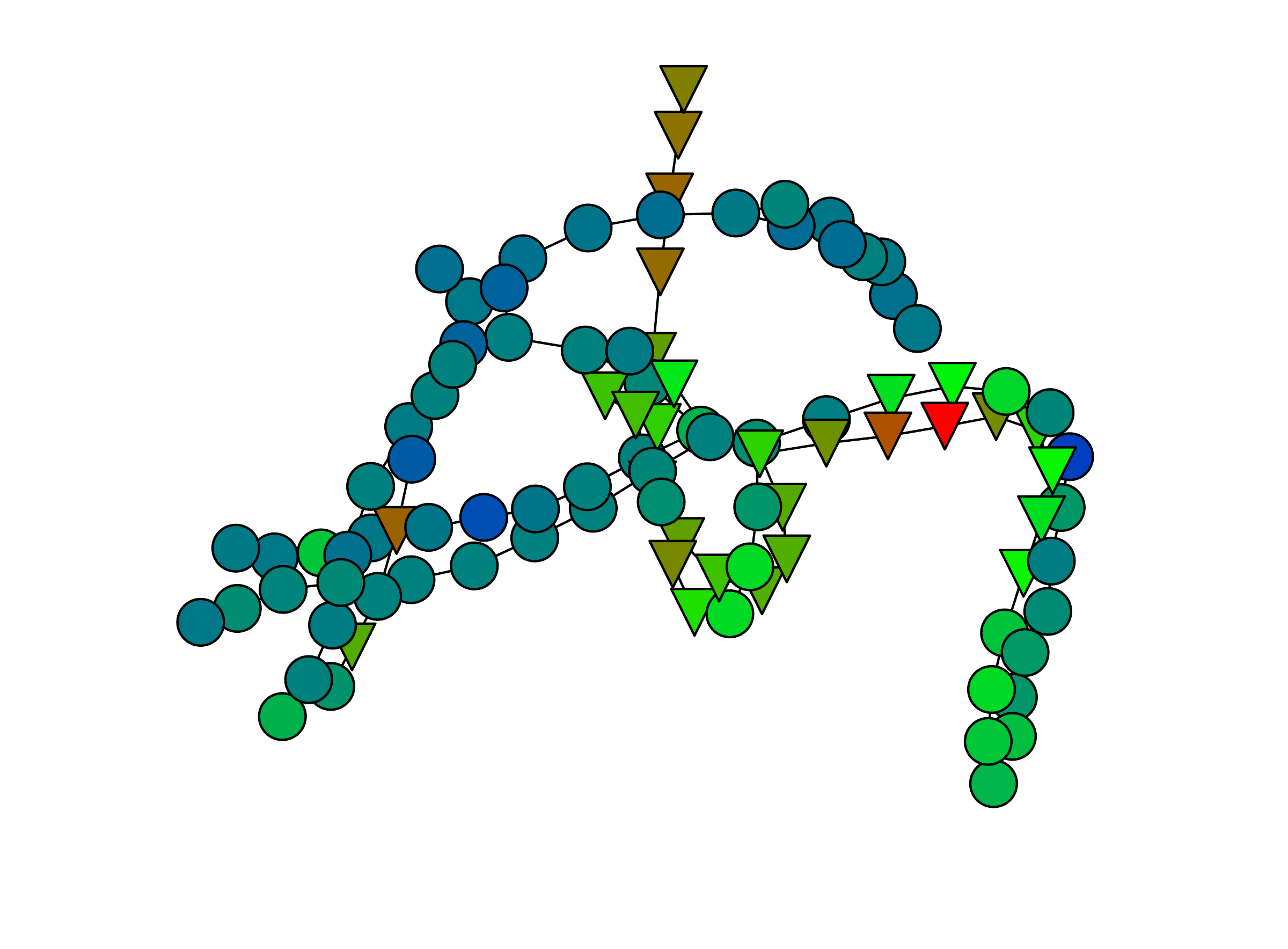}
}
\subfloat[IV \label{fig::dynamic_signal_traffic_alpha_0_3}]
{
\includegraphics[keepaspectratio, width=0.11\textwidth,trim={3cm 2cm 3cm 1cm},clip]{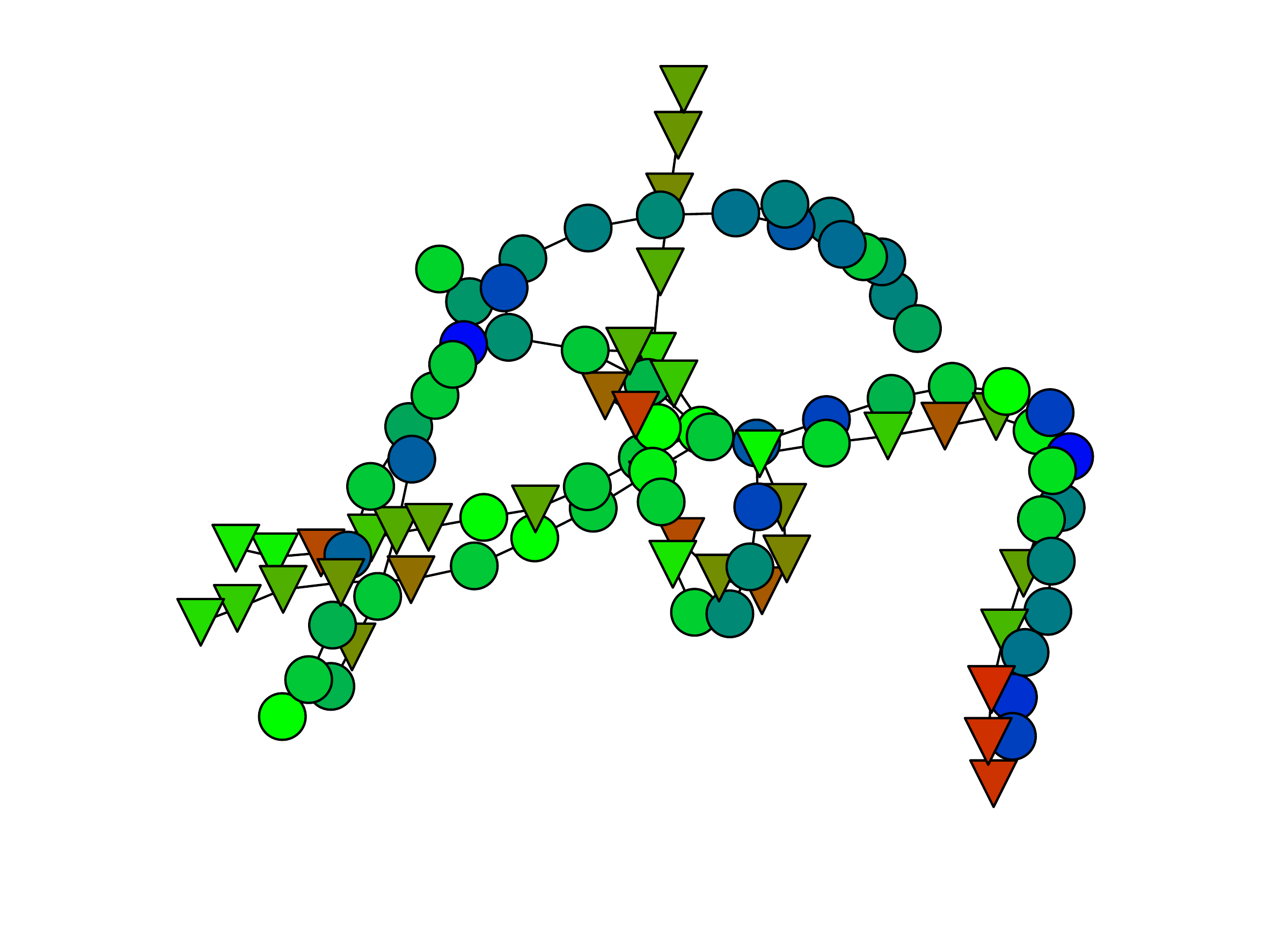}
}

\subfloat[I \label{fig::dynamic_signal_traffic_alpha_2_0}]
{
\includegraphics[keepaspectratio, width=0.11\textwidth,trim={3cm 2cm 3cm 1cm},clip]{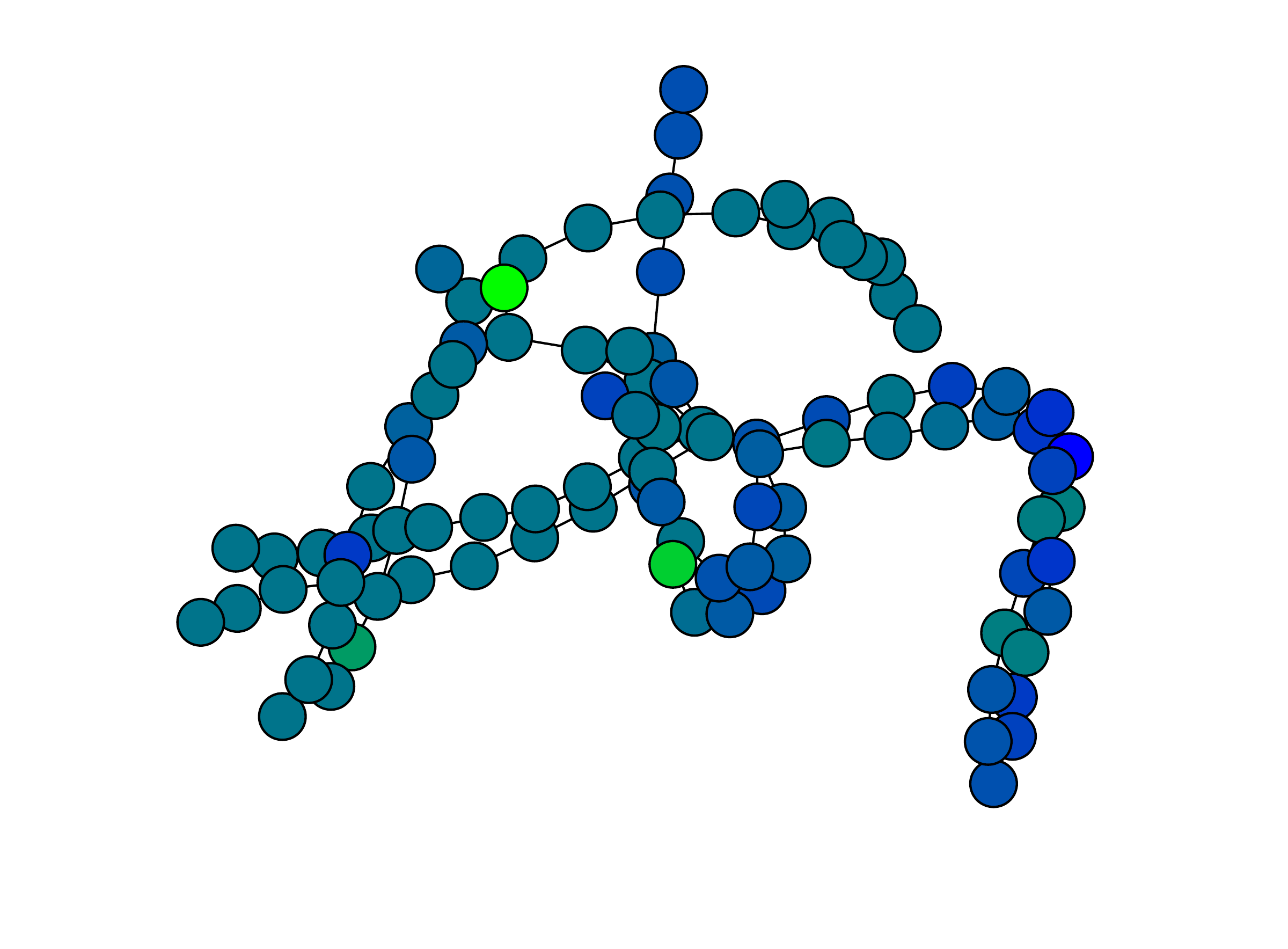}
}
\subfloat[II]
{
\includegraphics[keepaspectratio, width=0.11\textwidth,trim={3cm 2cm 3cm 1cm},clip]{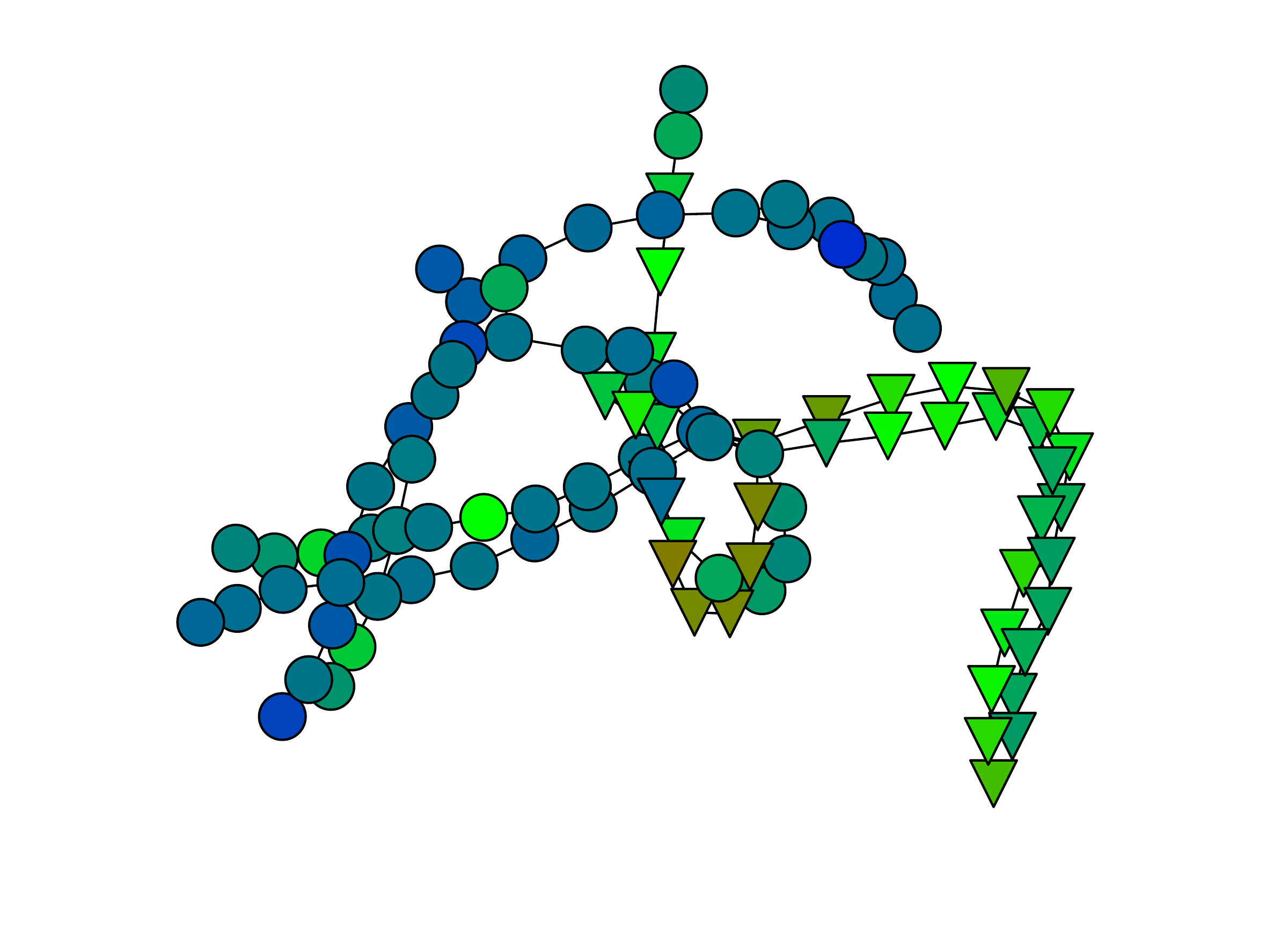}
}
\subfloat[III]
{
\includegraphics[keepaspectratio, width=0.11\textwidth,trim={3cm 2cm 3cm 1cm},clip]{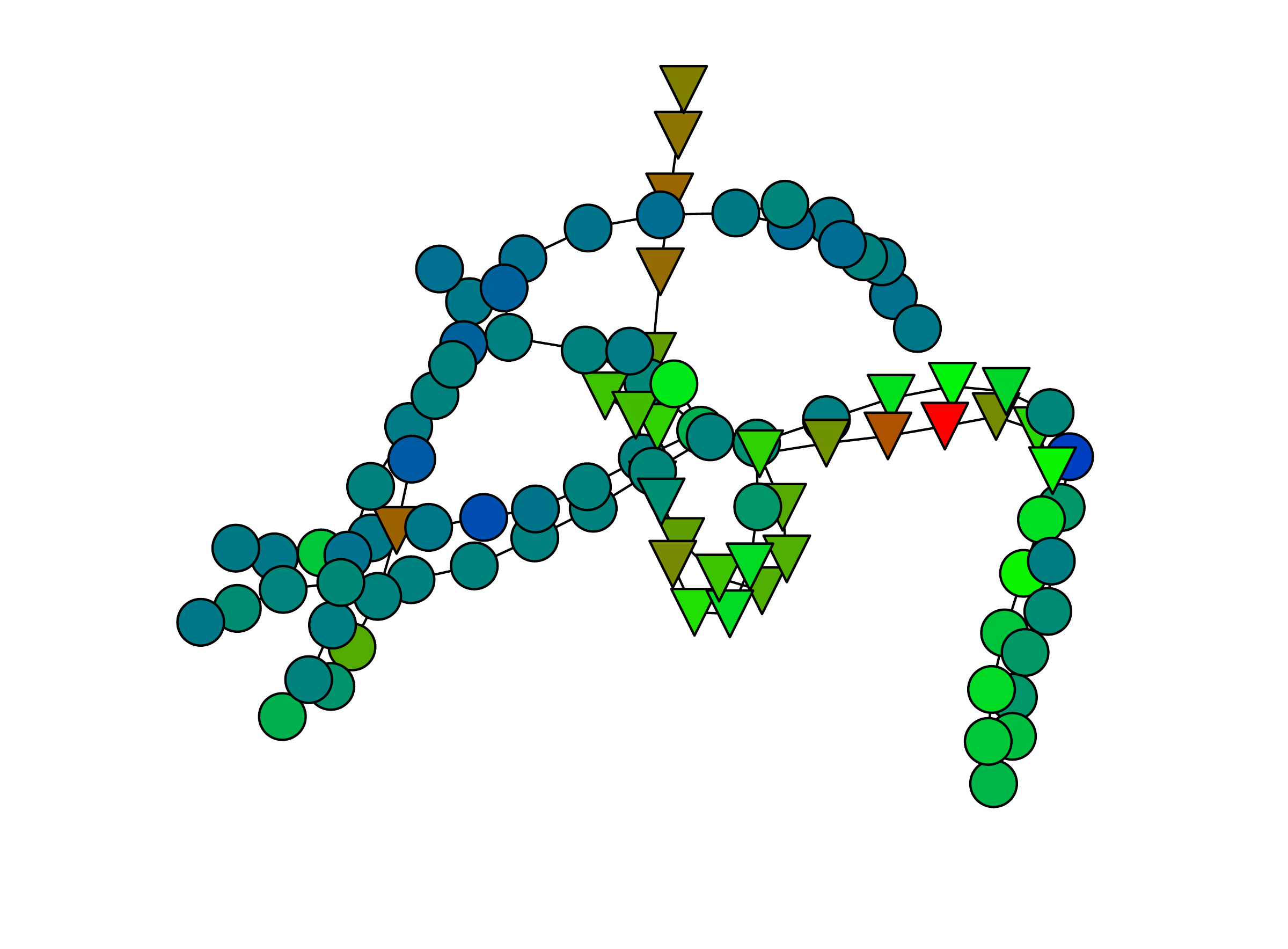}
}
\subfloat[IV \label{fig::dynamic_signal_traffic_alpha_2_3}]
{
\includegraphics[keepaspectratio, width=0.11\textwidth,trim={3cm 2cm 3cm 1cm},clip]{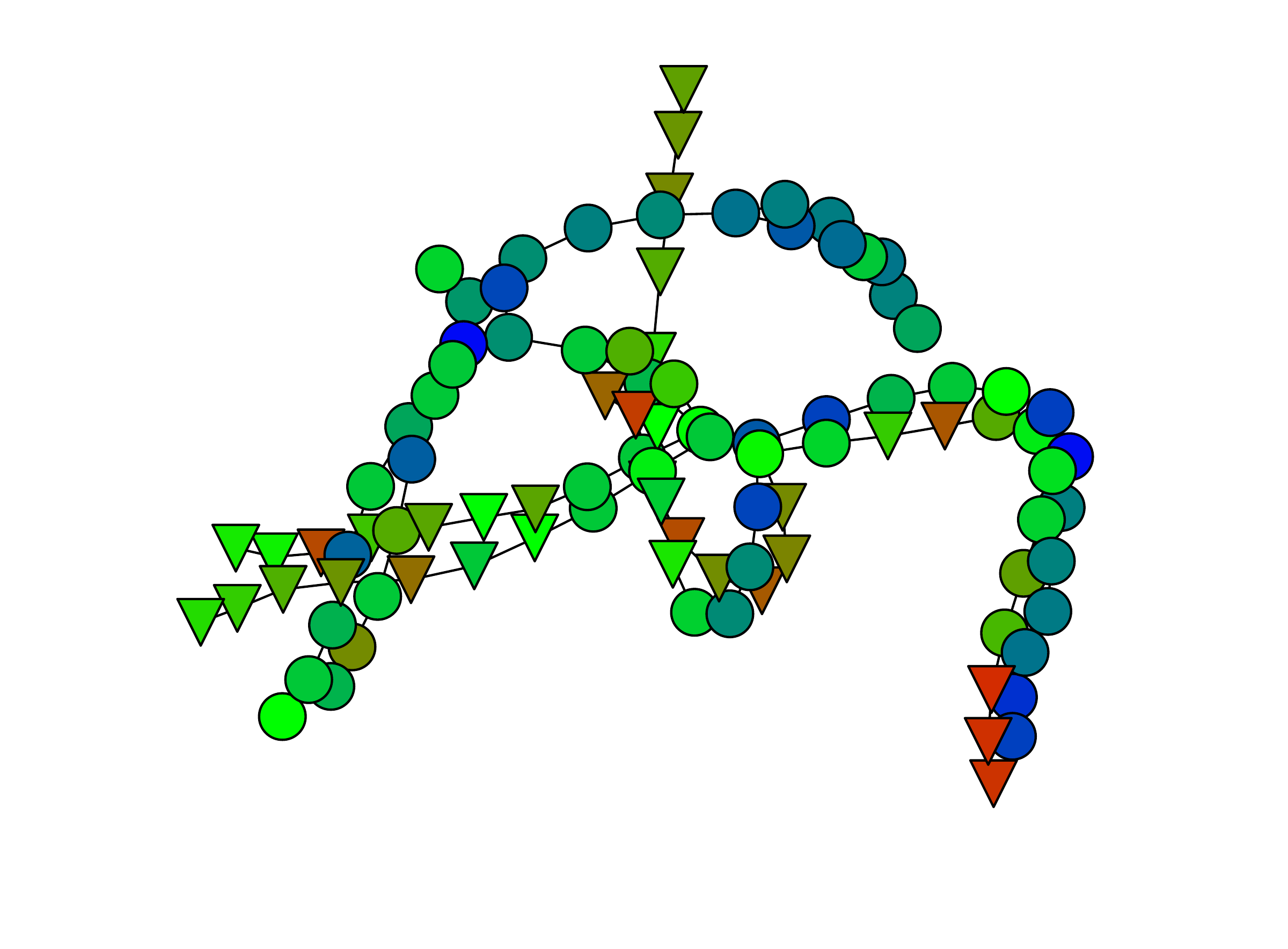}
}

\subfloat[I \label{fig::dynamic_signal_traffic_beta_2_0}]
{
\includegraphics[keepaspectratio, width=0.11\textwidth,trim={3cm 2cm 3cm 1cm},clip]{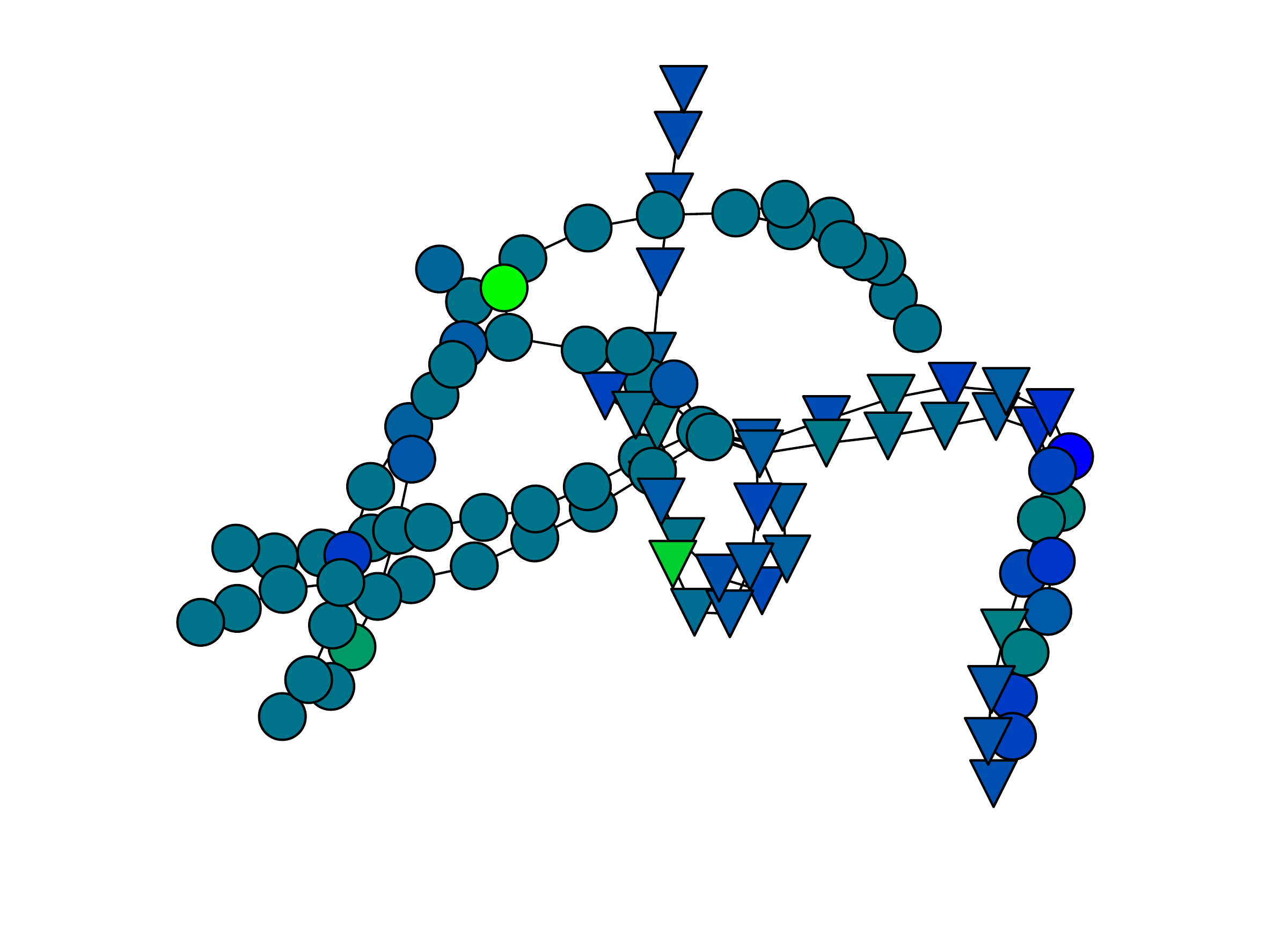}
}
\subfloat[II]
{
\includegraphics[keepaspectratio, width=0.11\textwidth,trim={3cm 2cm 3cm 1cm},clip]{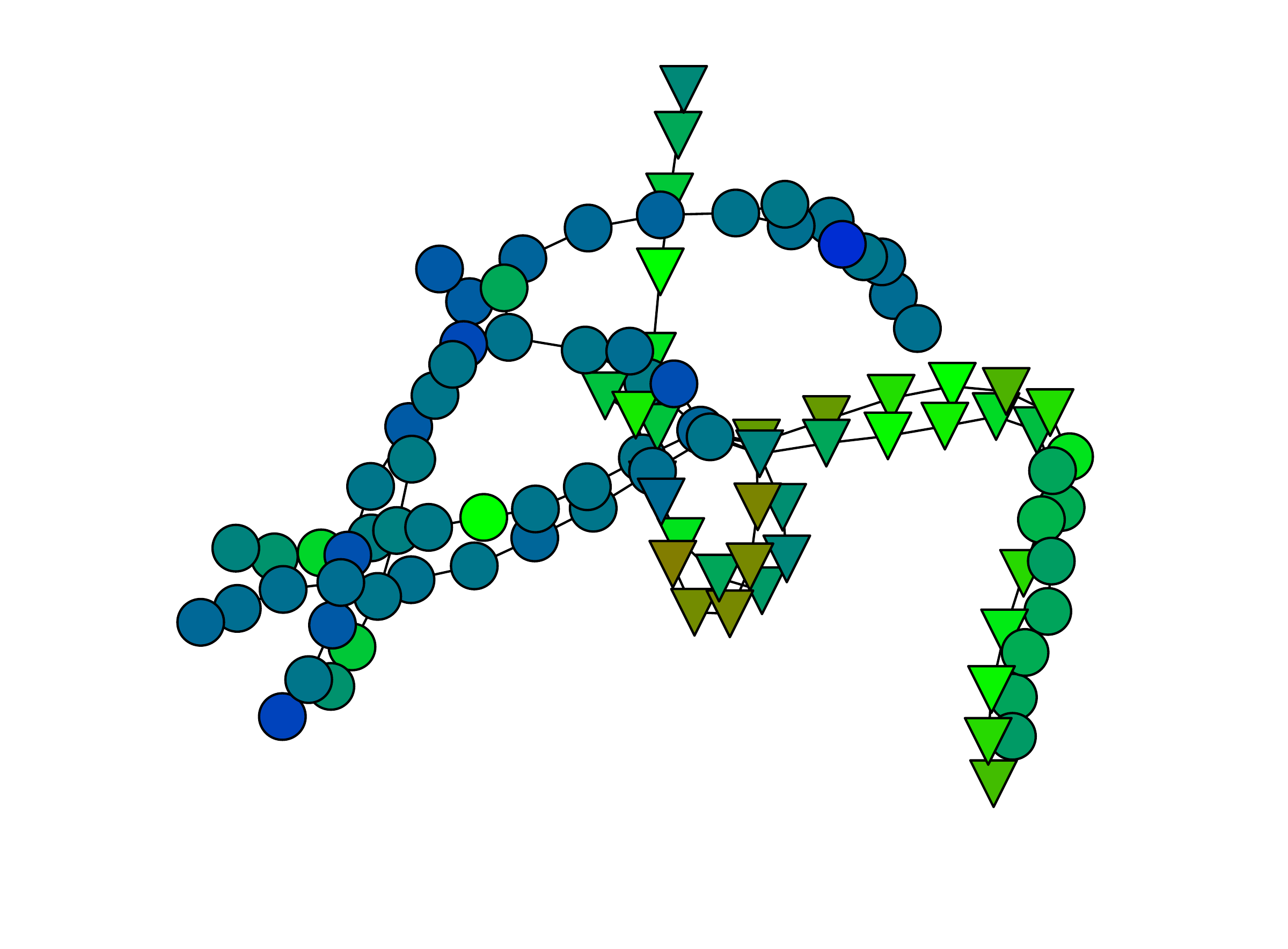}
}
\subfloat[III]
{
\includegraphics[keepaspectratio, width=0.11\textwidth,trim={3cm 2cm 3cm 1cm},clip]{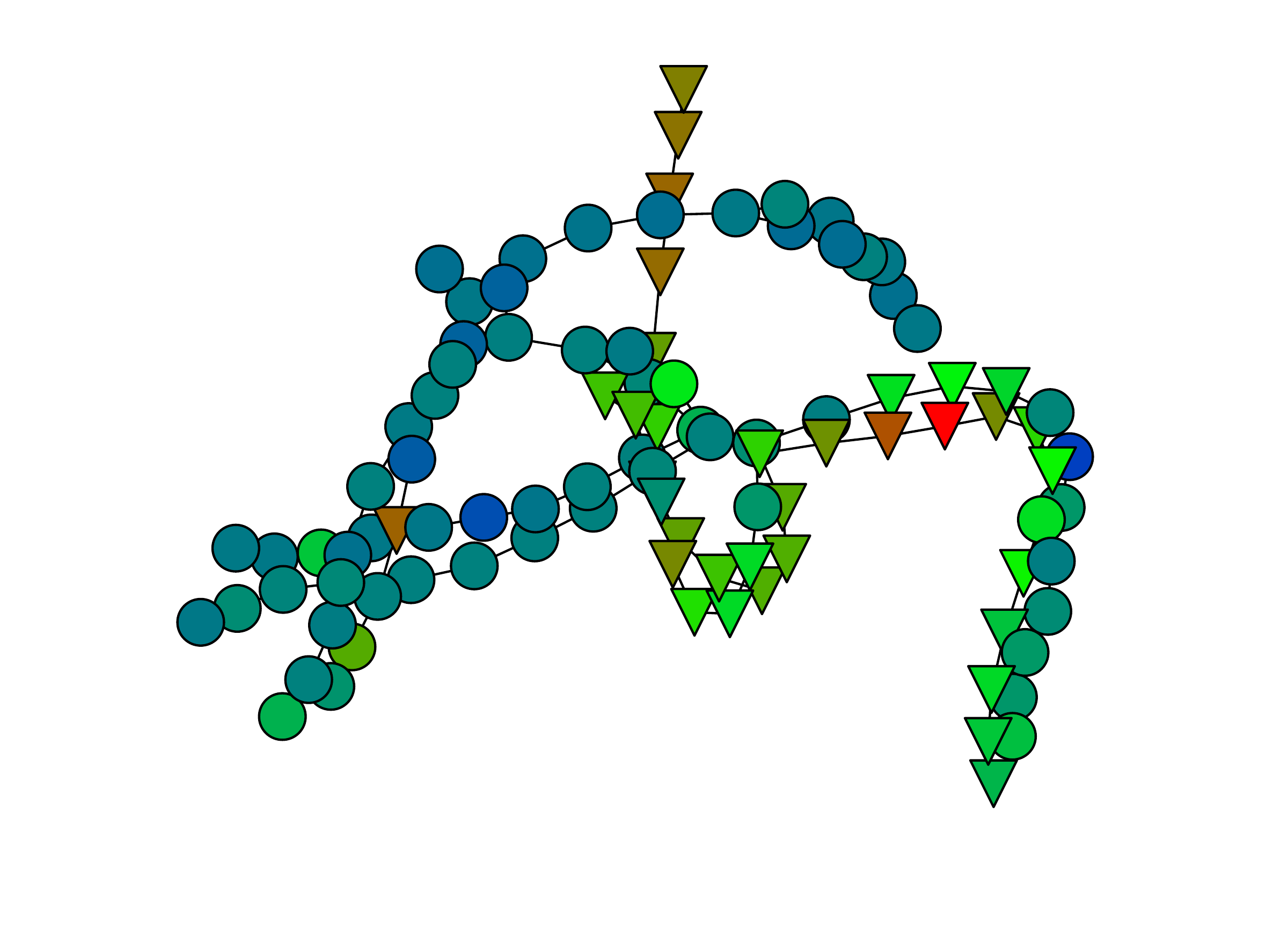}
}
\subfloat[IV \label{fig::dynamic_signal_traffic_beta_2_3}]
{
\includegraphics[keepaspectratio, width=0.11\textwidth,trim={3cm 2cm 3cm 1cm},clip]{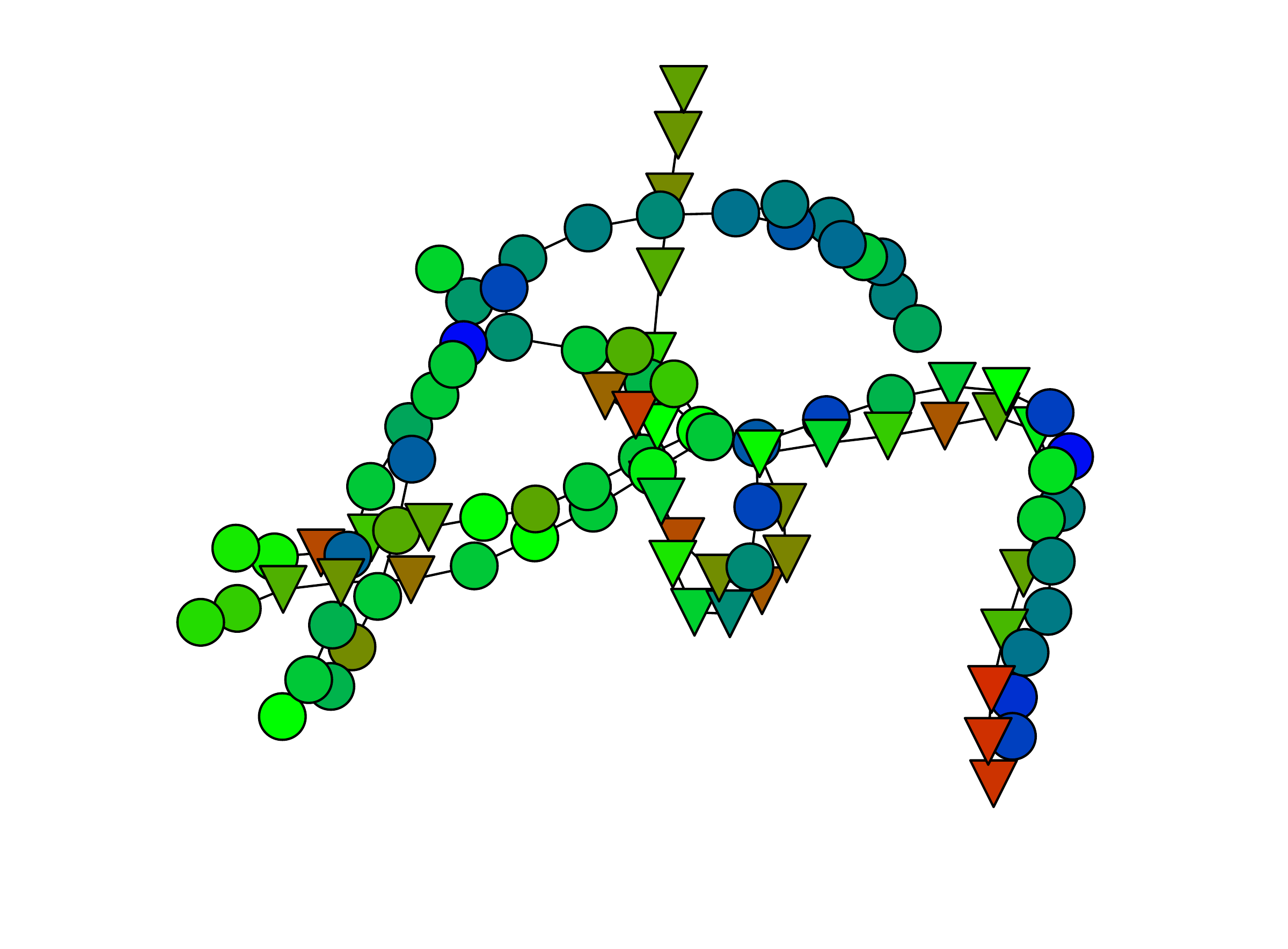}
}
\caption{Wavelet cut of a 4-snapshot dynamic traffic network with vehicle speeds as a signal. Vertex colors correspond to speed values (red for high and blue for low) and shapes indicate the partitions for 3 different settings: $\alpha=0.$ and $\beta=1$ (a-d, no network effect), $\alpha=200.$ and $\beta=1.$ (e-h, large network effect with low smoothness), and $\alpha=200.$ and $\beta=10.$ (i-l, large network effect and high smoothness) \label{fig::dynamic_signal_traffic}.}
\end{figure} 
\begin{figure}[ht!]
\centering
\subfloat[Traffic \label{fig::signal_compression_traffic}]{
\includegraphics[keepaspectratio, width=0.23\textwidth]{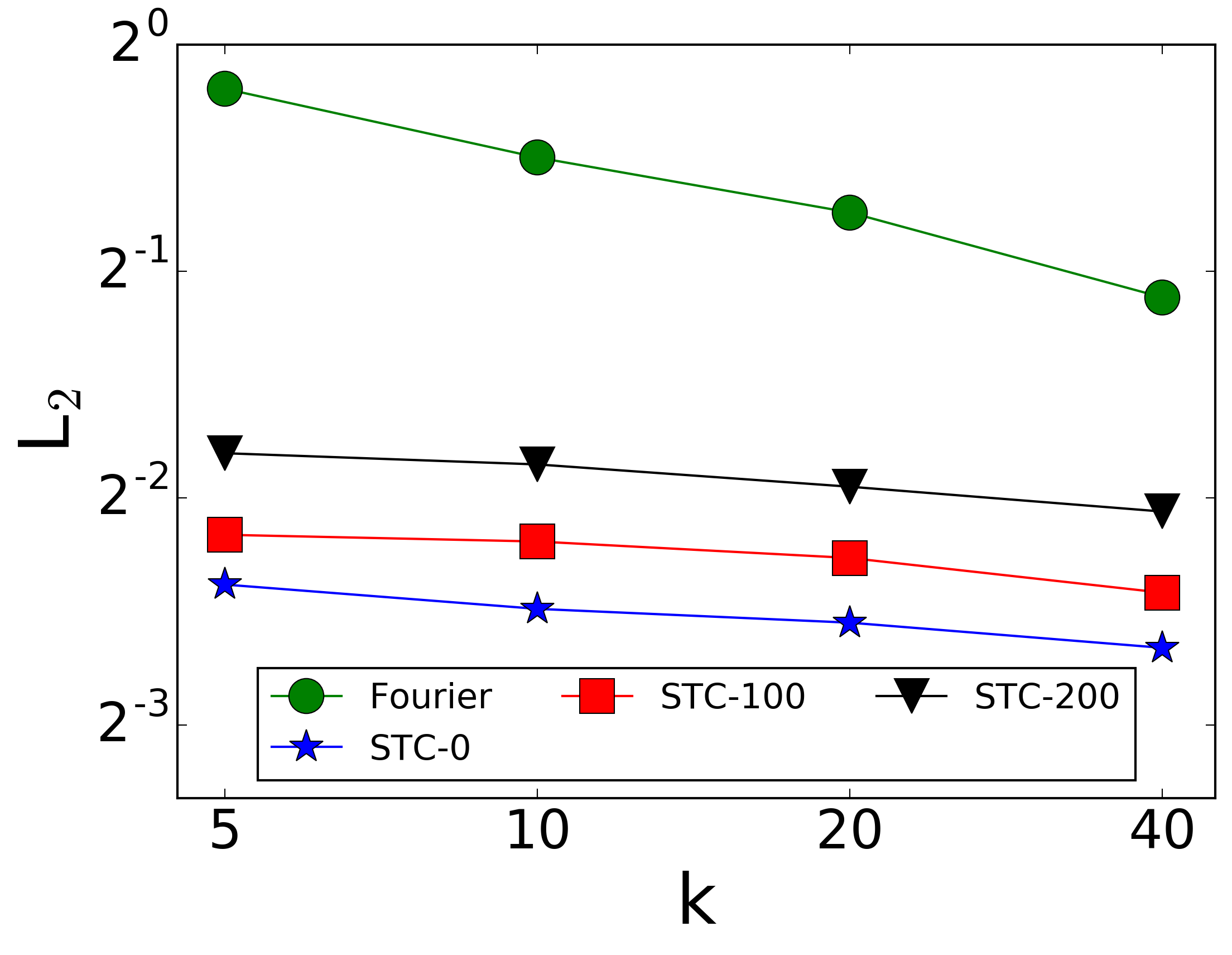}
}
\subfloat[School-heat \label{fig::signal_compression_heat}]{
\includegraphics[keepaspectratio, width=0.23\textwidth]{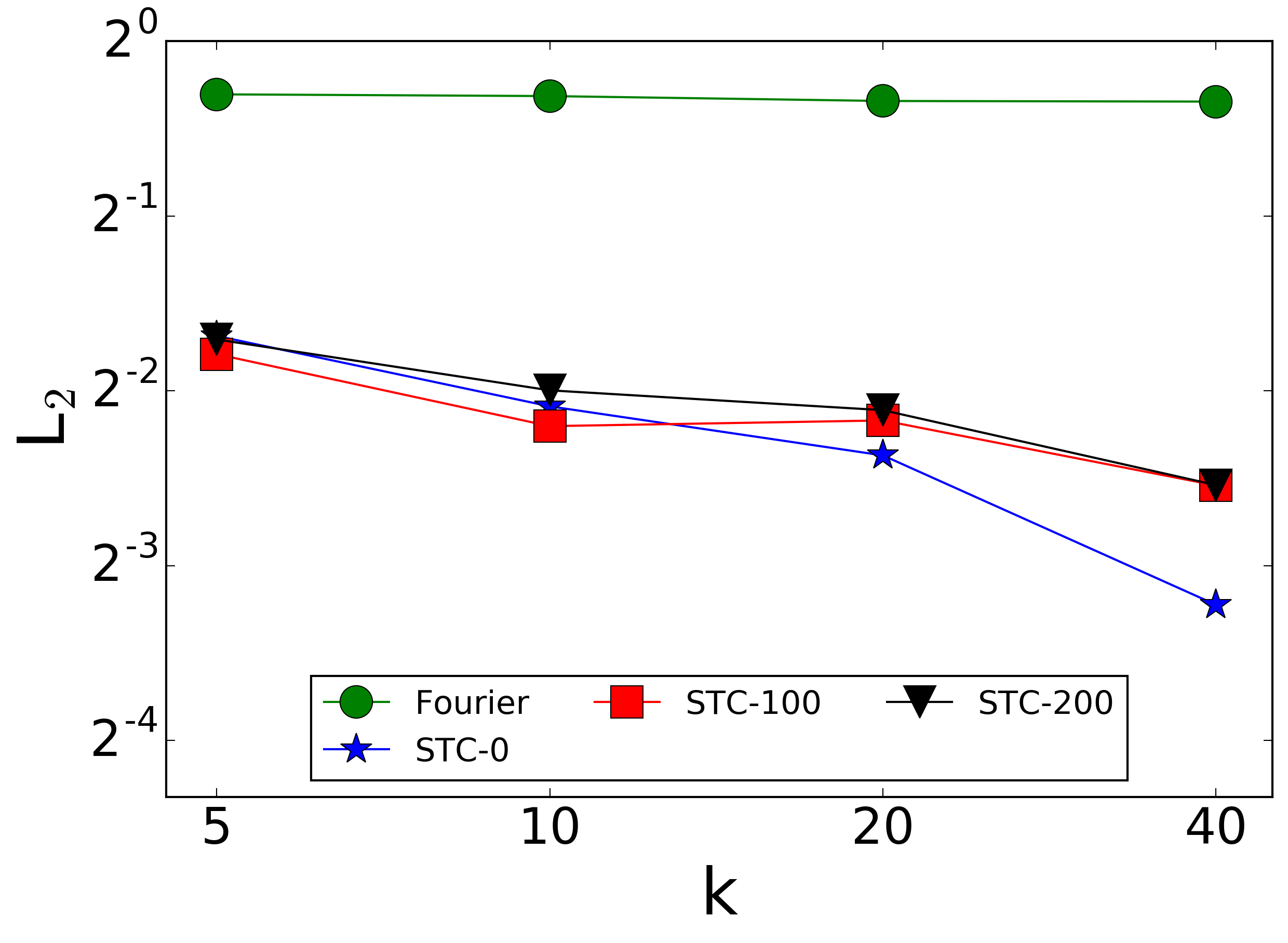}
}
\caption{$L_2$ error with different representation sizes $k$ for Graph Fourier and our approach while setting the regularization parameter $\alpha$ to $200$, $100$, and $0$. \label{fig::signal_compression}}
\end{figure}

Community detection results, for $2$ and $5$ communities, are shown in Table \ref{table::comm_detec_results}. For \textit{School} (\ref{table::comm_detec_school}), both \textit{GenLovain} and our methods found the same communities ($\beta=0.25$) when $k=2$, outperforming \textit{FacetNet} in all the metrics. However, for $k=5$, different communities were discovered by the methods, with \textit{Sparsest} and \textit{Normalized Cuts} achieving the best results in terms of sparsity and n-sparsity, respectively. Our methods also achieve competitive results in terms of modularity. Similar results were found using \textit{DBLP} ($\beta=0.5$), as shown in Table \ref{table::comm_detec_dblp}, although \textit{Sparsest} and \textit{Normalized Cuts} switch as the best method for each other's metric in some settings. This is possible because our algorithms are approximations (i.e. not optimal). We illustrate the communities found in \textit{School} ($k=2$, same as single cut) and \textit{DBLP} ($k=5$) in Figures \ref{fig::example_primary_school} and \ref{fig::dyn_comm_dblp}, respectively.

\subsection{Signal Processing on Graphs}

We finish our evaluation with the analysis of dynamic signals on graphs. In Figure \ref{fig::dynamic_signal_traffic}, we illustrate three dynamic wavelets for \textit{Traffic} discovered using our approach under different settings. First, in Figures \ref{fig::dynamic_signal_traffic_alpha_0_0}-\ref{fig::dynamic_signal_traffic_alpha_0_3}, we consider cuts that take only the graph signal into account by setting both the regularization parameter $\alpha$ and the smoothness parameter $\beta$ to $0$, which leads to a cut that follows the traffic speeds but has many edges and is not smooth. Next (Figures \ref{fig::dynamic_signal_traffic_alpha_2_0}-\ref{fig::dynamic_signal_traffic_alpha_2_3}), we increase $\alpha$ to $200$, producing a much sparser cut that is still not smooth. Finally, in Figures \ref{fig::dynamic_signal_traffic_beta_2_0}-\ref{fig::dynamic_signal_traffic_beta_2_3}, we increase the smoothness $\beta$ to $10$, which forces most of the vertices to remain in the same partition despite of speed variations.

We also evaluate our approach in signal compression, which consists of computing a compact representation for a dynamic signal. As a baseline, we consider the \textit{Graph Fourier} scheme \cite{shuman2013emerging} applied to the temporal graph (i.e. the multiplex graph that combines all the snapshots). The size of the representation ($k$) is the number of partitions and the number of top eigenvectors for our method and \textit{Graph Fourier}, respectively. Figures \ref{fig::signal_compression_traffic} and \ref{fig::signal_compression_heat} show the compression results in terms of $L_2$ error using a fixed representation size $k$ for the \textit{Traffic} and \textit{School-heat} datasets, respectively. We vary the value of the regularization parameter $\alpha$, which controls the impact of the network structure over the wavelets computed, for our method. As expected, a larger value of $\alpha$ leads to a higher $L_2$ error. However, even for a high regularization, our approach is still able to compute wavelets that accurately compress the signal, outperforming the baseline.

\section{Conclusion}

This paper studied cut problems in temporal graphs. Extensions of two existing graph cut problems, the sparsest and normalized cuts, by enforcing the smoothness of cuts over time, were introduced. To solve these problems, we have proposed spectral approaches based on multiplex graphs that compute relaxed versions of temporal cuts as eigenvectors. Scalable versions of our solutions using divide-and-conquer and low-rank matrix approximation were also presented. In order to compute cuts that take into account also graph signals, we have extended graph wavelets to the dynamic setting. Experiments have shown that our temporal cut algorithms outperform the baseline methods in terms of quality and are competitive regarding running time. Moreover, temporal cuts enable the discovery of dynamic communities and the analysis of dynamic processes on graphs.

This work opens several lines for future investigation: (i) temporal cuts, as a general framework for solving problems involving dynamic data, can be applied in many scenarios, we are particularly interested to see how our method performs in computer vision tasks; (ii) Perturbation Theory can provide deeper theoretical insights into the properties of temporal cuts  (see \cite{taylor2015eigenvector,sole2013spectral}); finally, (iii) we want to study \textit{Cheeger inequalities} \cite{chung1997spectral} for temporal cuts, as means to better understand the performance of our algorithms.

\textbf{Acknowledgment}. Research was sponsored by the Army Research Laboratory and was accomplished under Cooperative Agreement Number W911NF-09-2-0053 (the ARL Network Science CTA). The views and conclusions contained in this document are those of the authors and should not be interpreted as representing the official policies, either expressed or implied, of the Army Research Laboratory or the U.S. Government. The U.S. Government is authorized to reproduce and distribute reprints for Government purposes notwithstanding any copyright notation here on.
\clearpage

\bibliographystyle{abbrv}
\bibliography{dynpros}  % sigproc.bib is the name of the Bibliography in this case
\clearpage
\appendix

\textbf{Proof of Lemma \ref{lemm::temporal_cut}}
\begin{proof}
Since $\mathcal{L}$ is the Laplacian of $\chi(\mathcal{G})$:
\begin{equation}
\begin{split}
\textbf{x}^{\intercal}\mathcal{L}\textbf{x} &= \sum_{(u,v) \in \mathcal{E}}(\textbf{x}[u]-\textbf{x}[v])^2W(u,v) \\
					& = \sum_{t=1}^m \sum_{(u,v) \in E_t} (\textbf{x}[u]-\textbf{x}[v])^2W(u,v)\\
					&\quad +\beta \sum_{t=1}^{m-1}\sum_{v \in V} (\textbf{x}[v_t]-\textbf{x}[v_{t+1}])^2\\
					& = 4\sum_{t=1}^m |(X_t,\overline{X}_t)|+4 \sum_{t=1}^{m-1}\Delta(X_t,\overline{X}_{t+1})\\
\end{split}
\nonumber
\end{equation}

Regarding the denominator:
\begin{equation}
\begin{split}
\textbf{x}^{\intercal}\mathcal{C}\textbf{x} &= \sum_{t=1}^m \sum_{u \neq v} (\textbf{x}[v_t]-\textbf{x}[u_t])^2\\
					& = 4\sum_{t=1}^m |X_t||\overline{X}_t|
					%\qedhere
\end{split}
\nonumber
\end{equation}
\end{proof}

\textbf{Proof of Lemma \ref{lemm::pseudoinverse_square_root_C}}
\begin{proof}
The spectrum of $C$ is $(\textbf{e}_1,\lambda_1)=(\textbf{1}_n,0)$ and $\lambda_2 = \ldots = \lambda_n = n$ for any vector $\textbf{e}_i \perp \textbf{1}_n$. As a consequence, the spectrum of $\mathcal{C}$ is in the form: $\lambda_1 = \ldots = \lambda_m = 0$ and $\lambda_{m+1} = \ldots \lambda_{nm} = n$ for any vector $\textbf{e}_i \perp \textbf{span}\{e_1, \ldots e_m\}$. We factorize $\mathcal{C}$ as $U\Lambda U^{\intercal}=n\sum_{t=m+1}^{mn}\textbf{e}_i.\textbf{e}_i^{\intercal}$, where $U$ and $\Lambda$ are eigenvector and eigenvalue matrices of $\mathcal{C}$. Similarly, we can factorize $\mathcal{C}^+=\frac{1}{n}\sum_{t=m+1}^{mn}\textbf{e}_i.\textbf{e}_i^{\intercal} = \frac{1}{n^2}\mathcal{C}$. Taking the square-root, $(\mathcal{C}^+)^{\frac{1}{2}} = \frac{1}{n}\mathcal{C}^{\frac{1}{2}} = \frac{1}{\sqrt{n}}\mathcal{C}$.
\end{proof}

\textbf{Proof of Lemma \ref{lemm::cut_relaxation_one}}
\begin{proof}
From Lemma \ref{lemm::temporal_cut}, after relaxing the constraint:
\begin{equation}
\textbf{x}* = \argmin_{\textbf{x} \in [-1,1]^{nm}, \textbf{x}\mathcal{C}\textbf{x} > 0} \frac{\textbf{x}^{\intercal}\mathcal{L}\textbf{x}}{\textbf{x}^{\intercal}\mathcal{C}\textbf{x}}
\label{eqn::relax_ratio}
\end{equation}
By performing the substitution $\textbf{y} = \mathcal{C}^{\frac{1}{2}}\textbf{x}$, we get:
\begin{equation}
\textbf{y}* = \argmin_{\mathcal{C}\textbf{y} \neq 0} \frac{\textbf{y}^{\intercal}(\mathcal{C}^+)^{\frac{1}{2}}\mathcal{L}(\mathcal{C}^+)^{\frac{1}{2}}\textbf{y}}{\textbf{y}^{\intercal}\textbf{y}}
\label{eqn::ratio_y}
\end{equation}
This is related to the variational characterization of the spectrum of $(\mathcal{C}^+)^{\frac{1}{2}}\mathcal{L}(\mathcal{C}^+)^{\frac{1}{2}}$, with the constraint that no solution be in the null space of $\mathcal{C}$. From Lemma \ref{lemm::pseudoinverse_square_root_C}, $(\mathcal{C}^+)^{\frac{1}{2}}\mathcal{L}(\mathcal{C}^+)^{\frac{1}{2}} = \frac{1}{n} \mathcal{CLC}$. Multiplying a matrix by a scalar does not change its eigenvectors. It holds that $\textbf{y}^{\intercal}(\mathcal{C}^+)^{\frac{1}{2}}\mathcal{L}(\mathcal{C}^+)^{\frac{1}{2}}\textbf{y} = 0$ if $\mathcal{C}\textbf{y} = \textbf{0}_{nm}$. Given that $\mathcal{G}$ is connected, our solution is the vector associated with the ($m$$+$$1$)-th smallest eigenvalue.
\end{proof}

\textbf{Proof of Lemma \ref{lemm::C_commutes}}
\begin{proof}
For any real $nm \times nm$ matrix $\mathcal{A}$, an eigenvector of $\mathcal{A}$ is also an eigenvector of $\mathcal{C}$ (see proof for Lemma \ref{lemm::pseudoinverse_square_root_C}). Therefore, $\mathcal{A}$ and $\mathcal{C}$ are simultaneously diagonalizeable, which is a sufficient condition for their commutability.
\end{proof}

\textbf{Proof of Theorem \ref{thm::norm_diff_matrix}}
\begin{proof}
Based on Lemmas \ref{lemm::temporal_cut} and \ref{lemm::pseudoinverse_square_root_C}, we can characterize the Rayleigh ratio involving the matrix $\mathcal{C}$:

\begin{equation}
\frac{\textbf{x}^{\intercal}\mathcal{C}\textbf{x}}{\textbf{x}^{\intercal}\textbf{x}} = 
\begin{cases}
0, & \text{if} \sum_t |X_t||\overline{X}_t| = 0\\
n, & \text{otherwise}.
\end{cases}
\label{eqn::rayleigh_C}
\end{equation}

Upper bounding the eigenvalues of a Laplacian matrix \cite{anderson1985eigenvalues}:
\begin{equation}
0 \leq \frac{\textbf{x}^{\intercal}\mathcal{L}\textbf{x}}{\textbf{x}^{\intercal}\textbf{x}} \leq 2(n+2\beta)
\label{eqn::eig_L}
\end{equation}

Combining \ref{eqn::rayleigh_C} and \ref{eqn::eig_L}:

\begin{equation}
\frac{(\textbf{x}*)^{\intercal}\mathcal{C}(\textbf{x}*)}{(\textbf{x}*)^{\intercal}(\textbf{x}*)} = n
\label{eqn::eign_C}
\end{equation}

as this is the only way to guarantee a strictly positive ratio in Equation \ref{eqn::diff_matrix}. Based on  the spectrum of $\mathcal{C}$:

\begin{equation}
\textbf{x}* = \argmin_{\textbf{x} \in [-1,1]^{nm}, \textbf{x}\mathcal{C}\textbf{x} > 0} \frac{\textbf{x}^{\intercal}\mathcal{L}\textbf{x}}{\textbf{x}^{\intercal}\textbf{x}}
\label{eqn::const_eigen}
\end{equation}

Equations \ref{eqn::relax_ratio} and \ref{eqn::const_eigen} are related to an \textit{eigenvalue problem} and a \textit{generalized eigenvalue problem}, respectively, and can be written as follows:
\begin{equation}
\mathcal{L}\textbf{x} = \lambda \mathcal{C}\textbf{x}, \qquad \mathcal{L}\textbf{y} = \lambda' \textbf{y}
\end{equation}
where $\lambda$ and $\lambda'$ have minimum values and $\textbf{x}\mathcal{C}\textbf{x}, \textbf{y}\mathcal{C}\textbf{y}> 0$. Also, from \ref{eqn::eign_C}, we know that $\mathcal{C}\textbf{y}=n\textbf{y}$. Using Lemma \ref{lemm::C_commutes}:

\begin{equation}
\mathcal{C}\mathcal{L}\textbf{y} = \mathcal{L}\mathcal{C}\textbf{y} = n\mathcal{L}\textbf{y} = \lambda'\mathcal{C}\textbf{y}
\end{equation}

Thus, $\mathcal{L}\textbf{y} = (\lambda'/n)\mathcal{C}\textbf{y}$ is a corresponding solution (same eigenvector) to the generalized problem. This implies that: 
\begin{equation}
\textbf{x}* = \argmin_{\textbf{x} \in [-1,1]^{nm}, \textbf{x}\mathcal{C}\textbf{x} > 0} \frac{\textbf{x}^{\intercal}\mathcal{L}\textbf{x}}{\textbf{x}^{\intercal}\mathcal{C}\textbf{x}}
\end{equation}
\end{proof}

\textbf{Proof of Lemma \ref{lemm::norm_temporal_cut}}
\begin{proof}
The numerator of the ratio is the same as in Lemma \ref{lemm::temporal_cut}. Regarding the denominator:
\begin{equation}
\begin{split}
\textbf{x}^{\intercal}\mathcal{D}^{\frac{1}{2}}\mathcal{C}\mathcal{D}^{\frac{1}{2}}\textbf{x} &= \sum_{t=1}^m \textbf{x}_{(t)}^{\intercal}D_t^{\frac{1}{2}}CD_t^{\frac{1}{2}}\textbf{x}_{(t)}
\nonumber
\end{split}
\end{equation}
where $C = nI-\textbf{1}_{n\times n}$ and $\textbf{x}_{(t)}$ and $D_t$ are the sub-vector and degree matrix, respectively, associated with $G_t$.
\begin{equation}
\begin{split}
\textbf{x}^{\intercal}\mathcal{D}^{\frac{1}{2}}\mathcal{C}\mathcal{D}^{\frac{1}{2}}\textbf{x} = \sum_{t=1}^m \textbf{x}_{(t)}^{\intercal}C\mathcal{D}_t\textbf{x}\\
\nonumber
\end{split}
\end{equation}

As in \cite[Theorem 4]{silva2016graph}, $(\textbf{x}_{(t)}C)[v] = -2|\overline{X}_t|$ if $v\in X_t$ and $(\textbf{x}_{(t)}C)[v] = -2|\overline{X}_t|$, otherwise. Thus,
\begin{equation}
\begin{split}
\textbf{x}^{\intercal}\mathcal{D}^{\frac{1}{2}}\mathcal{C}\mathcal{D}^{\frac{1}{2}}\textbf{x} & = \sum_{t=1}^m \left(|X_t|\sum_{v \in \overline{X}_t}deg(v)+|\overline{X}_t|\sum_{v \in X_t}deg(v)\right)\\
& =\sum_{t=1}^m |X_t|vol(\overline{X}_t) + |\overline{X}_{t}|vol(X_t)
\end{split}
\nonumber
\end{equation}
As a consequence, $\textbf{x}^{\intercal}\mathcal{D}^{\frac{1}{2}}\mathcal{C}\mathcal{D}^{\frac{1}{2}}\textbf{x}$ is maximized when $vol(X_t)$ and $vol(\overline{X}_t)$ are balanced over time. 
\end{proof}

\textbf{Proof of Lemma \ref{lemm::norm_cut_relaxation_one}}
\begin{proof}
The proof is similar to Lemma \ref{lemm::cut_relaxation_one}. By performing the substitution $\textbf{y}=\mathcal{C}^{\frac{1}{2}}\mathcal{D}^{\frac{1}{2}}\textbf{x}$ on the ratio from Lemma \ref{lemm::norm_temporal_cut}, we get Equation \ref{eqn::norm_cut_relaxation_one}. We can also use the fact that $\textbf{y}^{\intercal}(\mathcal{C}^+)^{\frac{1}{2}}(\mathcal{D}^+)^{\frac{1}{2}}$ $\mathcal{L}(\mathcal{D}^+)^{\frac{1}{2}}(\mathcal{C}^+)^{\frac{1}{2}}\textbf{y} = 0$ if $\mathcal{C}\textbf{y} = \textbf{0}_{nm}$ to show that the solution is the ($m$$+$$1$)-th smallest eigenvector.
\end{proof}

\textbf{Proof of Theorem \ref{thm::norm_diff_matrix}}
\begin{proof}
Similar to Theorem \ref{thm::norm_diff_matrix}. Based on the Rayleigh ratio of the matrix $\mathcal{C}$ and the upper bound from Equation \ref{eqn::eig_L}:
\begin{equation}
\textbf{x}* = \argmin_{\textbf{x} \in [-1,1]^{nm}, \textbf{x}\mathcal{C}\textbf{x} > 0} \frac{\textbf{x}^{\intercal}\mathcal{L}\textbf{x}}{\textbf{x}^{\intercal}\textbf{x}}
\label{eqn::const_norm_eigen}
\end{equation}

The ratio from Lemma \ref{lemm::norm_temporal_cut} and Equation \ref{eqn::const_norm_eigen} are related to the following \textit{eigenvalue problem} and \textit{generalized eigenvalue problem}, respectively:
\begin{equation}
\mathcal{L}\textbf{x} = \lambda \mathcal{D}^{\frac{1}{2}}\mathcal{C}\mathcal{D}^{\frac{1}{2}}\textbf{x}, \qquad \mathcal{L}\textbf{y} = \lambda' \mathcal{D}\textbf{y}
\end{equation}

We can also apply Lemma \ref{lemm::C_commutes} to show that:
\begin{equation}
\textbf{x}* = \argmin_{\textbf{x} \in [-1,1]^{nm}, \textbf{x}\mathcal{C}\textbf{x} > 0} \frac{\textbf{x}^{\intercal}\mathcal{L}\textbf{x}}{\textbf{x}^{\intercal}\mathcal{D}^{\frac{1}{2}}\mathcal{C}\mathcal{D}^{\frac{1}{2}}\textbf{x}}
\end{equation}
\end{proof}

\textbf{Proof of Theorem \ref{thm::divide_and_conquer}}
\begin{proof}
We prove the theorem by showing that $\mathcal{M}_S$ and $\mathcal{Q}$ are similar matrices under the change of basis $\mathcal{U}^{\intercal}$ and thus $\mathcal{M} = (\mathcal{U}^{\intercal})^{-1}\mathcal{Q}\mathcal{U}^{\intercal}$. Let's define a matrix $\mathcal{B}$ as follows:
\begin{equation}
\mathcal{B} = \beta \begin{pmatrix} I & -I & 0 & \ldots & 0 \\ -I& 2I & -I& \ldots & 0\\ \vdots &  & \ddots & \ldots & -I\\ 0 & 0 & \ldots & -I& I \end{pmatrix}\nonumber
\end{equation}
Also, let $M_i = 3(n-2\beta)(nI-\textbf{1}_{n\times n})-L_i$. Because $L_i$ is symmetric, $\mathcal{U}^{-1}=\mathcal{U}^{\intercal}$. For an eigenvector matrix $\mathcal{U}$, $\mathcal{U}\mathcal{U}^{\intercal}$ is an $nm\times nm$ identity matrix $\mathcal{I}$. We rewrite $\mathcal{M}_S$ as:
\begin{equation}
\begin{split}
\mathcal{M}_S&= \textbf{diag}(M_1, M_2 \ldots M_m) - \mathcal{B}\\
			&=\mathcal{U}\mathbf{\Lambda}\mathcal{U}^{\intercal} - \mathcal{I}\mathcal{B}\mathcal{I}\\
			&=\mathcal{U}\mathbf{\Lambda}\mathcal{U}^{\intercal} - \mathcal{U}\mathcal{U}^{\intercal}\mathcal{B}\mathcal{U}\mathcal{U}^{\intercal}\\
			&=\mathcal{U}(\mathbf{\Lambda}- \mathcal{U}^{\intercal}\mathcal{B}\mathcal{U})\mathcal{U}^{\intercal}\\
			&=(\mathcal{U}^{\intercal})^{-1}\mathcal{Q}\mathcal{U}^{\intercal} 
\end{split}
\nonumber
\end{equation}
\end{proof}

\textbf{Proof of Lemma \ref{lemm::complement_matrix}}
\begin{proof}
The spectrum of $C$ is $(\textbf{e}_1,\lambda_1)=(\textbf{1}_n,0)$ and $\lambda_2 = \ldots = \lambda_n = n$ for any vector $\textbf{e}_i \perp \textbf{1}_n$. As $M_i$ is also a Laplacian, it follows that $\lambda_1=0$ and $e_1=e_1^L$. Also, by definition $L.e_i^L=\lambda_i^L.e_i^L$, and thus $(3(n+2\beta)C-L)e_i^L=(3(n+2\beta)n-\lambda_i)e_i^L$ for $i>0$.
\end{proof}

\textbf{Proof of Theorem \ref{thm::dyn_wavelet}}
\begin{proof}
Let $\textbf{x}_{(t)}$ be the part of $\textbf{x}$ corresponding to snapshot $t$ and $\mathcal{S}_{(t,h)}$ be the block of $\mathcal{S}$ corresponding to the dissimilarities between vertices in snapshots $t$ and $h$, respectively. We define a vector $\textbf{z} = \mathcal{C}\textbf{x}$. From \cite[Theorem 4]{silva2016graph}, we know that $\textbf{z}_{(t)b}$, the $b$-th position of vector $\textbf{z}_{(t)}$, can take two possible values: $-2|\overline{X}_t|$ if $\textbf{x}_{(t)b}=-1$ and $2|X_t|$, otherwise. Therefore, $\textbf{x}_{(t)}^{\intercal}C\mathcal{S}_{(t,h)}C\textbf{x}_{(h)}$ is equal to the following quadratic form of the matrix $\mathcal{S}_{(t,h)}$:

\begin{equation}
\begin{split}
\textbf{z}_{(t)}^{\intercal}\mathcal{S}_{(t,h)}\textbf{z}_{(h)}& = \sum_{u\in V}\sum_{v\in V} (\textbf{f}[u]-\textbf{f}[v])^2\textbf{z}_{(t)u}\textbf{z}_{(h)v}\\
& = 4\sum_{u\in X_t}\sum_{v\in X_h} (\textbf{f}[u]-\textbf{f}[v])^2|\overline{X}_t||\overline{X}_h|\\
& + 4\sum_{u\in \overline{X}_t}\sum_{v\in \overline{X}_h} (\textbf{f}[u]-\textbf{f}[v])^2|X_t||X_h|\\
& - 4\sum_{u\in \overline{X}_t}\sum_{v\in X_h} (\textbf{f}[u]-\textbf{f}[v])^2|X_t||\overline{X}_h|\\
& - 4\sum_{u\in X_t}\sum_{v\in \overline{X}_h} (\textbf{f}[u]-\textbf{f}[v])^2|\overline{X}_t||X_h|\\
& = -8(|X_t|\sum_{v\in \overline{X}_t}\textbf{f}[v]-|\overline{X}_t|\sum_{u\in X_t}\textbf{f}[u])\\
& \times (|X_h|\sum_{v\in \overline{X}_h}\textbf{f}[v]-|\overline{X}_h|\sum_{u\in X_h}\textbf{f}[u])
\end{split}
\end{equation}
$\textbf{x}^{\intercal}\mathcal{C}\mathcal{S}\mathcal{C}\textbf{x} = \sum_{t=1}^m \textbf{z}_{(t)}^{\intercal}\mathcal{S}_{(t,t)}\textbf{z}_{(t)} + \sum_{t=1}^{m-1}\textbf{z}_{(t)}^{\intercal}\mathcal{S}_{(t,t+1)}\textbf{z}_{(t+1)}$, which gives the numerator of Equation \ref{eqn::dyn_wavelet_energy}. The denominator is as in Lemma \ref{lemm::temporal_cut}.
\end{proof}

\textbf{Performance Results}

\begin{figure*}[ht!]
\centering
\subfloat[\#Vertices \label{fig::perf_syn_data_sparse_vertices}]{
\includegraphics[keepaspectratio, width=0.24\textwidth]{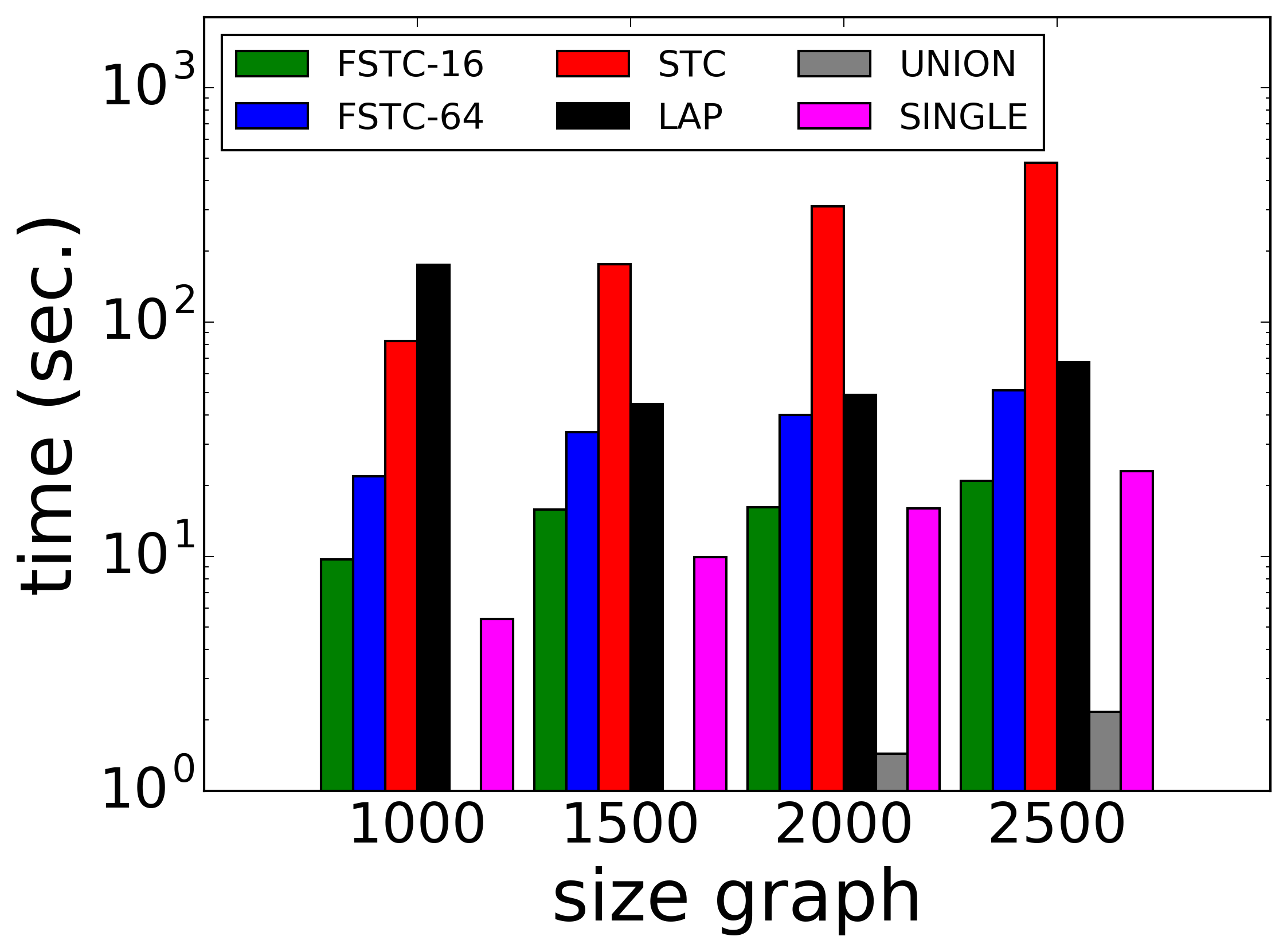}
}
\subfloat[Density]{
\includegraphics[keepaspectratio, width=0.24\textwidth]{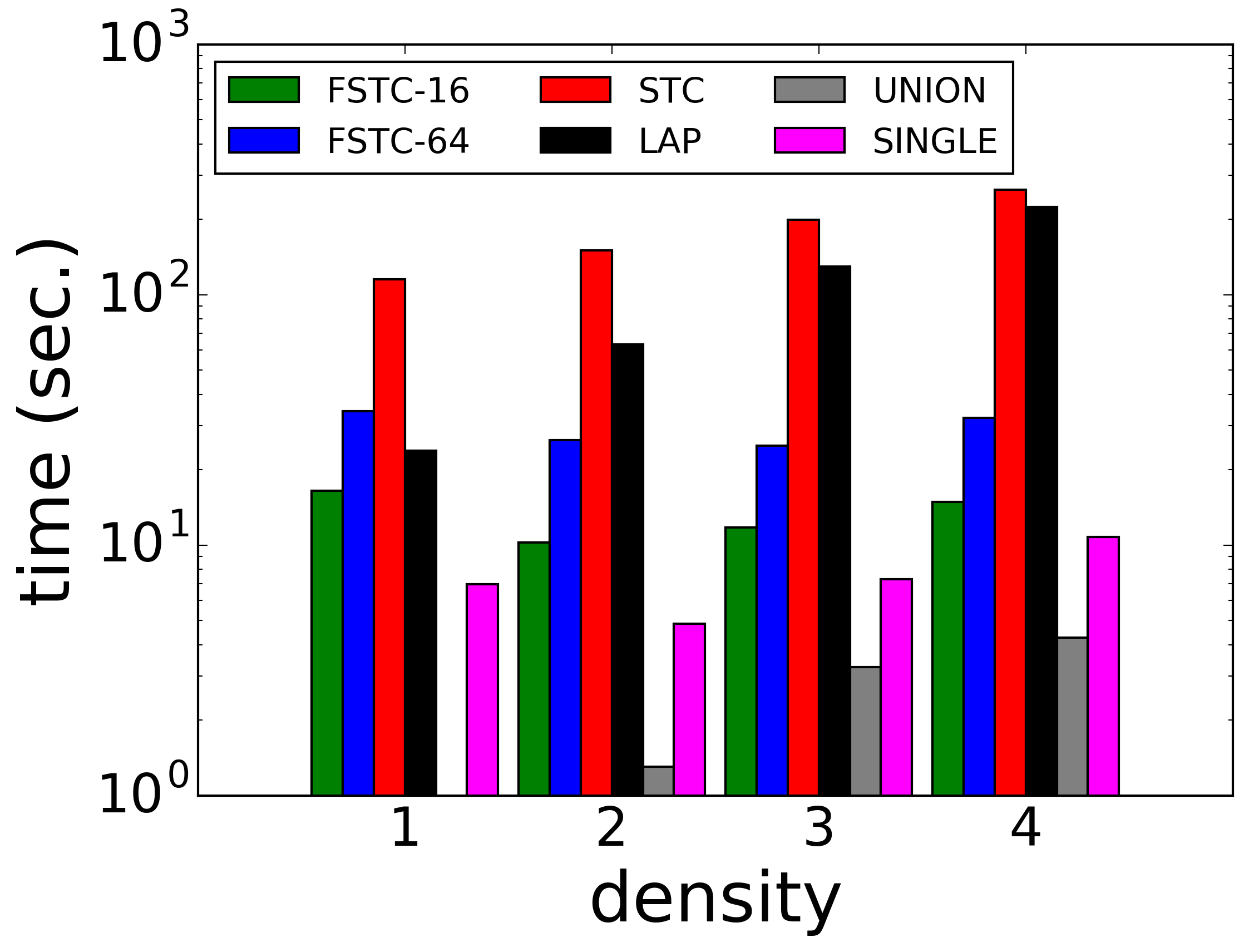}
}
\subfloat[\#Snapshots]{
\includegraphics[keepaspectratio, width=0.24\textwidth]{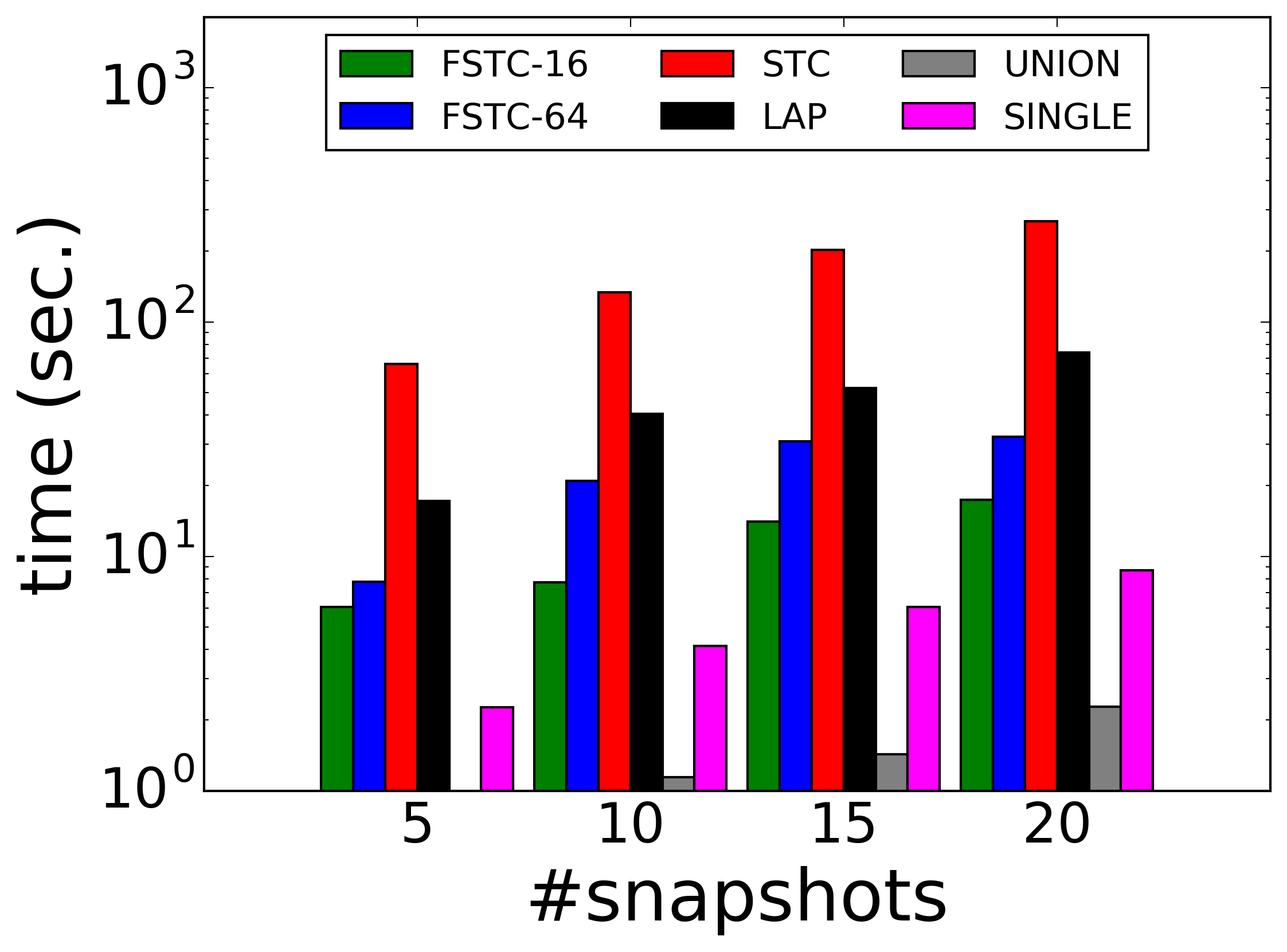}
}
\subfloat[Rank \label{fig::perf_syn_data_sparse_rank}]{
\includegraphics[keepaspectratio, width=0.24\textwidth]{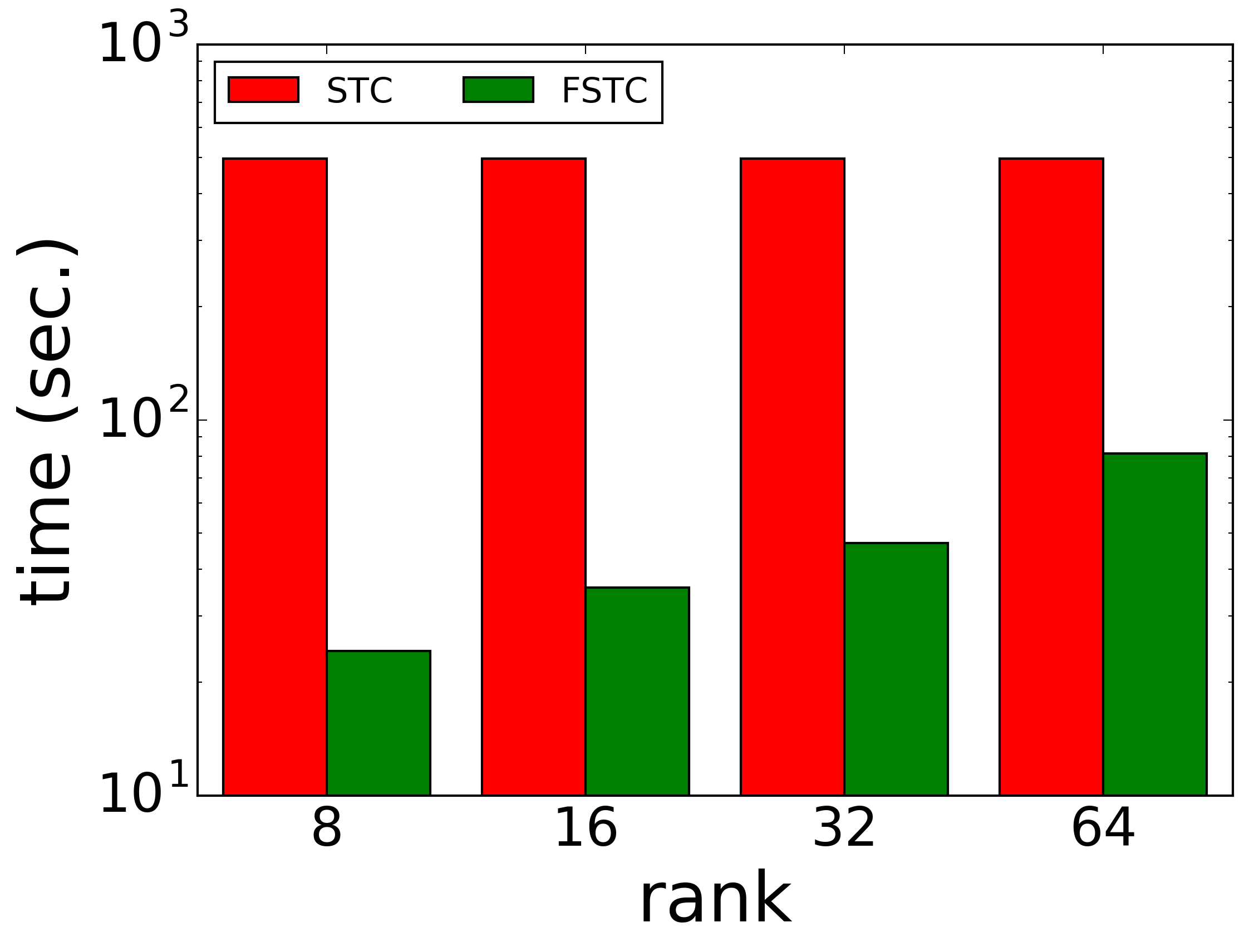}
}

\subfloat[\#Vertices \label{fig::perf_syn_data_norm_vertices}]{
\includegraphics[keepaspectratio, width=0.24\textwidth]{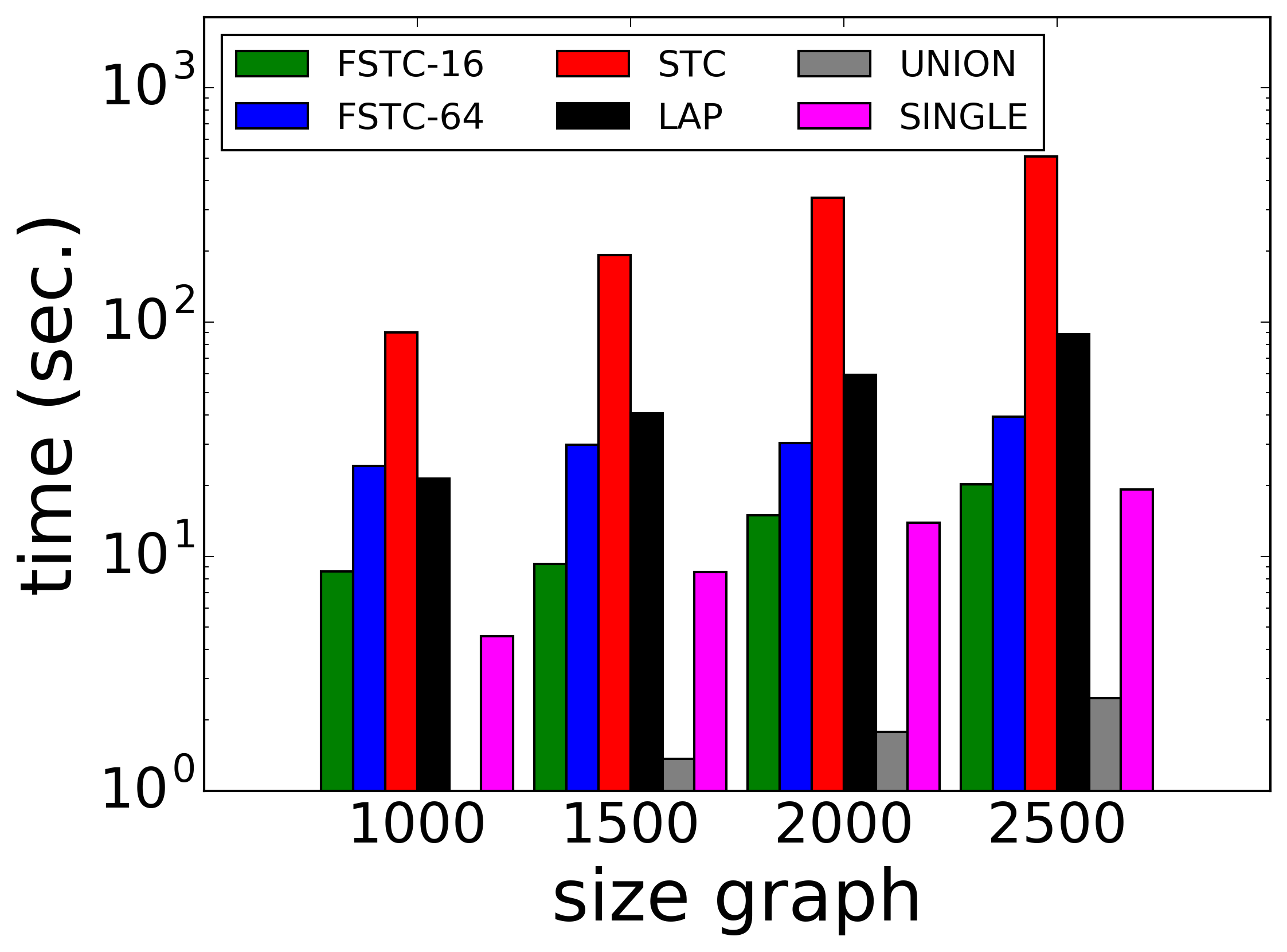}
}
\subfloat[Density]{
\includegraphics[keepaspectratio, width=0.24\textwidth]{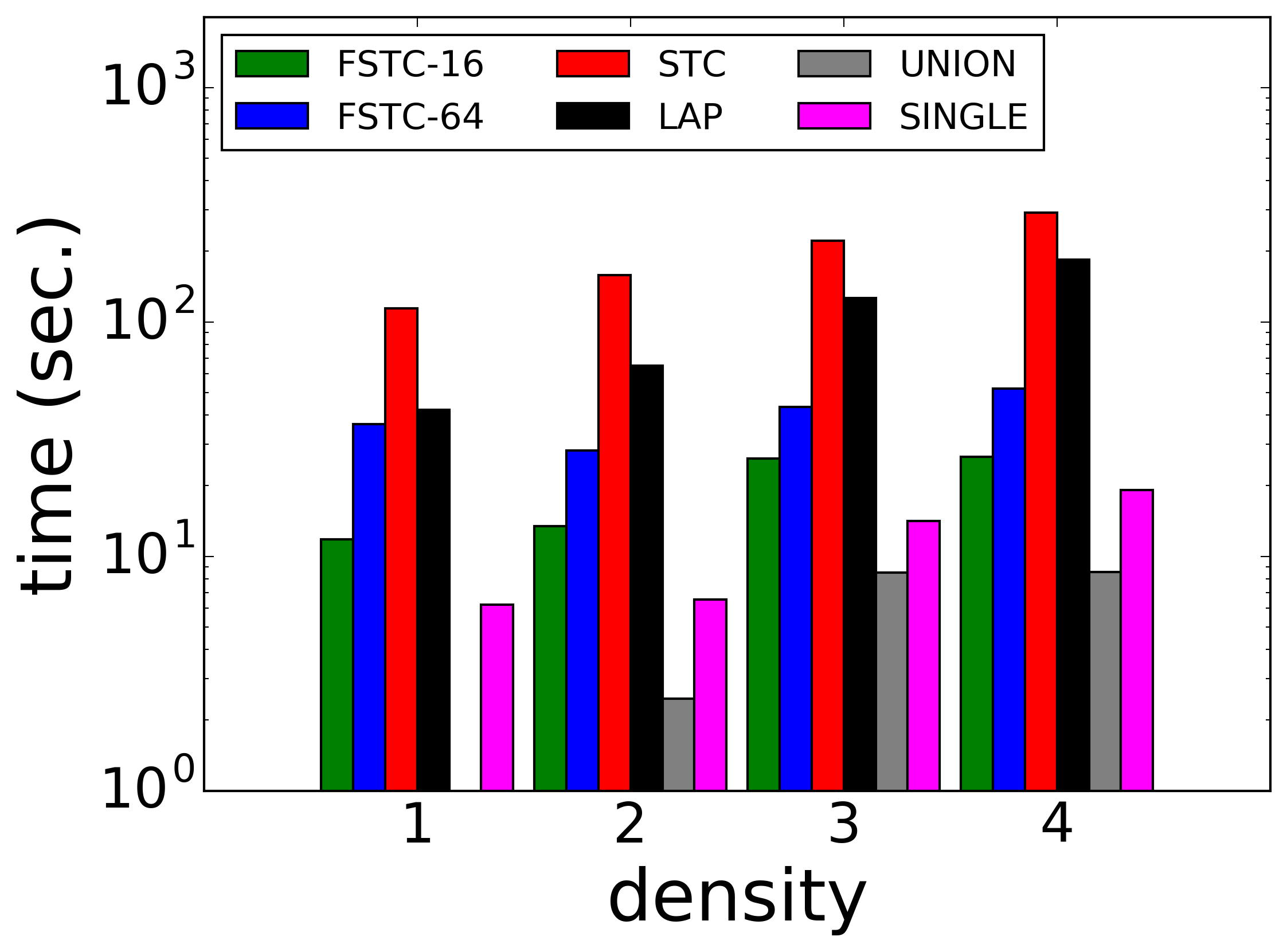}
}
\subfloat[\#Snapshots]{
\includegraphics[keepaspectratio, width=0.24\textwidth]{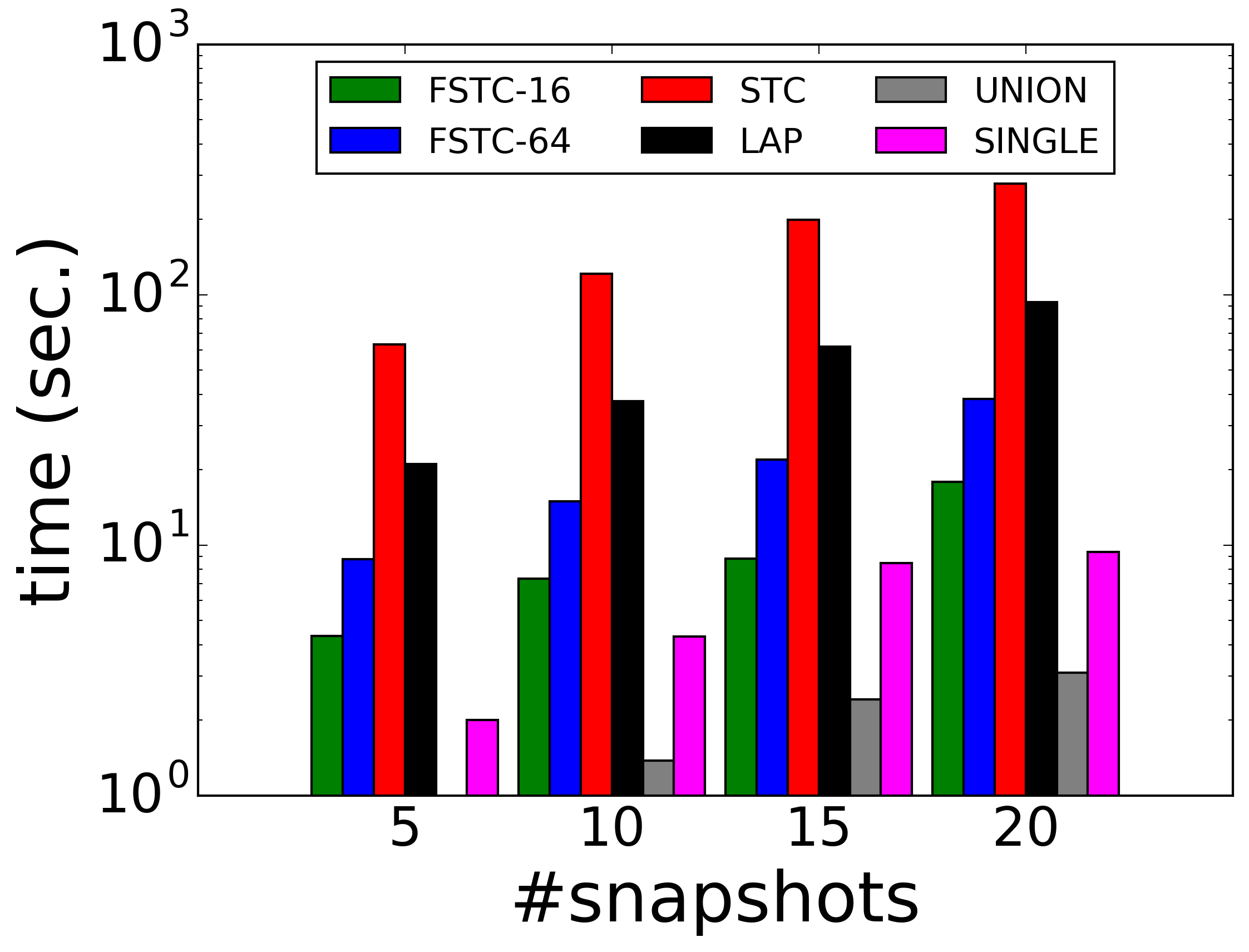}
}
\subfloat[Rank \label{fig::perf_syn_data_norm_rank}]{
\includegraphics[keepaspectratio, width=0.24\textwidth]{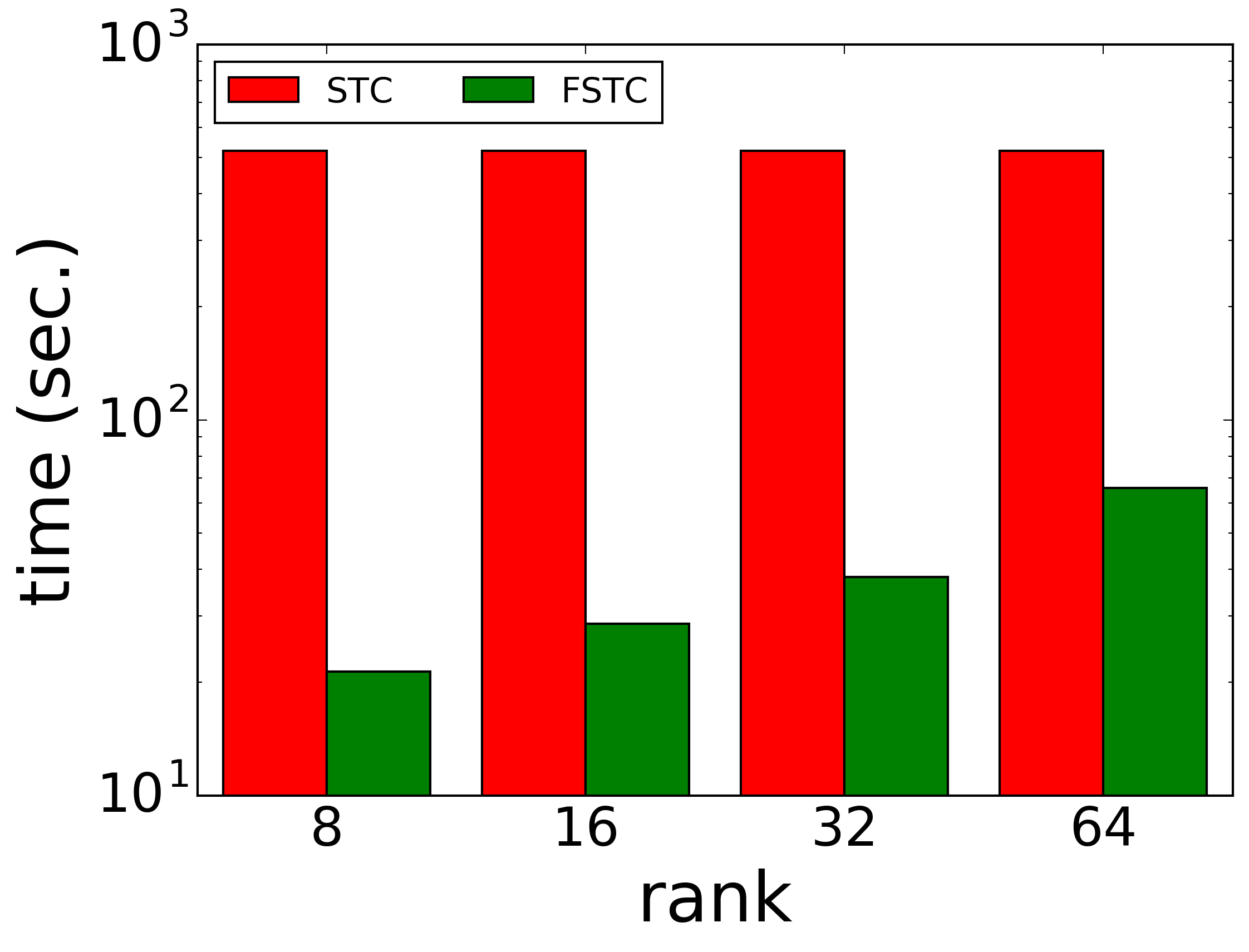}
}
\caption{Running time results for sparsest (a-d) and normalized (e-h) cuts on synthetic data. \label{fig::perf_syn_data}}
\end{figure*}

Figure \ref{fig::perf_syn_data} shows the performance results, in terms of running time, using synthetic data for sparsest (Figure \ref{fig::perf_syn_data_sparse_vertices}-\ref{fig::perf_syn_data_sparse_rank}) and normalized (Figures \ref{fig::perf_syn_data_norm_vertices}-\ref{fig::perf_syn_data_norm_rank}) cuts. We vary the number of vertices, density (or grid connectivity), number of snapshots, and also the rank of FSTC. Similar conclusions can be drawn for both cut problems. \textit{UNION} is the most efficient method, as it operates over an $n\times n$ matrix. On the other hand, \textit{STC} and \textit{LAP}, which operate over $nm\times nm$ matrices, are the most time consuming methods. \textit{STC} is even slower than \textit{LAP}, due to its denser matrix. \textit{SINGLE} and \textit{FSTC} achieve similar performance, with running times close to \textit{UNION}'s times. Figures \ref{fig::perf_syn_data_sparse_rank} and \ref{fig::perf_syn_data_norm_rank} illustrate how the rank $r$ of the matrix approximation performed by \textit{FSTC} enables significant performance gains compared to \textit{STC}. \\

\textbf{Synthetic heat process over the School network}

\begin{figure*}[ht!]
\centering
\subfloat[I]{
\includegraphics[keepaspectratio, width=0.32\textwidth]{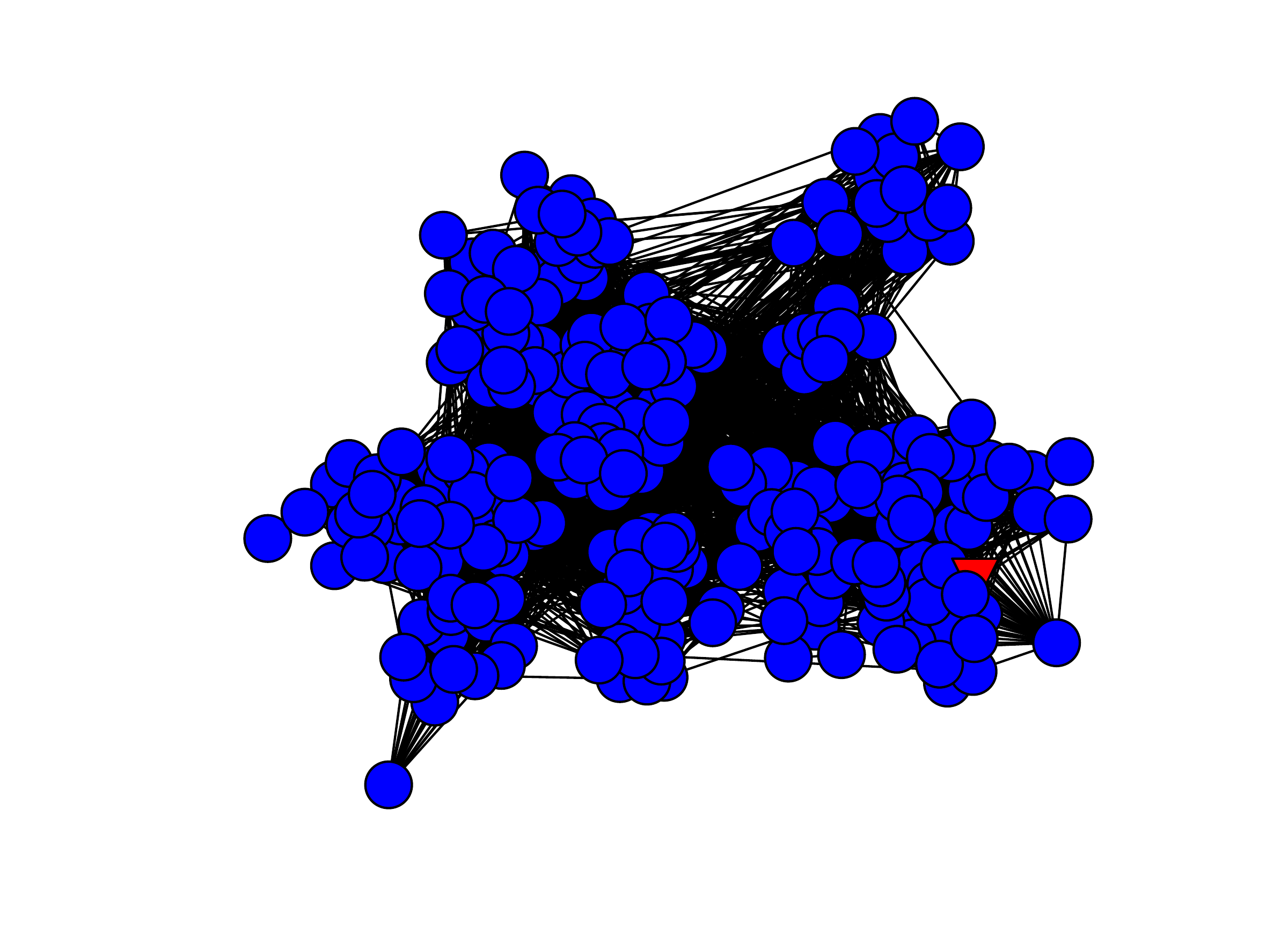}
}
\subfloat[II]{
\includegraphics[keepaspectratio, width=0.32\textwidth]{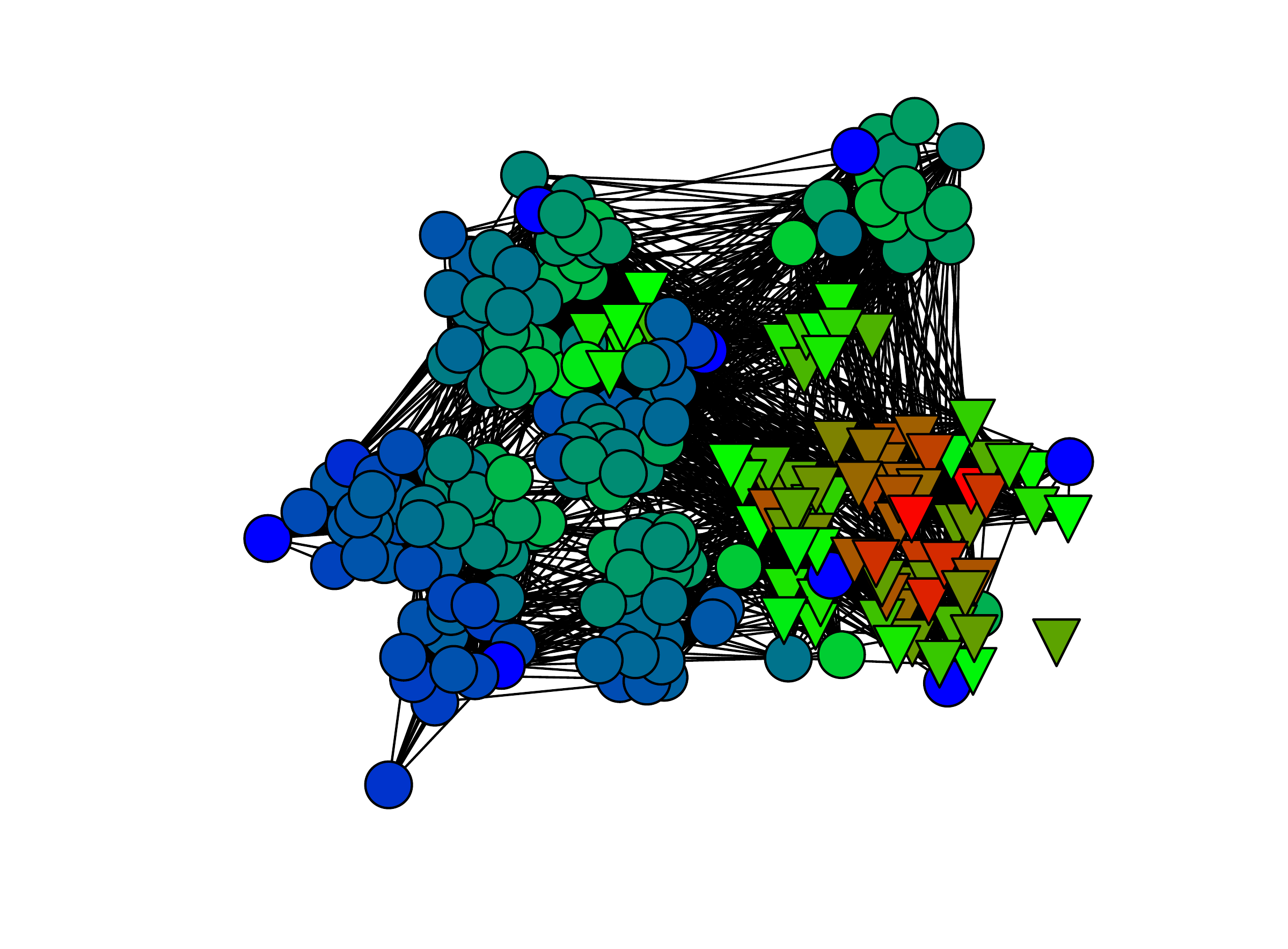}
}
\subfloat[III]{
\includegraphics[keepaspectratio, width=0.32\textwidth]{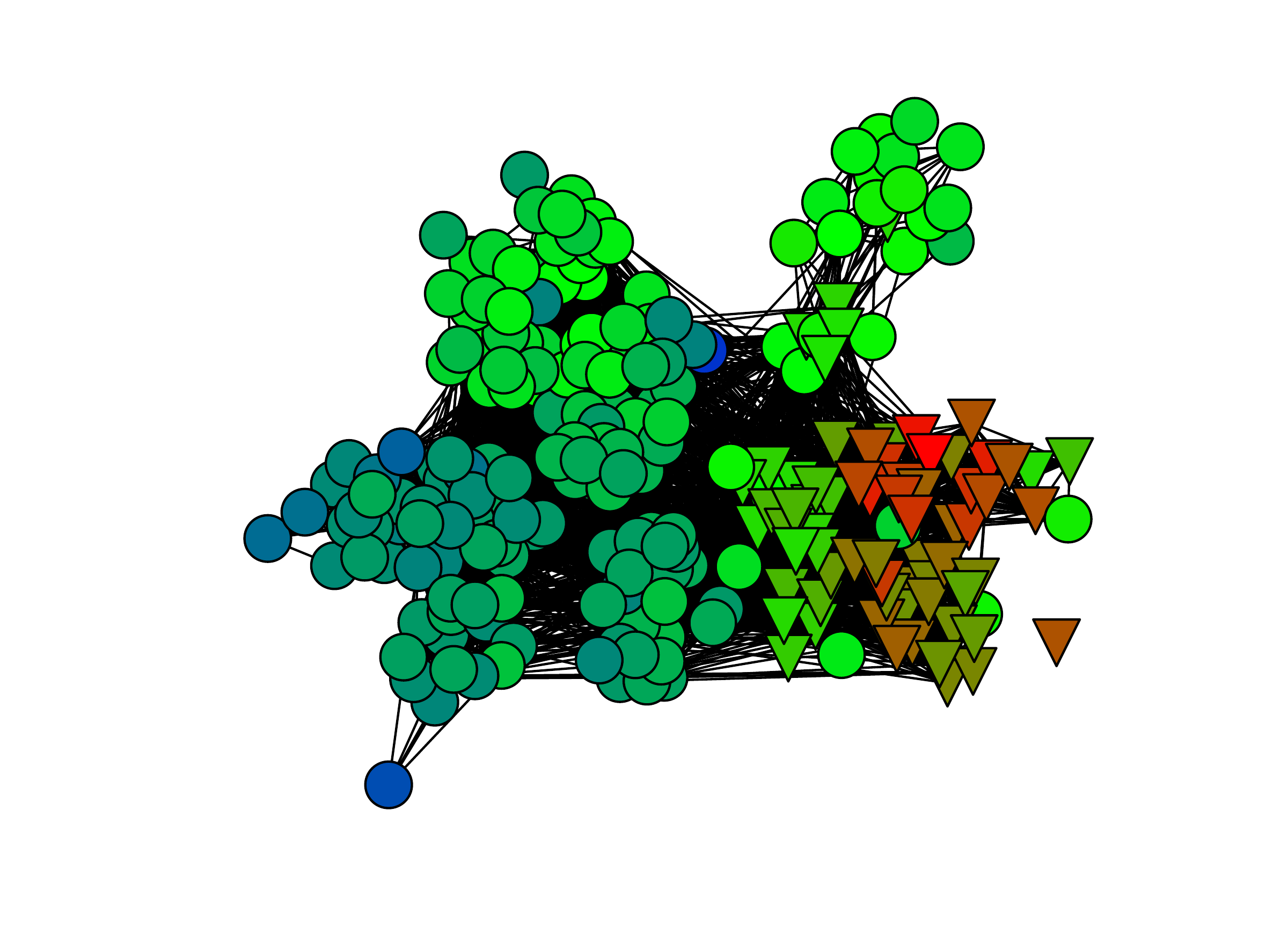}
}
\caption{\textit{School-heat} dataset and associated temporal cut (vertex shapes). The dynamic graph signal is generated based on a heat process over the dynamic \textit{School} network structure. Vertex colors correspond to signal values (red for high and blue for low). \textit{Better seen in color.}\label{fig::school-heat}}
\end{figure*}

Figure \ref{fig::school-heat} shows the \textit{School-heat} dataset. Given Laplacians $L_0, L_1$ and $L_2$, associated with the three snapshots, the initial signal $\textbf{f}_{(0)}$ is set to $|V|$ for an arbitrary starting vertex and to $0$ for the remaining vertices. For $t>0$ the signal is computed using the heat equation:

\begin{equation}
\textbf{f}_{(t)} = e^{-tL_t}\textbf{f}_{(t-1)}
\end{equation}

%Appendix A
%\balancecolumns % GM June 2007
% That's all folks!
\end{document}